\documentclass[prb,letterpaper, twocolumn,showpacs,amsmath,amsfonts,amssymb,preprintnumbers]{revtex4-1}
\pdfoutput=1
\usepackage[T1]{fontenc}\usepackage[latin1]{inputenc}
\usepackage{dcolumn,color,booktabs,microtype,afterpage}
\usepackage[pdftex]{graphicx}
\usepackage[charter]{mathdesign}
\renewcommand{\figurename}{Fig.}
\renewcommand{\tablename}{Table}
\makeatletter\renewcommand{\fnum@figure}[1]{\figurename~\thefigure~(color online).}\makeatother
\makeatletter\renewcommand{\fnum@table}[1]{\tablename~\thetable.}\makeatother
\usepackage[colorlinks,plainpages=false,linkcolor=blue,urlcolor=blue,citecolor=blue,pdfpagemode=UseNone,pdfstartview=FitBH]{hyperref}

\newcommand{ \eg }{\mbox{e.\:g.}}
\newcommand{ \ie }{\mbox{i.\:e.}}
\newcommand{ \etal }{\mbox{et al.}}

\newcommand{\kf }{\mbox{$k_{\rm f}$}}
\newcommand{\ki }{\mbox{$k_{\rm i}$}}

\newcommand{\QQ }{\mbox{$\mathbf Q$}}
\newcommand{\qaf }{\mbox{$\mathbf Q_{\text{AFM}}$}}

\newcommand{\imChiQE }{\mbox{$\chi''(\mathbf Q,\omega)$}}
\newcommand{\imChiQEl }{\mbox{$\chi_{\ell}''(\mathbf Q,\omega)$}}
\newcommand{\imChiQEu }{\mbox{$\chi_{u}''(\mathbf Q,\omega)$}}

\newcommand{\imChiQInt }{\mbox{$\chi''(\omega)$}}

\newcommand { \astar } {\mbox{$a^*$}}
\newcommand { \bstar } {\mbox{$b^*$}}
\newcommand { \cstar } {\mbox{$c^*$}}
\newcommand { \aastar } {\mbox{$\mathbf{a^*}$}}
\newcommand { \bbstar } {\mbox{$\mathbf{b^*}$}}
\newcommand { \ccstar } {\mbox{$\mathbf{c^*}$}}
\newcommand { \deltaIC } {\mbox{$\Delta_{\rm IC}$}}

\newcommand{ \deltaSC } {\mbox{$\Delta_{\rm SC}$}}

\newcommand{\ybco }{\mbox{YBa$_2$Cu$_3$O$_{6+x}$}}

\newcommand{\ybcosix }{\mbox{YBa$_2$Cu$_3$O$_{6}$}}
\newcommand{\ybcosixfour }{\mbox{YBa$_2$Cu$_3$O$_{6.45}$}}

\newcommand{\ybcosixsix }{\mbox{YBa$_2$Cu$_3$O$_{6.6}$}}
\newcommand{\ybcosixeight }{\mbox{YBa$_2$Cu$_3$O$_{6.85}$}}

\newcommand{\lsbco }{\mbox{La$_{2-x}$(Sr,Ba)$_x$CuO$_4$}}

\newcommand{\lsco }{\mbox{La$_{2-x}$Sr$_x$CuO$_4$}}

\newcommand{\bscco }{\mbox{Bi$_2$Sr$_2$CaCu$_2$O$_{8+\delta}$}}

\newcommand{\tc} {\mbox{$T_{\rm c}$}}

\newcommand{\EE } {\mbox{$\omega$}}
\newcommand{\Eres } {\mbox{$\omega_{\rm r}$}}
\newcommand{\ores } {\mbox{$\omega_{\rm r}$}}

\newcommand{\ph} {\mbox{$p_{\rm h}$}}
\newcommand{\kb} {\mbox{$k_{\rm B}$}}
\usepackage{amssymb}
\usepackage{amsmath}

\begin{document}

\title{Neutron scattering study and analytical description of the spin excitation spectrum of twin-free YBa$_2$Cu$_3$O$_{6.6}$}

\author{V. Hinkov}
\email[email: ]{V.Hinkov@fkf.mpg.de}
\affiliation{Max-Planck-Institut für Festk\"orperforschung, Heisenbergstra{\ss}e 1, 70569 Stuttgart, Germany.}

\author{B. Keimer}
\affiliation{Max-Planck-Institut für Festk\"orperforschung, Heisenbergstra{\ss}e 1, 70569 Stuttgart, Germany.}

\author{A. Ivanov}
\affiliation{Institut Laue-Langevin, 6 Rue Jules Horowitz, F-38042 Grenoble cedex 9, France}

\author{P. Bourges}
\affiliation{Laboratoire L\'{e}on Brillouin, CEA-CNRS, CEA-Saclay, 91191 Gif-sur-Yvette, France}

\author{Y. Sidis}
\affiliation{Laboratoire L\'{e}on Brillouin, CEA-CNRS, CEA-Saclay, 91191 Gif-sur-Yvette, France}

\author{C. D. Frost}
\affiliation{ISIS Facility, Rutherford Appleton Laboratory, Chilton, Didcot OX11 0QX, UK}

\begin{abstract}
We present a comprehensive inelastic neutron scattering study of the magnetic excitations in twin-free YBa$_2$Cu$_3$O$_{6.6}$ ($T_\text{c}=61$\,K) for 5\,K\,$\leq T \leq$\,290\,K. Taking full account of the instrumental resolution, we derive analytical model functions for the magnetic susceptibility $\chi''(\mathbf Q,\omega)$ at $T = 5$\,K and 70\,K in absolute units. Our models are supported by previous results on similar samples and are valid at least up to excitation energies of $\omega = 100$ meV. The detailed knowledge of $\chi''(\mathbf Q,\omega)$  permits quantitative comparison to the results of complementary techniques including angle-resolved photoemission spectroscopy (ARPES), as demonstrated in Dahm {\it et al.}, Nature Phys. {\bf 5}, 217, (2009). Based on accurate modeling of the effect of the resolution function on the detected intensity, we determine important intrinsic features of the spin excitation spectrum, with a focus on the differences above and below \tc. In particular, at $T=70$\,K the spectrum exhibits a pronounced twofold in-plane anisotropy at low energies, which evolves towards fourfold rotational symmetry at high energies, and the related dispersion is ``Y''-shaped. At $T= 5$ K, on the other hand, the spectrum develops a continuous, downward-dispersing ``resonant'' mode with weaker in-plane anisotropy. We understand this topology change as arising from the competition between superconductivity and the same electronic liquid-crystal state as observed in \ybcosixfour. We discuss our data in the context of different theoretical scenarios suggested to explain this state.
\end{abstract}

\pacs{\vspace{-0.2em}74.72.Gh 75.40.Gb 78.70.Nx 74.20.Mn \vspace{-0.5em}}

\maketitle

\section{Introduction}

In the recent years, substantial progress has been made in the characterization of fermionic quasiparticles in metallic and superconducting cuprates. While quantum oscillation experiments on both overdoped and underdoped cuprates document the presence of Landau quasiparticles on at least some segments on the Fermi surface, \cite{Doiron-LeyraudTaillefer07, BanguraHussey08, YellandCooper08, HelmGross09} a variety of spectroscopic experiments including angle-resolved photoemission spectroscopy (ARPES), \cite{Borisenko06, ZabolotnyyBorisenko07, ZabolotnyyBorisenko07a, *[{For a review, see }] [{}] DamascelliShen03} scanning tunneling spectroscopy (STS) \cite{LeeDavis06, Jenkins09}, and optical spectroscopy \cite{Hwang07, Heumen09} demonstrate that these quasiparticles are strongly renormalized by interactions with low-energy bosonic modes. A detailed description of this fermion-boson coupling may hold the key to a quantitative understanding of high-temperature superconductivity. However, in order to assess the viability of boson-exchange pairing mechanisms in general \cite{Anderson07, Khatami09} as well as the strength of particular electron-boson pairing channels, the bosonic spectral functions must be determined as accurately as the fermionic ones. For spin excitations and phonons, the bosonic excitations currently under discussion as the main candidates for strong coupling to fermions, this can be accomplished by inelastic neutron scattering. The neutron scattering data can then be cross-correlated with the results of complementary probes that provide direct information on the fermionic excitations.

Here we report a comprehensive inelastic magnetic neutron scattering study of spin excitations in the underdoped high-temperature superconductor \ybcosixsix\ (hole concentration $\ph \sim 0.12$, superconducting transition temperature $\tc=61$\,K), which is suitable as a model compound for several reasons. First, unlike the widely studied Bi$_2$Sr$_2$Ca$_{n-1}$Cu$_{n}$O$_{2(n+2)+\delta}$ system, the crystal structure of \ybcosixsix\ does not include incommensurate lattice modulations. Moreover, its electron system is less affected by chemical disorder than those of many other high-temperature superconductors, as demonstrated by exceptionally narrow nuclear-magnetic-resonance lines,\cite{BobroffAlloul02} and by the fact that quantum oscillations were observed in \ybco\ crystals with somewhat lower doping levels.\cite{Doiron-LeyraudTaillefer07} Crystallographic twin boundaries, a common type of microstructural defect that disturbs the electron system and obscures detection of anisotropies in the CuO$_2$ planes, can be effectively eliminated in \ybcosixsix. Finally, the doping level of \ybcosixsix\ is high enough such that the effect of superconductivity on the spin dynamics is clearly apparent, but low enough to ascertain that the spin excitations are sufficiently intense to allow neutron scattering measurements with high signal-to-background ratio.

A series of prior neutron scattering studies has already provided valuable information about the overall layout of the spin excitation spectrum of \ybcosixsix. \cite{Arai99, Hayden04, HinkovKeimer07} In particular, these studies showed that the spectrum in the superconducting state exhibits the so-called hourglass dispersion in which upward- and downward-dispersing branches meet and form the commensurate resonance peak at the wave vector characterizing antiferromagnetism in the undoped parent compounds. Similar hourglass dispersions were also reported for optimally-doped \ybco,\cite{PailhesBourges03, ReznikIsmer08} and other families of hole-doped cuprates including \lsbco, \cite{ChristensenMcMorrow04, Tranquada04} and \bscco. \cite{FauqueSidis07, XuGu09} However, since the small neutron scattering cross section requires crystals with volumes in the cm$^3$ range, most of the earlier experiments on \ybcosixsix\ were carried out on specimens that contain parasitic phases as well as microcracks and other microstructural defects, and are hence not suitable for measurements with most other experimental methods.\cite{*[{We remark that the high energy scale of the resonance peak and of the superconducting gap \deltaSC\ is a challenge for the most commonly used thermal three-axis spectrometers and is one of the reasons why the present work was not possible before. In the recently discovered iron-arsenide superconductors, the resonance mode is observed between $\sim5$ and $15$\,meV, and one can more conveniently map \imChiQE\ up to $\sim5\deltaSC$ on a thermal spectrometer, see }] [{}] InosovPark10} Since samples grown in different laboratories often vary in their physical properties even if their compositions are nominally identical, this situation has severely limited any quantitative comparison of neutron scattering data to the results of complementary probes. In order to overcome these difficulties, we have performed neutron scattering experiments on mosaics consisting of small \ybcosixsix\ crystals of superior quality. Specimens grown under identical conditions have turned out to be suitable for investigation by a variety of complementary experimental techniques \cite{PailhesBourges03, HinkovKeimer04, Borisenko06, FauqueBourges06, PailhesBourges06, ZabolotnyyBorisenko07, ZabolotnyyBorisenko07a, WaskeHess07, HinkovKeimer07, InosovBorisenko07, HinkovHaug08, WhiteHinkov09, DahmHinkov09, HaugHinkov09, MaitiFink09}.

Following this strategy, we have recently taken the first steps towards a quantitative cross-correlation of neutron scattering and ARPES spectra measured on the same \ybcosixsix\ crystals. \cite{DahmHinkov09} An analytical fitting function for the imaginary part of the wave-vector and energy dependent spin susceptibility, \imChiQE, determined by neutron scattering proved to be an essential ingredient in this study. We expect this approach to be more generally useful for attempts to determine the coupling between spin excitations and fermionic quasiparticles from spectroscopic data, and to asses its role in the formation of the superconducting state. In this article, we hence provide a comprehensive description of the spin dynamics of \ybcosixsix\ at temperatures above and below \tc. We provide a full account of the influence of the instrumental resolution function on the intensity and shape of the features observed in the neutron spectra, and we test various fitting models against the full experimental data set. The outcome is an analytical function for \imChiQE\ in absolute units that reliably reproduces all available experimental data up to at least $\EE \sim 100$ meV.

The second major motivation for our study derives from theoretical work that has been inspired by the hourglass dispersion. The scenarios proposed to describe this unusual excitation mode range from itinerant-electron models in which the hourglass is due to an excitonic collective mode of a $d$-wave superconductor,\cite{YamaseMetzner06, EreminBennemann07} to collective excitations of a variety of ordered states with localized electrons that compete with $d$-wave superconductivity, including site-centered and bond-centered stripes, \cite{AndersenHedegard05, YaoCarlson06, UhrigSchmidt04, VojtaUlbricht04, SeiboldLorenzana05}, spin spirals\cite{Lindgard05, MilsteinSushkov08, PardiniSushkov08, Sushkov09} and models which take into account both itinerant excitations and local spin rotations.\cite{Zeyher10} These competing states break both the translational and the rotational $C_4$ symmetry of the CuO$_2$ square lattice and thus generate Bragg reflections in neutron scattering experiments. Bragg peaks characteristic of incommensurate magnetic order have indeed been observed at low temperatures in \lsbco\ in wide range centered around $x \sim p \sim 1/8$,\cite{Tranquada95, Fujita02, Tranquada05, Chang08} and in deeply underdoped \ybco.\cite{HinkovHaug08} The fact that no such peaks are present in \ybcosixsix\ indicates that for this composition fluctuating stripe\cite{VojtaVojta06} or fully itinerant approaches\cite{Eschrig06, YamaseMetzner06} might be a better starting point.

Recent neutron scattering experiments on untwinned \ybco\ crystals with a somewhat lower doping level ($x = 0.45$, $\ph \sim 0.085$) have provided evidence of a collective ordering phenomenon besides static magnetic order and $d$-wave superconductivity. At this doping level, static magnetic order vanishes upon heating above $\sim 2$ K, but low-energy incommensurate spin fluctuation are observed up to much higher temperatures. Since the incommensurability selects a unique axis in the CuO$_2$ plane and exhibits a sharp onset at $T \sim 150$ K, these data have been interpreted in terms of an ``electronic liquid crystal'' phase that breaks the $C_4$ rotational symmetry down to $C_2$, but conserves the translational symmetry of the lattice.\cite{KivelsonFradkin98} This phase can be viewed from a localized-electron perspective as an array of fluctuating stripes or spirals with a spontaneously selected propagation vector in the CuO$_2$ planes, \cite{KivelsonFradkin98, KivelsonTranquada03, VojtaVojta06} or from an itinerant-electron perspective as a Pomeranchuk state with a spontaneous Fermi-surface deformation.\cite{YamaseKohno00, HalbothMetzner00, YamaseKohno00a, OganesyanKivelson01, JakubczykMetzner09} The small orthorhombic distortion of the \ybco\ crystal structure serves as an aligning field for the propagation vector that permits direct measurements of the in-plane anisotropy with volume-integrating experimental probes.  Strongly temperature dependent in-plane anisotropies of the electrical conductivity \cite{AndoSegawa02,LeeBasov04} and the Nernst effect \cite{DaouTaillefer10} are also consistent with electronic liquid-crystal behavior in \ybco, as are neutron scattering results for strongly underdoped La$_{1.96}$Sr$_{0.04}$CuO$_4$.\cite{*[{See }] [{. In contrast to \ybco, however, the incommensurability in \lsco\ at this doping level, and the orthorhombic distortion, are along the diagonal direction.}] Matsuda08}

In a recent brief communication, \cite{HinkovKeimer07} we had shown that the spin dynamics of \ybcosixsix\ at temperatures above \tc\ also exhibits a strong in-plane anisotropy reminiscent of electronic liquid-crystal behavior, consistent with the observation of strongly anisotropic charge transport at this doping level, and that the crossover to the more isotropic dynamics below \tc\ involves a change in the topology of the spin excitation spectrum. Here we fully document this topology change, with particular focus on the effect of the instrumental resolution function. We expect that the comprehensive description of the spin dynamics above and below \tc, including the analytical fitting functions, will prove useful as a basis for comparison to theoretical work on the spin dynamics of electronic liquid crystals \cite{KeeKimChung03, KaoKee05, VojtaVojta06, KimKivelson08,YamasePRB80, SunKim10, PardiniSushkov08, Sushkov09, *[{For a review, see also }] [{}] Vojta09} and $d$-wave superconductors \cite{DahmTewordt95a, DemlerZhang95, BulutScalapino96, MillisMonien96, AbanovChubukov99, KaoSi00, Onufrieva02, BrinckmannLee02, YamaseMetzner06}, and on the interplay between these two ordering phenomena.

The paper is structured as follows. In section \ref{section:experimentalDetails} we describe the experimental method and technical details. In \ref{subsec:definitons} we define the physical quantities we have measured and the notation we adopt. In \ref{subsec:setup} we discuss the experimental setup, with particular emphasis on aspects relevant for the evaluation of the instrumental resolution. We further explain the choice of the scattering geometry. In \ref{subsec:dataAnalysis} we give details of the data analysis, the background correction, and the way different parts of the data measured under different conditions were brought to the same scale. The characteristic properties of the \ybcosixsix\ sample we have used are summarized in \ref{subsec:sample}. In section \ref{sec:results} we show our experimental results. First, in \ref{subsec:overview} we give a qualitative overview of the most salient features of the spectrum, already apparent in the raw data. Then we present data the below (\ref{subsec:lowE}) and above (\ref{subsec:highE})) the resonance energy $\Eres \sim 38.5$ meV in more detail, both for the superconducting and the normal states. In \ref{subsec:dispersion} we discuss the dispersion relations in both states. In \ref{subsec:model}, finally, we show the \imChiQE\ models which we found to fit best our data in the superconducting and the normal state, respectively. At the end of \ref{subsec:model} we discuss the validity range of our models and compare to the data previously reported. In section \ref{sec:discussion}, we discuss the implications of our data for different proposed states with broken $C_4$ symmetry and the connection to the electronic liquid-crystal state observed in more strongly underdoped \ybcosixfour.

\section{Experimental Method and Technical Details}
\label{section:experimentalDetails}

\begin{figure*}
\begin{minipage}[]{0.33\textwidth}
\textbf{a)}\includegraphics[width=0.8\textwidth]{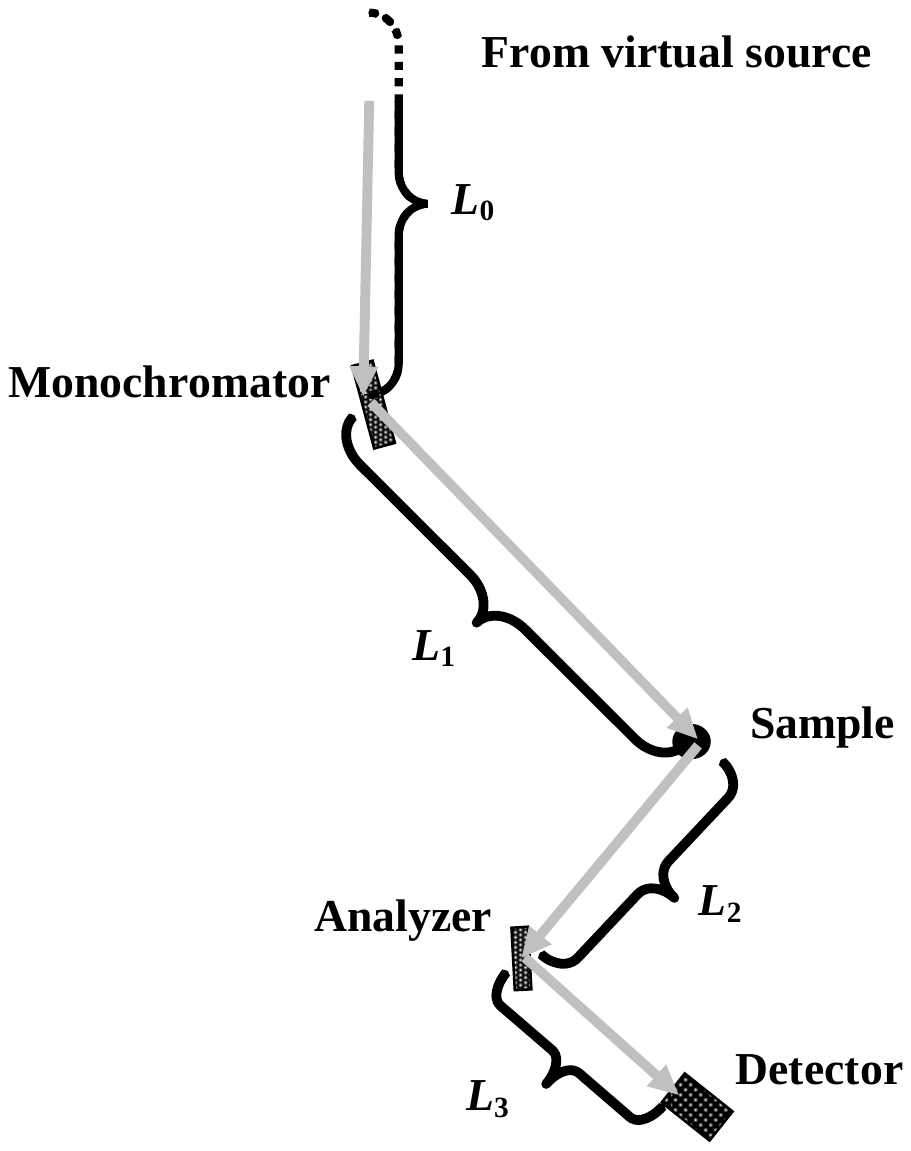}
\end{minipage}
\begin{minipage}[]{0.33\textwidth}
\begin{tabular}{|c|c|c|c|}\hline
\multicolumn{3}{|c|}{$L_0$, dist. virt. source -- mono } & 228 \\\hline
\multicolumn{3}{|c|}{$L_1$, dist. mono -- sample } & 228 \\\hline
\multicolumn{3}{|c|}{$L_2$, dist. sample -- analyzer } & 130 \\\hline
\multicolumn{3}{|c|}{$L_3$, dist. analyzer -- detector} & 65 \\\hline\hline
element & width & height & thickn. \\\hline
source  & 3     & 16     & --      \\\hline
monochr. & 23   & 19.7   & 0.2     \\\hline
analyzer & 23   & 10     & 0.2     \\\hline
detector & 3.5  & 8      & --      \\\hline\hline
\multicolumn{4}{|c|}{monochromator and analyzer:}\\
\multicolumn{4}{|c|}{PG, (002)-reflection, 3.355 \AA}\\\hline
\end{tabular}
\end{minipage}
\begin{minipage}[]{0.31\textwidth}
\textbf{b)}\includegraphics[width=0.99\textwidth]{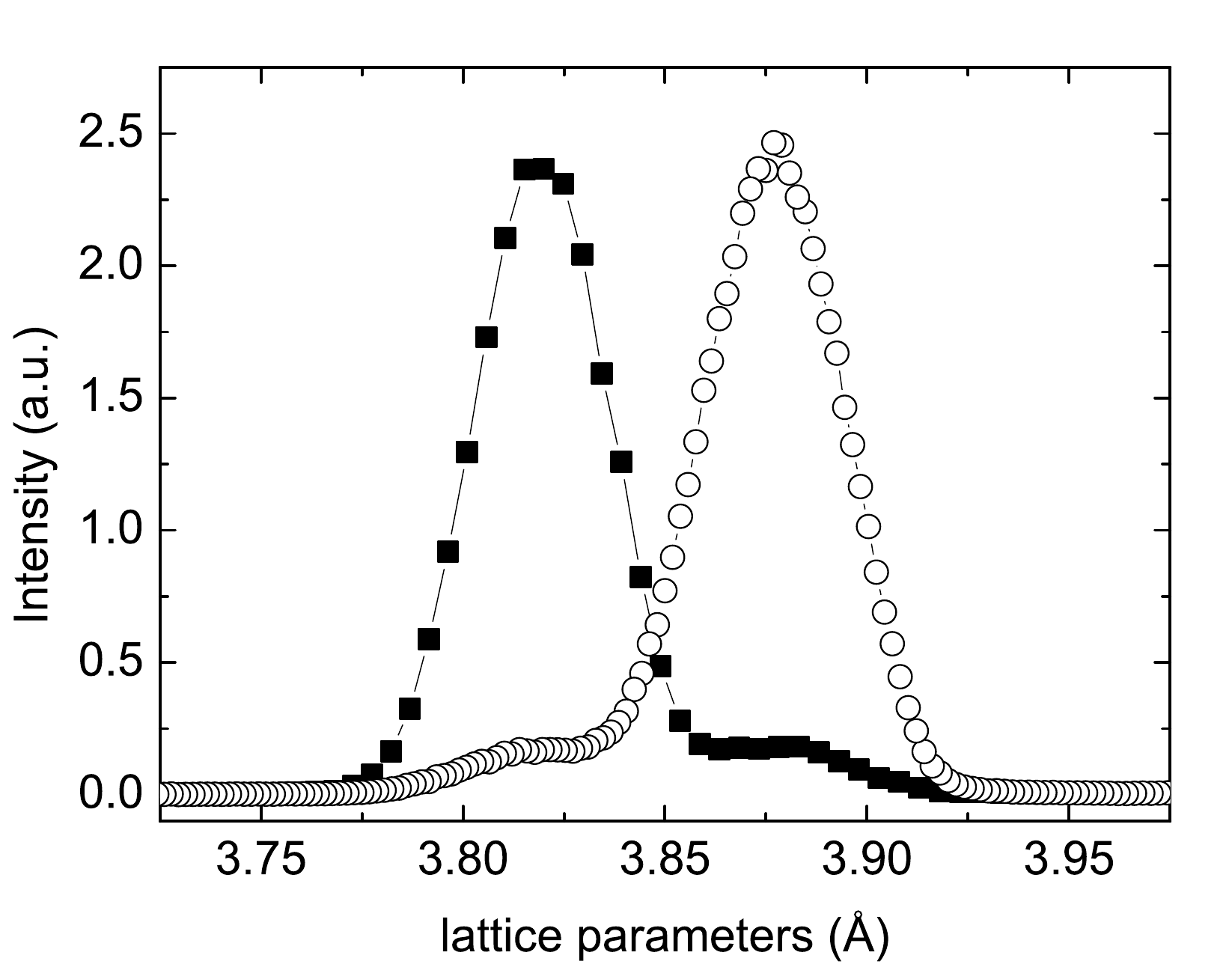}
\end{minipage}
\caption{\label{fig:setup} Experimental setup and sample characteristics. a) The ``W-configuration'' of the IN8 spectrometer at ILL, top view. The angle names follow the ILL nomenclature. The table shows the sizes of the components and distances between them and further setup details. All measures are in cm. The angles $A2$, $A4$ and $A6$ are defined to be zero if the incoming and outgoing beam at the corresponding component are parallel.  The virtual source is defined by an ``empty collimator'' ($w=3 \text{ cm}$, $h=16\text{ cm}$ and $l=18\text{ cm}$), placed between the real source and the monochromator. The distance between the latter two is 810 cm. b) Longitudinal scans through the (200) and (020) structural Bragg reflections demonstrating a twin-domain population ratio of 94:6.}
\end{figure*}

\subsection{Neutron scattering definitions and notation}
\label{subsec:definitons}

The differential neutron scattering cross section is related to the spin-spin correlation function $S(\QQ,\omega)$ in the following way (in the general case of magnetic contributions to the scattering, $S(\QQ,\omega)$  is a tensor):\cite{Squires}
\begin{eqnarray}
\label{eq:crossSection}
\frac{d^2\sigma}{d\Omega\,dE} &  = &  (\gamma r_0)^2\, N \frac{\kf}{\ki}f^2(\QQ)e^{-2W(\scriptsize{\QQ})} \nonumber \\
 & \times & \sum_{\alpha\,\beta}\left( \delta_{\alpha\beta}-\frac{Q_\alpha Q_\beta}{Q^2}\right) S_{\alpha\beta}(\QQ,\omega).
\end{eqnarray}
Here, $r_0$ is the classical electron radius, $\gamma=1.913$, $N$ is the number of spins in the system, $\mathbf{k}_\text{i}$ and $\mathbf{k}_\text{f}$  are the wave vectors of the incident and scattered neutron, $\QQ=\mathbf{k}_\text{f}-\mathbf{k}_\text{i}$, and $\alpha$ and $\beta$ are the components of $S$ and $Q$. Further, we adopt the notation that $\omega$ is the energy \emph{lost} by the neutron (in our notation, $\hbar$ is 1). The Debye-Waller factor $e^{-2W(\scriptsize{\QQ})}$ tends to suppress the intensity at higher $Q$, but within the error it is not important at the temperatures we work and will be neglected. The magnetic form factor $f(\QQ)$, which represents the Fourier transform of the unpaired electrons in orbitals contributing to the scattering (in our case, in the Cu$^{2+}$ $x^2-y^2$ orbital), suppresses the intensity at higher $\QQ$ as well. It is taken into account as described in \ref{subsec:dataAnalysis}.

The imaginary part of the magnetic susceptibility tensor $\chi_{\alpha \beta}''(\QQ, \omega)$ , which defines the response of the system, is the more straight-forward quantity for theoretical considerations. Hence, the model functions derived in \ref{subsec:model} represent this quantity. Susceptibility and correlation function are related by
\begin{equation}
\chi_{\alpha \beta}''(\QQ, \omega) = \pi (g\mu_\text{\tiny{B}})^2 [1-\exp(-\omega\beta)] S_{\alpha \beta}(\QQ,\omega)
\end{equation}
where $\beta=1/\kb T$. At the temperatures for which we develop our analytical models in \ref{subsec:model}, the detailed balance factor $1+n(\omega)\equiv1/[1-\exp(-\omega\beta)] $ can be neglected above $\sim 10$\,meV, \ie\ essentially for all energies we have studied, and therefore, $S\propto\chi''$.

In the absence of magnetic long range order and a preferred spin orientation, $\sum_{\alpha\,\beta}(\delta_{\alpha \beta}-Q_{\alpha}Q_{\beta}/Q^2)\chi''_{\alpha \beta}(\QQ,\omega)=2\,\text{Tr}(\chi''_{\alpha \beta})/3$. Using the isotropic susceptibility $\chi''\equiv\text{Tr}(\chi''_{\alpha \beta})$, the scattering cross section becomes\cite{Squires, FongKeimer00}
\begin{equation}
\frac{d^2\sigma}{d\Omega\,dE} = 2 (\gamma r_0)^2\, N \frac{\kf}{\ki}f^2(\QQ)e^{-2W}\frac{1+n(\omega)}{\pi(g\mu_\text{\tiny{B}})^2}\chi''(\QQ,\omega).
\end{equation}

 Here, we adopt the notation that  $\omega$ is the energy lost by the neutron (we set $\hbar$ to 1), while $\QQ$ is the wave-vector transfer to the neutron. $\QQ=(H,K,L)$ is given in units of the reciprocal lattice vectors \aastar, \bbstar\ and \ccstar\ (r.l.u., as usual omitted in the following), where $a=2\pi/\astar=3.82\,\text{\AA}^{-1}$, $b=2\pi/\bstar=3.875\,\text{\AA}^{-1}$ and $c=2\pi/\cstar=11.73\,\text{\AA}^{-1}$. As in the parent compound \ybcosix, magnetic excitations are centered around the antiferromagnetic wavevector $\qaf=(0.5, 0.5, L)$, and the equivalent points $\left([2h+1]\;0.5, [2k+1]\;0.5, L\right)$ in higher Brillouin zones.

The \ybco\ structure comprises two antiferromagnetically coupled CuO$_2$ layers separated by $d=3.285\,\text{\AA}$ (bilayers), while the distance between adjacent bilayers is much larger and the coupling is three orders of magnitude smaller. Using a Heisenberg model for the parent compound, it can be shown that the intra-bilayer coupling gives rise to acoustic and optical magnetic excitations, whose intensity is modulated by $G_\text{odd}^2$ and $G_\text{even}^2$, where $G_\text{odd}=\sin(\pi L d/c)$ and $G_\text{even}=\cos(\pi L d/c)$ are the acoustic and optical structure factors, respectively:\cite{TranquadaKeimer89}
\begin{equation}
\imChiQE=G^2_\text{odd}\, \chi''_\text{odd}(\QQ,\omega)  + G^2_\text{even}\, \chi''_\text{even}(\QQ,\omega)
\end{equation}
This bilayer modulation persists well into the overdoped regime,\cite{PailhesBourges03, PailhesBourges06} and even excitations exhibit a gap of about $30-40$\,meV in our doping range.\cite{FongKeimer00} In this study we will concentrate exclusively on odd excitations, which peak at $L\sim(2l+1)\times 1.7$. The even channel has been previously studied in some detail,\cite{FongKeimer00, PailhesBourges03, WooDai06, PailhesBourges06} however the determination of an analytical formula describing $\chi''_\text{even}(\QQ,\omega)$ needs to await future investigations.

\subsection{Inelastic neutron scattering setup}
\label{subsec:setup}

The technique of three-axes neutron scattering is described in Ref. \onlinecite{TripleAxisBook}. The signal measured in a scattering experiment is proportional to the differential scattering cross section convoluted with the instrumental resolution function. The resolution function depends in a complicated way on the properties of the instrumental setup.\cite{TripleAxisBook} In the following, we will describe these properties and refer to \ref{subsec:dataAnalysis} for details about the software used to perform the actual resolution analysis.

Most of the results presented here were obtained at the thermal three-axes spectrometer IN8 (Institute Laue-Langevin, Grenoble), while some measurements were done at the thermal spectrometer 2T (Laboratoire L{\'e}on Brilloiun, Gif-sur-Yvette, France) and at the time-of-flight (TOF) spectrometer MAPS (ISIS spallation source, Rutherford Appleton Laboratory, Chilton, UK).

In Fig.~\ref{fig:setup}a) we show a sketch of the IN8 setup, along with characteristics of the components and the distances between them. Both the monochromator and the analyzer consisted of an array of pyrolytic graphite (PG) crystals set for the (002) reflection, which allowed both horizontal and vertical focussing to maximize the neutron flux at the sample. We worked in the constant-\kf\ mode, and a PG filter was put in the path of the scattered beam to extinguish the contamination from higher-order neutrons. This filter is particularly well suited for measurements at $\kf=2.66{\,\text\AA}^{-1}$ and is slightly less efficient at $\kf=4.1$ and $4.5{\,\text\AA}^{-1}$. Further constraints to be taken into account are the kinematics of the neutron scattering process which restricts measurements at low  \kf\ to low energy transfers, and the fact that the resolution is better at low \kf, usually at the expense of a lower neutron intensity. Consequently, measurements were done at $\kf=2.66{\,\text\AA}^{-1}$ up to the resonance energy of $\Eres=38$\,meV which roughly coincides with the kinematically allowed range  (in the given BZ, see below) and also the range of most intense excitations (Fig~\ref{fig:qInt}). Above \Eres, the highest reasonable $\kf$ of $4.5{\,\text\AA}^{-1}$ was used, again dictated by kinematic restrictions at high energies.\footnote{For instance, the highest attainable energy in the BZ around (0.5, 1.5, 1.7) was $\sim60$\,meV due to the onset of parasitic intensity from the direct beam.}

A special feature of IN8 is its virtual-source configuration, employed to maximize neutron flux: The real source is circular, has a large diameter of 20\,cm and is 810\,cm away from the monochromator. In order to use the horizontal focussing mode of the monochromator without degrading the energy resolution, the distances $L_0$ and $L_1$ must be identical (Rowland condition), which is achieved by a slit appropriately placed between real source and monochromator, thus defining the virtual source. In principle, the Rowland condition must also be fulfilled for the horizontally focussing analyzer, however one can choose $L_3$ smaller if the horizontal extent of the detector is large enough to accept all the incoming neutrons of the beam focussed behind the detector plane.

A further requirement to fulfill the Rowland condition is the appropriate choice of the \ki\ (\kf) dependent focussing curvature to ensure that the Bragg angle is constant across the monochromator (analyzer) surface. Usually, the curvature is optimized empirically to obtain the best energy resolution at the instrument. For the resolution calculations (see also \ref{subsec:model}) we will use the following formulae:
\begin{equation*}
\frac{1}{R_\text{\tiny MH}} = \frac{\sin\theta_\text{\tiny M}}{L_1}; \quad
\frac{1}{R_\text{\tiny AH}} = \frac{\sin\theta_\text{\tiny A}}{L_2}
\end{equation*}
where $R_\text{\tiny MH}$ and $R_\text{\tiny AH}$ are the horizontal curvature radii and $\theta_\text{\tiny M}$ and $\theta_\text{\tiny A}$ are the Bragg angles at the monochromator (MH) and analyzer (AH), respectively.

Since to a first approximation the vertical focussing has no impact on the Bragg angle, only the vertical resolution is affected which, as usual, was quite relaxed and optimized for highest flux in our case. For the calculations we use the following formula for the vertical curvature radius:
\begin{equation*}
\frac{1}{R_\text{\tiny V}}=\left(\frac{1}{L_\text{bef}}+\frac{1}{L_\text{aft}} \right)\frac{1}{2 \sin\theta}
\end{equation*}
where $L_\text{bef}=L_0$ or $L_2$, $L_\text{aft}=L_1$ or $L_3$ and $\theta=\theta_\text{\tiny M}$ or $\theta_\text{\tiny A}$ for the monochromator and analyzer, respectively. The vertical analyzer focussing is fixed and optimized for an intermediate \kf\ of $\sim3.3\,\text{\AA}^{-1}$ which was taken into account in the resolution calculations.

Since we are interested in the spin excitations in the CuO$_2$ planes and hence need a high resolution there, it seems optimal to work in a configuration with the \astar- and \bstar-directions are in the horizontal plane. However, since we are probing excitations in the odd channel, whose intensity disappears at $L=0$, and its first maximum appears for $L\sim1.7$, see \ref{subsec:definitons}. Taking into account the further constraints -- to keep $Q$ small to gain intensity due to the Cu$^{2+}$ magnetic form factor and to put \ccstar\ as close to vertical as possible -- we measured most scans along \aastar\ and \bbstar\ around $\qaf=(0.5,-1.5,-1.7)$ and $(1.5,0.5,1.7)$, respectively. This entails a scattering plane in which the \ccstar-axis is tilted by $\sim19^\circ$ from the vertical direction.

An important objective of this study is to see how the excitations evolve around \qaf\ in the two distinct directions \aastar\ and \bbstar. Thus, to eliminate any remaining uncertainties related to details of the resolution calculations, we decided to perform the scans in these two directions under identical resolution conditions. This was achieved by performing the \aastar- and \bbstar-scans around the two equivalent  \qaf\ indicated above. This change of scattering planes corresponds to a rotation of the resolution ellipsoid by $90^\circ$ w.\,r.\,t. the $a$-$b$-plane. To avoid the necessity of taking the sample out of the cryostat for this change, we developed a special holder with two cradles which allowed us to ``pre-tilt'' the sample by $\sim10^\circ$ around the two perpendicular directions \aastar\ and \bbstar, respectively, before mounting it into a standard orange cryostat. The cryostat itself, fixed to the sample table,  can only be tilted by about $\pm10^\circ$ around the same directions -- our special holder effectively shifts the zero by $10^\circ$ and the accessible tilting range to $\sim0^\circ-20^\circ$ for both directions. This allowed us to access the necessary angle of $19^\circ$ and change between the \aastar- and \bbstar-configuration in a simple way.

\subsection{Data analysis}
\label{subsec:dataAnalysis}
At our doping level, the spectral weight of the spin excitations and the signal-to-background ratio are relatively large, so that at most energies a subtraction of a linear background provides sufficient correction (raw data are shown in Fig.~\ref{fig:raw}). At  energies  $\sim30$\,meV, there are sometimes spurious peaks, which, however, turned out to be temperature-independent in most cases. Since at these energies the spin excitations are strongly overdamped close to room temperature (see also Fig.~\ref{fig:raw}), the magnetic contribution to the neutron-scattering cross section is best represented by the data from which a background scan obtained at $T=250$\,K has been subtracted. For the sake of consistency, in several cases we perform this correction up to $\Eres=38$\,meV, \eg\ in the color maps of Figs. \ref{fig:mapsConv} and \ref{fig:mapsAlt}, or in the fits of Fig.~\ref{fig:allFits}. Naturally, due to the smooth background there (Fig.~\ref{fig:raw}e--h) this has no consequences for our analysis.

As mentioned, the energy ranges below and above \Eres\ were measured with different \kf\ and therefore under different resolution conditions. Furthermore, kinematical constraints forced us to measure scans at $\EE\geq70$\,meV in higher Brillouin zones (Fig.~\ref{fig:highE}) where the magnetic intensity is suppressed due to the Cu$^{2+}$ form factor (which is anisotropic\cite{ShamotoTranquada93, CasaltaWolf94, Walters09} and only known with limited precision). For a quantitative analysis, it is thus important to put measurements done under such different conditions onto the same intensity scale. To this end, we measured the resonance peak at $\Eres=38$\,meV with all \kf\ and in all Brillouin zones used in these experiments. We then fitted the scan profiles to the damped harmonic oscillator (DHO) model of the hourglass dispersion discussed in \ref{subsec:model}, taking full account of the instrumental resolution, and extracted corresponding scaling factors. This allowed us to obtain a very satisfactory fit of the measured scan profiles at all energies to the same DHO-model with the \emph{same}, globally valid parameter values (Figs. \ref{fig:allFits} and \ref{fig:allFits2}). This, and the fact that the energy evolution of the \QQ-integrated susceptibility of the model agrees well with the evolution of the \QQ-integrated susceptibility obtained directly from the data (Fig.~\ref{fig:qIntModel}) lends further support to the validity of our approach.

We used the program \emph{Rescal5}\footnote{Rescal5 is available on the web page of the ``Institut Laue-Langevin'', www.ill.eu.} to calculate the resolution function and convolute it with the model functions we develop in \ref{subsec:model}. Our resolution calculations were based on the Popovici method\cite{Popovici75} which takes full account of geometrical properties of the instrument components (sizes and distances). Since the program had not been used for setups with a virtual source before, we compared the obtained resolution function with the results of the program RESTRAX5.0.5\footnote{A description of the program can be found on the web site http://neutron.ujf.cas.cz/restrax.} which uses ray-tracing methods and is thus more flexible in implementing unusual geometries. The discrepancy between the two programs concerning the resolution width was below $10\%$ and justifies the usage of \emph{Rescal5}, which allows the import of data, definition of parameters and other things in a more transparent way.

\subsection{Characteristics of the \ybcosixsix\ sample}
\label{subsec:sample}
The sample on which the experiments were carried out is an array consisting of 180 co-aligned \ybcosixsix\ single crystals, with a total volume of $\sim450$\,mm$^3$ and a total mosaicity of less than $1.2^{\circ}$.  The crystals were detwinned by subjecting them to a uniaxial mechanical stress of $\sim 5 \times 10^7$\,N\,m$^{-2}$ along a $\langle 100 \rangle$ direction at $400\,^{\circ}$C in argon for at least two hours.\cite{LinBender92, Voronkova93} After detwinning, they exhibit superconducting transition temperatures (midpoint) of $\tc\sim61$\,K and transition widths $\Delta\tc\sim2$\,K, as determined by magnetometry for each crystal. A comparison of the area of the main (200) and (020) Bragg reflections to the area of the respective satellites due to the minority twin domains yields a twin-domain population ratio of 94:6 for the entire sample, Fig.~\ref{fig:setup}b). The doping level is $\sim0.12$ holes per planar copper.\cite{LiangBonn06, TallonBernhard95} Further preparation details are given in Refs. \onlinecite{HinkovKeimer04, HinkovKeimer07}.

\section{Results}
\label{sec:results}

\begin{figure}[t]
\includegraphics[width=0.85\columnwidth]{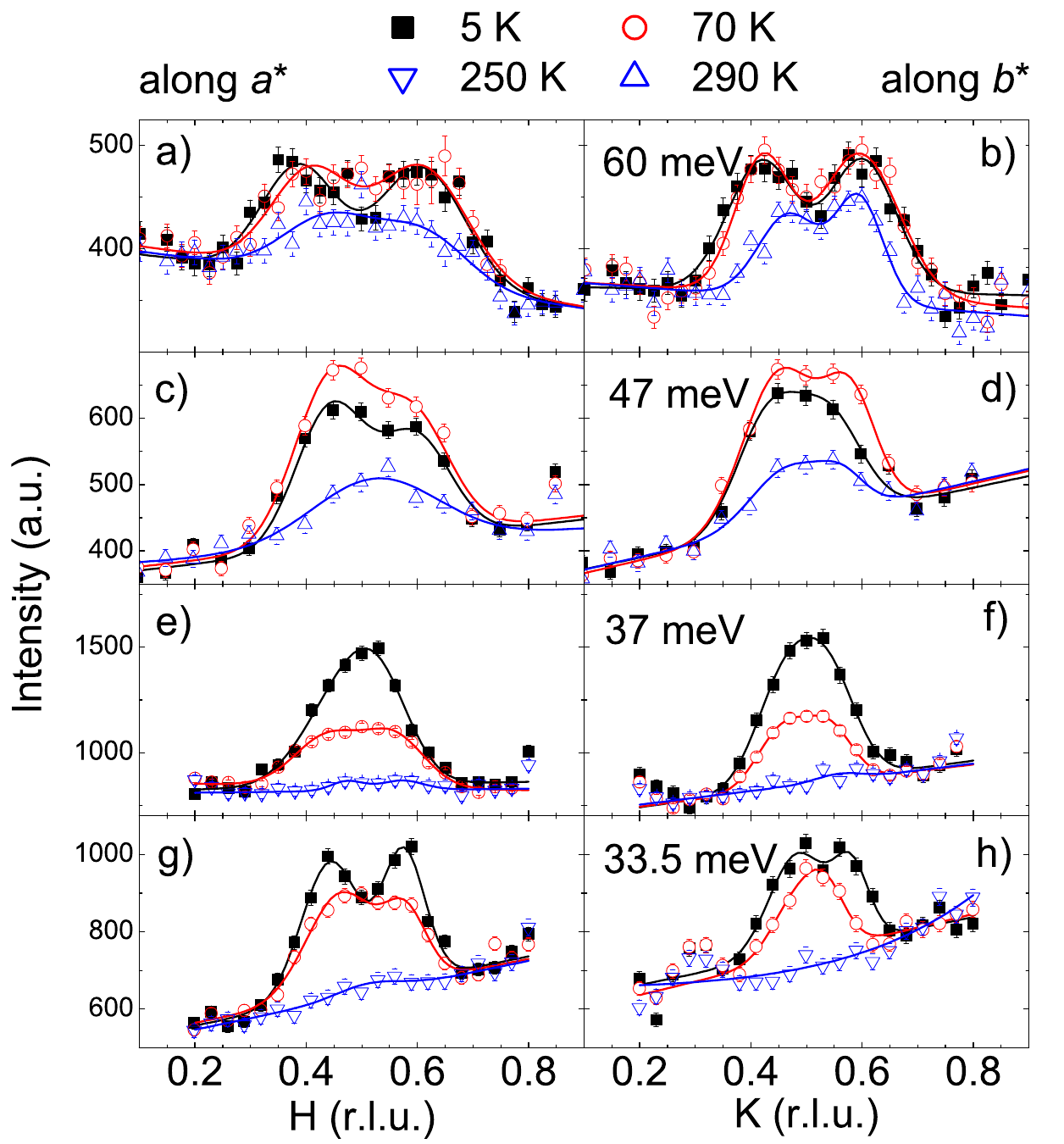}
\caption{\label{fig:raw} Energy evolution of the in-plane spin excitations around \qaf\ at different temperatures. The raw data are shown at 5\,K and 70\,K, while a constant BG has been subtracted at 250\,K and 290\,K to take into account the increase of multi-phonon scattering. Corrections for the Bose factor are small and were not applied. The final wave-vector \kf\ was fixed to $2.66\mbox{ \AA}^{-1}$ for $\EE\leq37$\,meV and to $4.5\mbox{\AA}^{-1}$ at higher energies. Scans along \astar\ and along \bstar\ were performed under identical resolution conditions. The data were previously shown in Ref. \onlinecite{HinkovKeimer07}.  \vspace{-1.3em}}
\end{figure}

\begin{figure}[b]
\begin{center}
\begin{tabular}{cc}
\includegraphics[width=0.42\columnwidth]{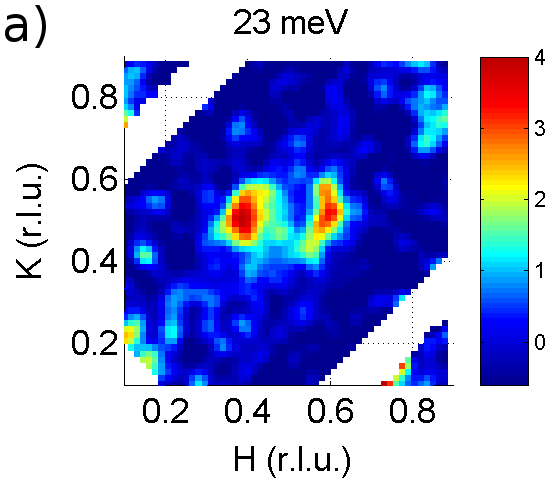}
&\includegraphics[width=0.42\columnwidth]{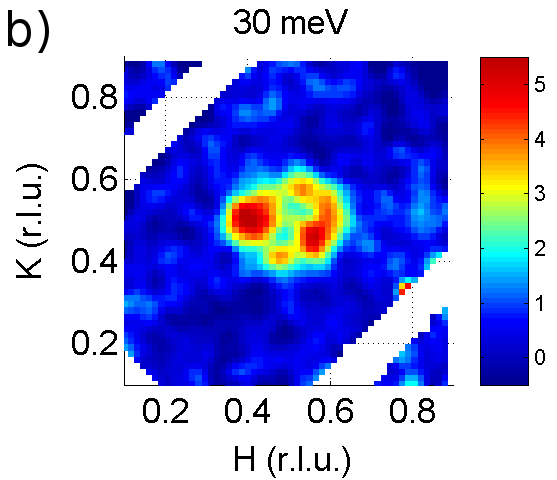}\\
\includegraphics[width=0.45\columnwidth]{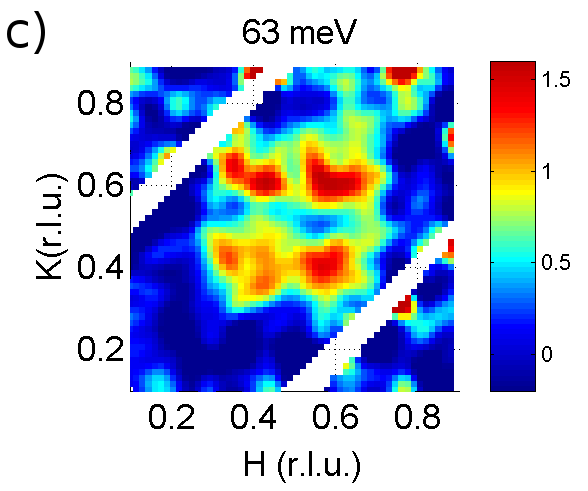}
&\includegraphics[width=0.45\columnwidth]{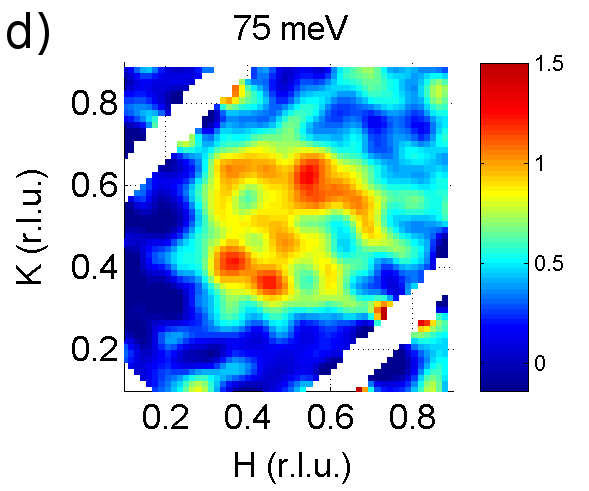}\\
\end{tabular}
\end{center}\vspace{-15pt}
\caption{\label{fig:TOF} Color representation of the in-plane magnetic intensity collected at the TOF spectrometer MAPS at ISIS for different energy transfers. The temperature was $3\mbox{ K}$. Intensity was integrated over an energy range of $\pm3.5$\,meV in a) and b), $\pm5$\,meV in c) and $\pm7.5$\,meV in d). The data represent acoustic excitations and were collected  at $L=1.6$ in a), at $2.1$ in  b) and at $4.7$ in c) and d), \ie\ close to one of the maxima at $L=1.7$ and 5.1. The incident energy was fixed to 120\,meV in a) -- c) and to 160\,meV in d). The source proton current was 170 $\mu$A and the Fermi-chopper rotation frequency was set to 250 Hz. A background quadratic in $Q$ was subtracted, and data are shown in arbitrary units.}
\end{figure}

This section is structured as follows: In \ref{subsec:overview} we give a brief qualitative overview of the most salient properties of the spectrum, already apparent in the raw three-axes (Fig.~\ref{fig:raw}) and time-of-flight (TOF) data (Fig.~\ref{fig:TOF}), and the energy evolution of the \QQ-integrated intensity (Fig.~\ref{fig:qInt}). In \ref{subsec:lowE} and \ref{subsec:highE} we discuss in more detail the in-plane anisotropy and temperature evolution at low energies ($\leq\Eres$, Figs. \ref{fig:lowE} and \ref{fig:tempDep}) and high energies ($>\Eres$, Fig.~\ref{fig:highE}), respectively. In \ref{subsec:dispersion}, we first describe a novel way to visualize the dispersion of data with strongly energy-dependent intensity in energy-momentum maps (Figs. \ref{fig:mapsConv} and \ref{fig:mapsAlt}). We then compare the dispersion topology below and above \tc\ both using these maps and a more conventional representation, namely plots of the energy evolution of the incommensurabilty \deltaIC, as obtained from Gaussian fits to the scan profiles (Fig.~\ref{fig:dispersion} and \ref{fig:dispDHO}). Based on this comparison and on a careful evaluation of the data in the vicinity of \Eres, we show that the hourglass dispersion with commensurate ``neck'' observed in the superconducting state is replaced by a Y-shaped dispersion: The latter is very steep and quasi one-dimensional below $\sim\Eres$ (Fig.~\ref{fig:topologyChange}), and exhibits an incommensurate geometry at any energy, including \Eres. Finally, in \ref{subsec:model} we develop empirical, analytical model functions for $\imChiQE$ based on a DHO model at $T=5$\,K, \ref{subsubsec:scstate}, and a combination of a DHO model and Gaussian profiles at $T=70$\,K, \ref{subsubsec:normalstate}. Despite the small number of free parameters, the models exhibit all the spectral features described in \ref{subsec:overview} -- \ref{subsec:dispersion}  and provide a very good fit to the experimental data (Figs. \ref{fig:qIntModel} -- \ref{fig:allFits2}), with the same parameter values valid at all energies.

\subsection{Overview}
\label{subsec:overview}
\begin{figure}[t]
\includegraphics[width=0.7\columnwidth]{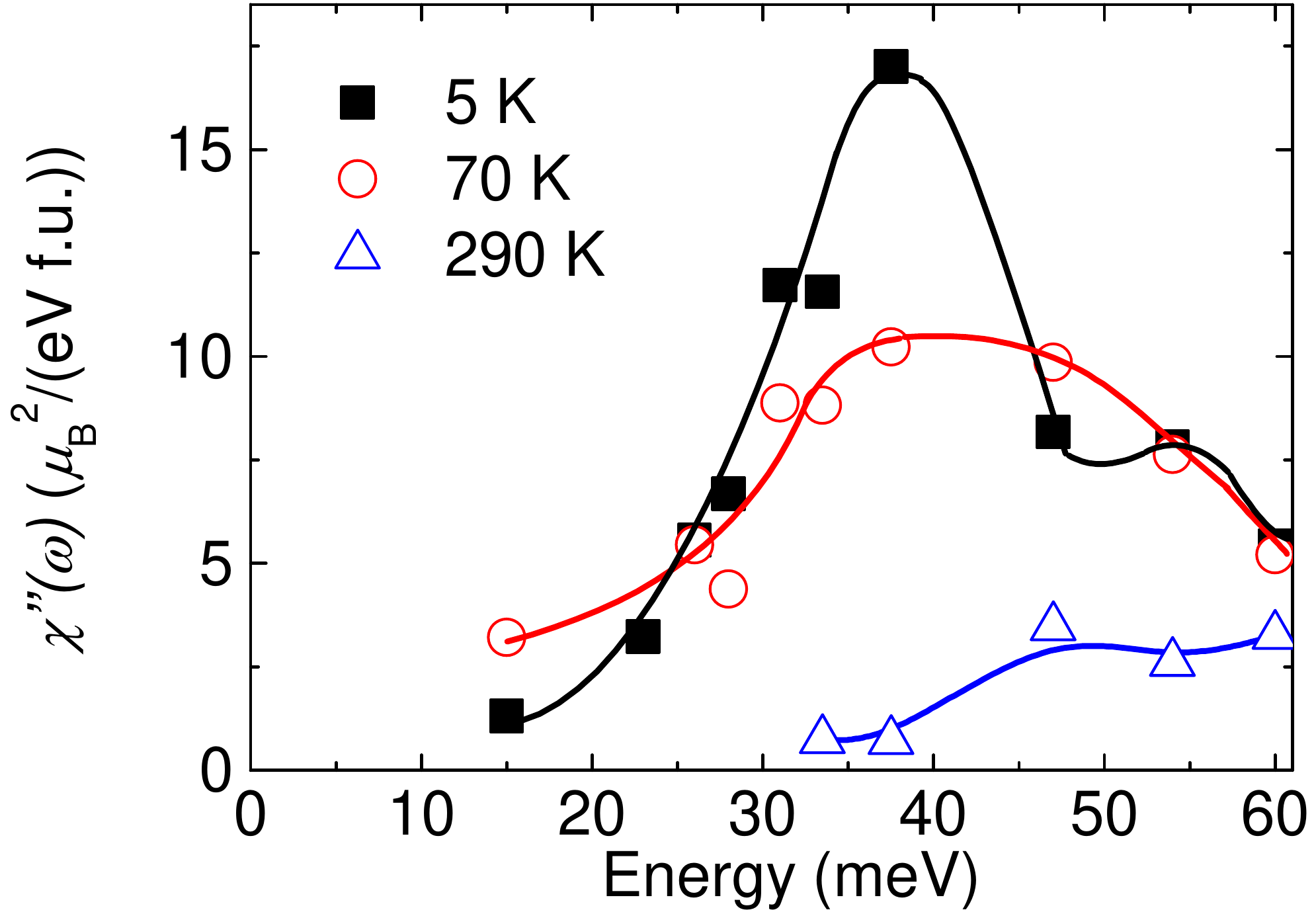}
\caption{\label{fig:qInt} Imaginary part of the \QQ-integrated (local) susceptibility as a function of the excitation energy at 5\,K, 70\,K and 290\,K.}
\end{figure}
In Fig.~\ref{fig:raw} we show unprocessed (``raw'') scan profiles for different \EE\ throughout the covered energy range, measured along the two principal directions \aastar\ and \bbstar. Several characteristic aspects of the data which we will investigate in quantitative detail later on are visible here on a qualitative level: Beginning deep in the superconducting state at $T=5$\,K, we observe, as previously reported, that the incommensurate (IC) branches at high\cite{BourgesFong00, Arai99, Hayden04} and low\cite{DaiMook98, MookDai98} energies merge to a single, commensurate resonance peak at $\Eres\sim38$\,meV. The resonance peak,\cite{*[{For a review, see }] [{}] SidisBourges07} originally discovered in \ybco,\cite{Rossat-MignodBourges91} was later also found in bilayer \bscco,\cite{FongKeimer99} single-layer Tl$_2$Ba$_2$CuO$_{6+\delta}$ (Ref. \onlinecite{HeBourges02}) and HgBa$_2$CuO$_{4+\delta}$,\cite{YuGreven10} and in electron-doped cuprates.\cite{ZhaoDai07, YuGreven08} The low- and high-energy branches, together with the resonance peak, form the ``hourglass'' dispersion with commensurate neck.\cite{Arai99, BourgesFong00, Hayden04, Tranquada04, HinkovKeimer07} Not only the peak intensity but also the  \QQ-integrated intensity  are maximal at the resonance energy \Eres\ (Fig.~\ref{fig:qInt}).\cite{Rossat-MignodBourges91} It is noteworthy that although there is a noticeable $a$-$b$-anisotropy at low energy, the signal is two-dimensional and incommensurate in both directions. Further properties become apparent from the constant-energy TOF maps in Fig.~\ref{fig:TOF}: First, the anisotropy increases with decreasing energy; second, the signal is nearly four-fold symmetric at high energies above \Eres;\cite{HinkovKeimer07, StockBuyers05} third, the orientation of the intensity maxima at high energies is rotated by $45^\circ$ with respect to the distribution at low energies.\cite{Hayden04, Tranquada04}

Upon raising $T$ to $70$\,K, \ie\ 10\,K above \tc, prominent changes occur at and below the resonance energy, while the high energies are hardly affected. This supports the notion that the resonance peak and the downward dispersing branch form a common, superconductivity-induced resonance mode.\cite{BourgesFong00} Notably, above \tc\ this mode is replaced by an incommensurate, nearly one-dimensional distribution.\cite{HinkovKeimer07} In the vicinity of \Eres\ and down to $\sim15$\,meV below \Eres, the \QQ-integrated intensity is higher in the superconducting state (Fig.~\ref{fig:qInt}), reflecting the appearance of the resonance mode. Fig.~\ref{fig:qInt} also indicates the opening of a superconductivity-induced gap below $\sim20$\,meV.\cite{StockBuyers04} Neglecting the redistribution in \QQ\ at constant energy for the moment, the resonance mode gathers its spectral weight  mostly from this gap and from a limited energy range around $\sim45$\,meV. Altogether, the impact of superconductivity is limited to $\EE\lesssim50-55$\,meV, and, concerning the excitation topology alone, even to $\EE\lesssim\Eres$ (see also Fig.~\ref{fig:mapsConv}e,f).

Finally, raising $T$ to room temperature eradicates any sign of coherent excitations for $\EE\lesssim\Eres$ and also notably suppresses the intensity at higher energies. At the same time, the incommensurability \deltaIC\ decreases.

\subsection{Low-energy excitations}
\label{subsec:lowE}
\begin{figure}[b]
\includegraphics[width=0.8\columnwidth]{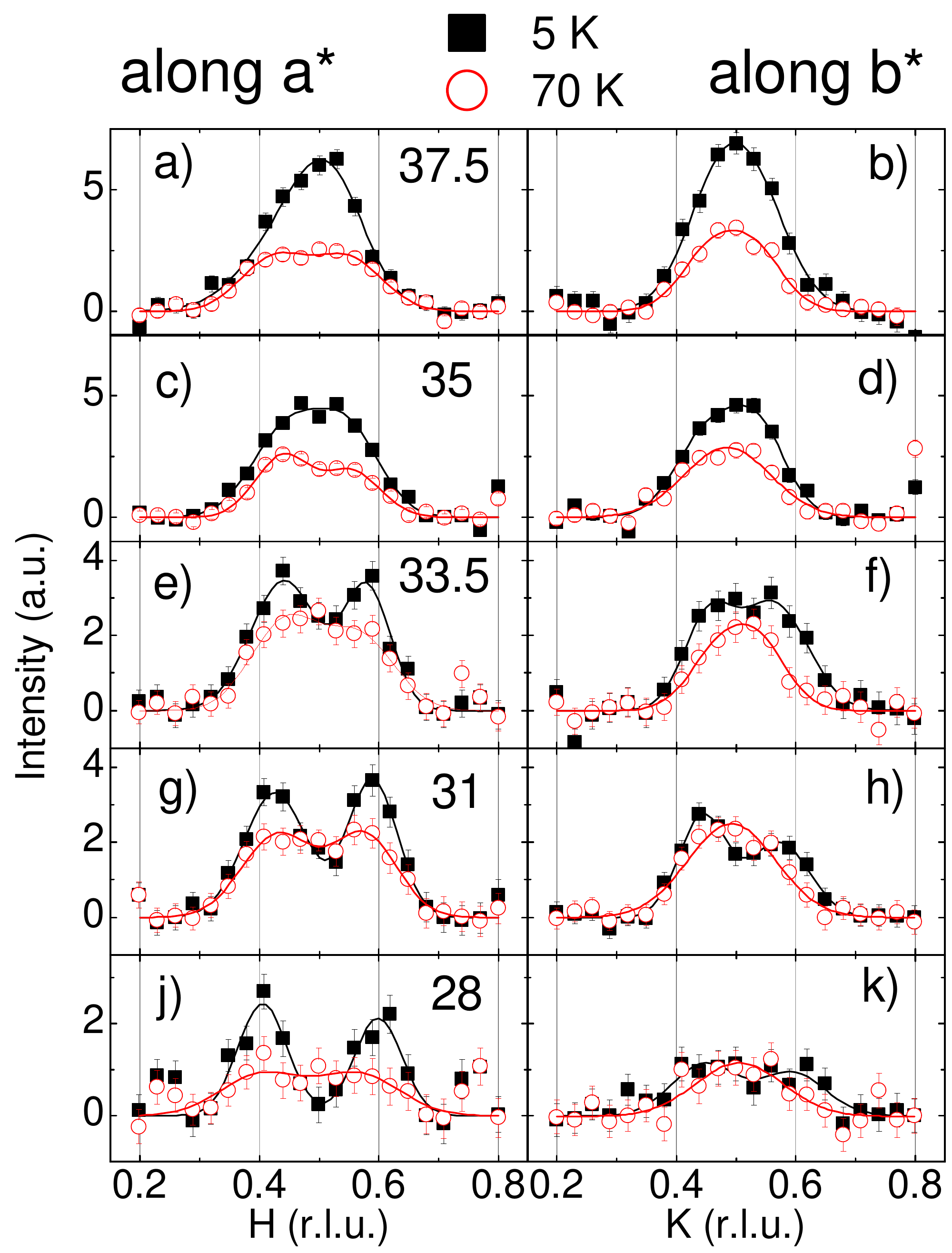}
\caption{\label{fig:lowE} Comparison of constant-energy scans at 70\,K (red circles) and at 5\,K (black squares). The scans along \astar\ are of the type $(H, -1.5, -1.7)$, those along \bstar\ of the type $(1.5, K, 1.7)$. Both datasets were collected under the same experimental conditions and normalized and BG corrected in the same way. The lines show Gaussian fits. The excitation energies denoted in the panels are in meV.}
\end{figure}

\begin{figure}[t]
\begin{center}
\includegraphics[width=0.95\columnwidth]{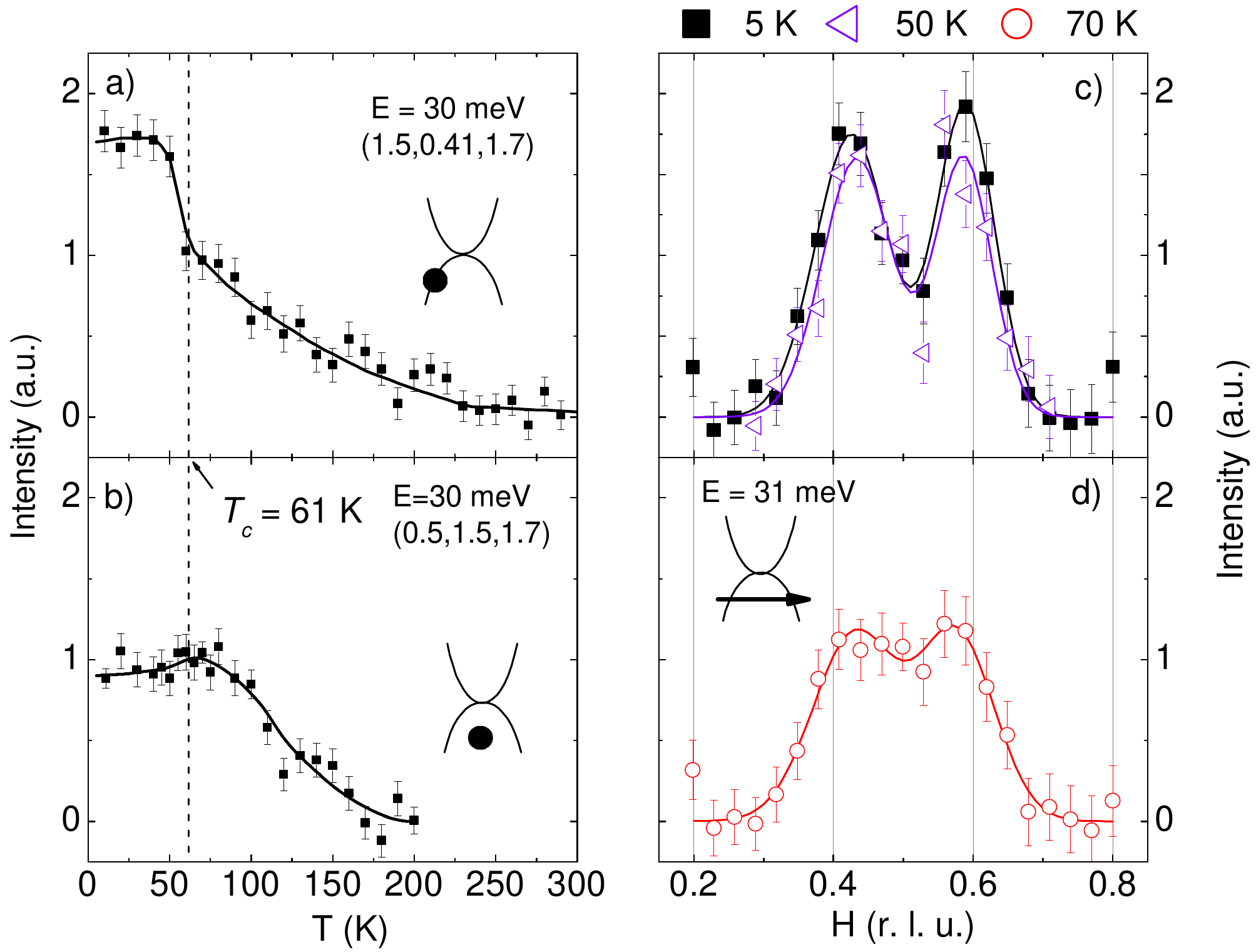}
\end{center}\vspace{-15pt}
\caption{\label{fig:tempDep} a), b) Temperature evolution of the magnetic intensity at two points of the downward dispersing mode ($\EE=30$\,meV), sketched in reference to the overall hourglass dispersion in the SC state. The values were obtained by subtracting a $Q$-linear background, determined by measuring points on both sides of the peak. c), d) $Q$-scan profiles at almost the same fixed energy as in a) and b) ($\EE=31$\,meV), measured at 5, 50 and 70\,K, respectively.}
\end{figure}

In Fig.~\ref{fig:lowE} we show $Q$-scan profiles in an extended energy range $\EE\leq\Eres$, below (5\,K) and above (70\,K) \tc. The data in the superconducting state exhibit a continuous, monotonic dispersion which is somewhat steeper along \bstar, and exhibits a marked in-plane intensity anisotropy,\cite{HinkovKeimer04, HinkovKeimer07} as first reported in Ref. \onlinecite{Mook00}.  When heating to 70\,K, we observe marked changes in the entire energy range: First, the in-plane distribution of the intensity changes, and we observe a quasi-one-dimensional signal in all measurements.  The data can be well fitted with two Gaussian peaks displaced in the \astar-direction from \qaf, whereas a single Gaussian is sufficient along \bstar\ and the fit is not significantly improved assuming an incommensurability in this direction as well. Remarkably, the incommensurability \deltaIC, \ie\ the displacement of the peaks from \qaf, hardly depends on energy (see also Fig.~\ref{fig:dispersion}), in particular there is no indication of a sharp, commensurate resonance peak at 70\,K. Altogether, this indicates that the 70\,K data cannot be reproduced by a simple temperature-induced broadening of the 5\,K data, which in turn implies that the dispersion topology is different below and above \tc. This is a central point which will be further elaborated and confirmed by our discussion of the overall dispersion in \ref{subsec:dispersion} and the fit to the analytical model in \ref{subsec:model} where resolution effects are taken into account.

Second, the wavevector-resolved representation of Fig.~\ref{fig:lowE} allows us to conclude that in addition to an intensity redistribution between different energies, already apparent in Fig.~\ref{fig:qInt}, there is also an in-plane redistribution at fixed energy (Fig.~\ref{fig:lowE}g--k; see also Fig.~\ref{fig:mapsConv}e--f).

Finally, more detailed measurements of the temperature evolution of scans through the downward-dispersing branch (Fig.~\ref{fig:tempDep}) show the same abrupt increase in intensity as the resonance peak (panel a and b).\cite{FongKeimer00,HinkovKeimer07} In addition, panels c) and d) show that the onset of superconductivity also abruptly changes the geometry, which then remains unaltered upon further cooling throughout the superconducting state.

\subsection{High-energy excitations}
\label{subsec:highE}

The excitations at high energies differ in several important aspects from those at low energies. First, as already mentioned in \ref{subsec:overview}, the in-plane geometry changes from a two-fold symmetric azimuthal distribution around \qaf\ with pronounced peaks in the \astar-direction below \Eres\ to a nearly four-fold symmetric distribution with pronounced peaks in both diagonal directions above \Eres\ (Fig.~\ref{fig:TOF}). Whereas the detailed model fits of the three-axes data in \ref{subsec:model} suggest a slightly steeper dispersion along \bstar\ than along \astar, the true intensity maxima are between these two directions and do not deviate from \{100\} within the error bars (Figs. \ref{fig:TOF} and \ref{fig:modelCuts}).

Second, except for a limited energy interval around 45\,meV, where the resonance peak gathers some of its spectral weight, the onset of superconductivity does not affect the spin excitations\cite{Hayden04} and in the \imChiQE\ model we will use the same function at both 5\,K and 70\,K.

\begin{figure}[b]
\includegraphics[width=0.99\columnwidth]{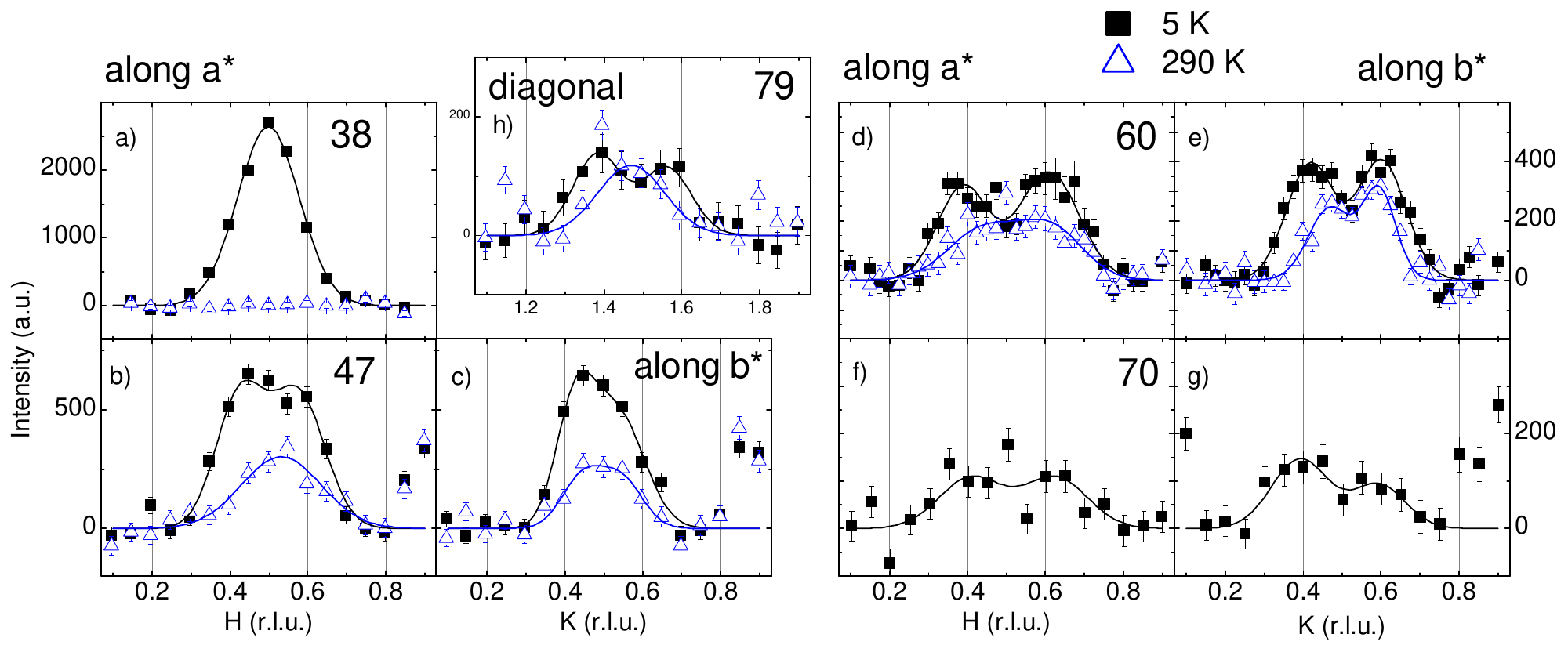}
\caption{\label{fig:highE} Energy evolution of the upward dispersing branch in the SC (5\,K) and the normal state (290\,K). The excitation energies denoted in the panels are in meV. Data were collected in the constant-\kf\ mode ($\kf=4.1\mbox{ \AA}^{-1}$ or $4.5\mbox{ \AA}^{-1}$). Due to restrictions arising from the kinematics of the neutron scattering process, scans at higher energies were taken in unusually high Brillouin zones. They are  of the type $(H,-2.5,-1.7)$ and $(2.5,K,1.7)$ at $70\mbox{ meV}$ and $(H,H,1.7)$ at $79\mbox{ meV}$. A $Q$-linear background has been subtracted. The lines result from Gaussian fits to the data. The intensities are not comparable among different energies.}
\end{figure}

Third, whereas we do not detect any intensity at room temperature at and below \Eres, above \Eres\ a progressively larger fraction of the intensity detected at 5\,K is already present at 290\,K. This is illustrated in Fig.~\ref{fig:highE}, where we show scan profiles up to 79\,meV. From Fig.~\ref{fig:qInt} it is clear that this fraction increases mainly because the low-temperature intensity decreases, while the high-temperature intensity does not change much with energy up to 60\,meV.

Finally, we observe that at room temperature the signal forms a broad maximum at \qaf\ which splits into separate peaks away from \qaf\ at lower temperatures, but above \tc. We stress again that the evolving geometry remains four-fold symmetric, in contrast to what we see at low energies. Overall, the change of the excitations with temperature is weaker than at low energies, in agreement with other compounds\cite{XuTranquada07, HinkovHaug08, Lipscombe09, XuGu09} and consistent with the idea of the universality of the upper branch.

A suggestive way to summarize the impact of superconductivity on the whole energy evolution is to define four energy intervals: The SC-spin-gap interval below $\sim20$\,meV, the resonance region between $\sim20$ and $\sim40$\,meV, the small ``gap'' interval around 45\,meV and the part above $\sim50-55$\,meV with no superconductivity-induced changes.

\subsection{Dispersion relations}
\label{subsec:dispersion}

\begin{figure*}[]
\begin{minipage}[]{0.6\textwidth}
\includegraphics[width=1\textwidth]{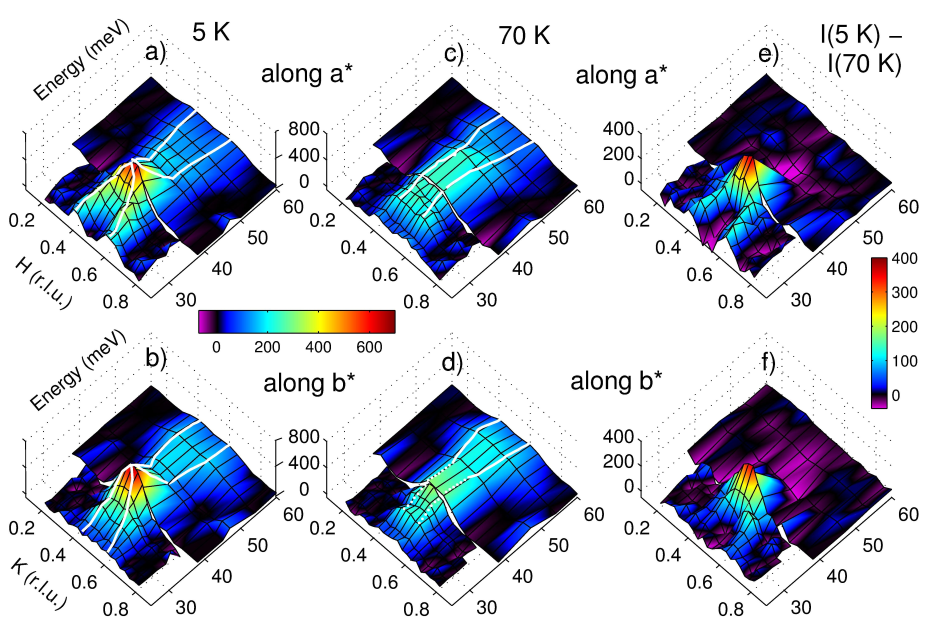}
\end{minipage}\vspace{-5pt}
\begin{minipage}[]{0.39\textwidth}
\caption{\label{fig:mapsConv} Color representation of the magnetic intensity between 26 and 60\,meV. Panels a),\,b) show the intensity at 5\,K, panels c),\,d) the intensity at $70$\,K\,$>\tc$ and panels e),\,f) the difference between 5 and 70\,K. The upper and lower rows are constructed from three-axes scans of the type $(H,-1.5,-1.7)$ and $(1.5,K,1.7)$, respectively. The BG was corrected by subtracting the intensity at 250\,K for $\EE<38$\,meV; in addition, the data were corrected for a $Q$-linear BG at all energies. The three axes represent transferred momentum $H$ or $K$ (r.l.u.), energy (meV) and magnetic intensity (a.u.).  Below 38\,meV, $\kf$ was fixed to $2.66\mbox{ \AA}^{-1}$, above 38\,meV to $4.5\mbox{ \AA}^{-1}$. Scans at 38\,meV were used to bring both energy ranges to the same scale. Crossings of black lines represent measured data points. White lines connect the fitted peak positions and dotted lines represent upper bounds on the incommensurability, as discussed in the text.}
\end{minipage}
\end{figure*}%

\begin{figure}[b]
\begin{center}
\includegraphics[width=0.9\columnwidth]{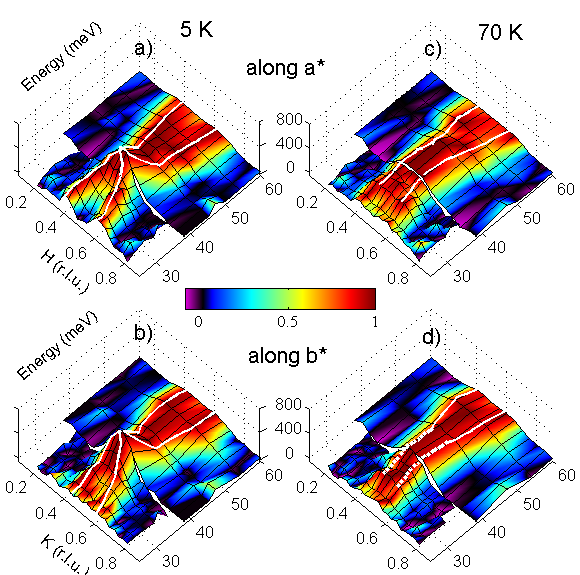}
\end{center}\vspace{-15pt}
\caption{\label{fig:mapsAlt} Color representation of the magnetic intensity between 26 and 60\,meV, based on the same data as in Fig.~\ref{fig:mapsConv}a)\,--\,d), but with a different color scale normalized to the peak intensity of each individual constant energy scan taken along the black lines. Further details are given in the caption of Fig.~\ref{fig:mapsConv}. The panels of this figure are reproduced from Ref. \onlinecite{HinkovKeimer07}.}
\end{figure}
Most of the data we show in this paper consist of or is based on constant-energy cuts through momentum space. There are several ways of representation, each of which stresses different aspects of the data. In the previous subsections we showed mainly single one-dimensional scan profiles or two-dimensional cuts which allow one to see the measured data in great detail, but obscure general trends occurring with changing energy: Overall intensity variations can be better visualized in a compact way in three-dimensional surface maps of \imChiQE, like the one shown in Fig.~\ref{fig:mapsConv}. In panels a) and b), the large intensity of the commensurate resonance peak at \Eres\ clearly stands out. However, the evolution of the \QQ-profile away from \Eres\ is still not reproduced very well. To obtain an even better representation of the dispersion, we can take advantage of a redundancy in the surface maps: The intensity is coded in both the color and the elevation. In Fig.~\ref{fig:mapsAlt} we show the same data as in Fig.~\ref{fig:mapsConv}a)--d), but whereas the elevation still represents the intensity, the color scale is now normalized to the peak intensity of each individual constant-energy scan taken along the black lines of constant energy. This relative color scale emphasizes the dispersive character at 5\,K in both the \astar- and \bstar-direction, with branches merging at the commensurate resonance peak

From Figs.~\ref{fig:mapsConv}c),\,d) and \ref{fig:mapsAlt}c),\,d) we see that the hourglass shape with commensurate resonance constriction is lost in the normal state at 70\,K and the intense peak at \Eres\ disappears. Instead, the quasi-one-dimensional signal consisting of two Gaussians along \astar\ (see \ref{subsec:lowE}) exhibits a nearly vertical dispersion, forming an overall ``Y''-shaped dispersion.

\begin{figure}[b]
\includegraphics[width=0.9\columnwidth]{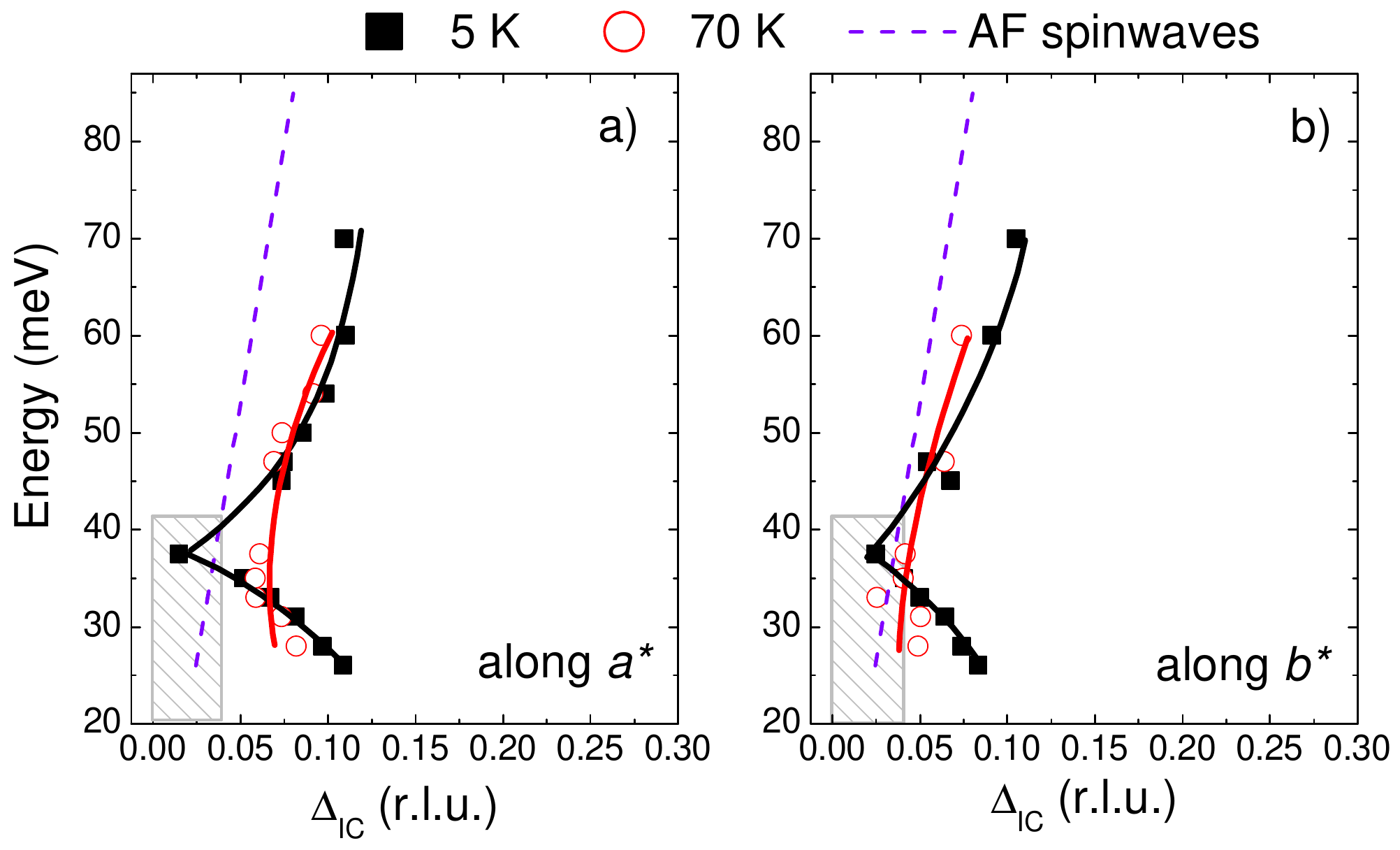}
\caption{\label{fig:dispersion} Dispersion relations along \astar\ and \bstar\ at 5 and 70\,K. Here, \deltaIC\ is obtained from half of the distance between the peak positions of two IC peaks. The solid lines are guides to the eye. The dashed line represents the dispersion of AF waves in insulating \mbox{YBa$_{2}$Cu$_{3}$O$_{6.15}$}.\cite{HaydenAeppli96} Incommensurabilities within the hatched area come from peaks
which lie so close together, that they cannot be discriminated from a single Gaussian peak. The error in the energy direction is determined by the energy
resolution and thus depends on \kf. Below $\Eres=38$\,meV it is of the order of 2 to 3\,meV, and above of the order of 6 to 8\,meV. The error in the $Q$-direction is mainly determined by the statistics of the scan (errors of the intensities at the distinct points of the scan) and by the question, how well both IC peaks can be separated. From the fits we obtain typically $\leq0.01\mbox{ r.l.u.}$ in the SC state for large \deltaIC\ and up to $\pm0.02\mbox{ r.l.u.}$ at 70\,K state close to the hatched region.}
\end{figure}

To further elaborate our observations regarding the changing signal topology, we show in Fig.~\ref{fig:dispersion} the dispersion relations at 5 and 70\,K as obtained from Gaussian fits to the scan profiles along \astar\ and \bstar. Clearly, the signal collapses to a single peak at \Eres\ in the superconducting state, and the plot confirms the finding that the signal exhibits a nearly vertical dispersion at 70\,K. The evaluation in \ref{subsubsec:normalstate} which considers the instrumental resolution shows that the data are in excellent agreement with a model consisting of two incommensurate Gaussian peaks with energy-independent \deltaIC\ below \Eres\ and an outward-dispersing damped harmonic oscillator mode above \Eres. However, we cannot exclude that at the lowest energies \deltaIC\ increases slightly, see also Fig.~\ref{fig:dispersion}.

In Fig.~\ref{fig:dispersion}, the right border of the hatched region defines an empirical lower bound for the resolvable incommensurability \deltaIC. While some fits converge towards a \deltaIC\ within the hatched region and thus seem to suggest a persisting incommensurability, they provide no significant improvement over fits with a single peak. This is only partly owed to resolution effects\footnote{At the instrumental conditions used up to $\sim\Eres$, the \QQ-resolution is lower than twice the hatched region.} -- another important reason is that $2\deltaIC$ becomes of the order of the intrinsic width of the peaks. Hence, at 70\,K the scan profiles along \bstar\ are indistinguishable from a single Gaussian around and below \Eres, and we can only define an upper limit for a potential incommensurability, as indicated by the dotted white lines in \ref{fig:mapsConv}d) and \ref{fig:mapsAlt}d). Our analysis in \ref{subsubsec:normalstate} and the fits in Fig.~\ref{fig:allFits} support the interpretation that the signal along \bstar\ arises from a single peak.

\begin{figure}[t]
\includegraphics[width=0.9\columnwidth]{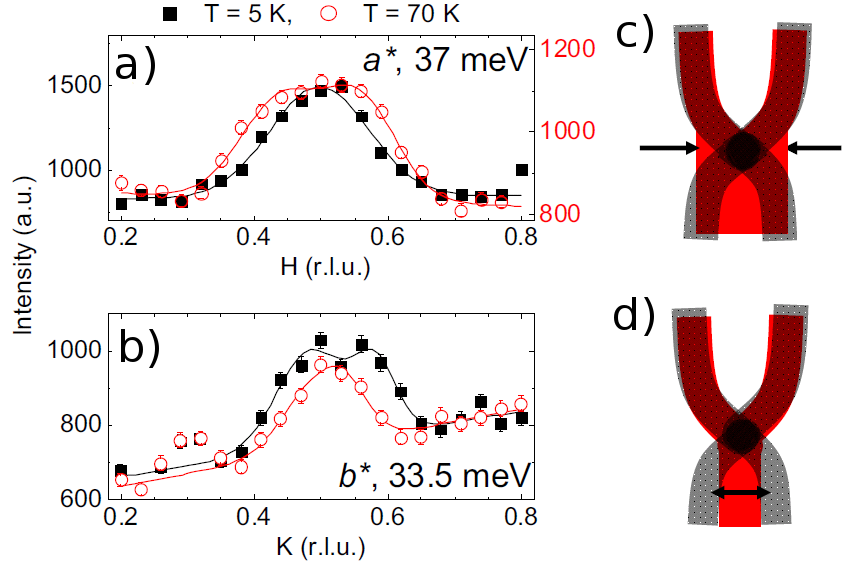}
\vspace{-4mm}
\caption{\label{fig:topologyChange} Evolution of constant-energy profiles (raw data) from $T=70$\,K$\,>\tc$ to $T=5$\,K in two characteristic energy regions. a) Scans along \astar\ at $\EE=\Eres$. b) Scans along \bstar\ at $\EE=33.5$\,meV$\,<\Eres$. The data were taken from Fig.~\ref{fig:raw}e) and h). At 37\,meV, the data at both temperatures were scaled to the same amplitude to allow a better comparison of their $Q$-widths. Panels c) and d) show schematic representations of the dispersion relations at 70\,K (red) and 5\,K (dotted grey) in the \astar- and \bstar-direction, respectively. Arrows indicate the most salient changes in the SC state. }
\end{figure}

In Fig.~\ref{fig:topologyChange}c) and d) we illustrate how the resonance forms out of the normal-state ``Y''-shaped dispersion: Along \astar, the main effect is the collapse of the incommensurate 70\,K signal to a single peak, the ``neck'' of the hourglass dispersion (Fig.~\ref{fig:topologyChange}a), accompanied by a moderate signal sharpening and increase of \deltaIC\ below \Eres. Along \bstar, on the other hand, the main effect is the splitting of the commensurate scan profiles into incommensurate peaks below \Eres, Fig.~\ref{fig:topologyChange}b), while there is no further sharpening at \Eres.

As shown in \ref{subsec:overview} and \ref{subsec:highE}, there is hardly any change across \tc\ in the high-energy branch. We thus next examine the difference $I_\text{diff}=I(5$\,K$)-I(70$\,K$)$, where $I(5$\,K$)$ and $I(70$\,K$)$ correspond to \imChiQE\ at 5 and 70\,K, respectively, Fig.~\ref{fig:mapsConv}e) and f). As might be expected, $I_\text{diff}$ is flat at high energies and negative around 50\,meV, see also Fig.~\ref{fig:qInt}. At low energies, a downward dispersing branch is observed in both directions. Due to the subtraction of the commensurate normal-state signal along \bstar, the incommensurability appears increased in Fig.~\ref{fig:mapsConv}f). $I_\text{diff}$ thus bears some resemblance to the signal observed in nearly optimally doped \ybcosixeight\ and other samples with higher doping,\cite{HinkovKeimer04, ReznikIsmer08} where the lower branch is much more prominent than the upper. Regarding the intensity difference is thus very instructive, since it suggests that only the lower branch should be associated with the resonance mode.

\subsection{Analytical formulae for {\imChiQE}}
\label{subsec:model}
\begin{figure}[b]
\includegraphics[width=0.75\columnwidth]{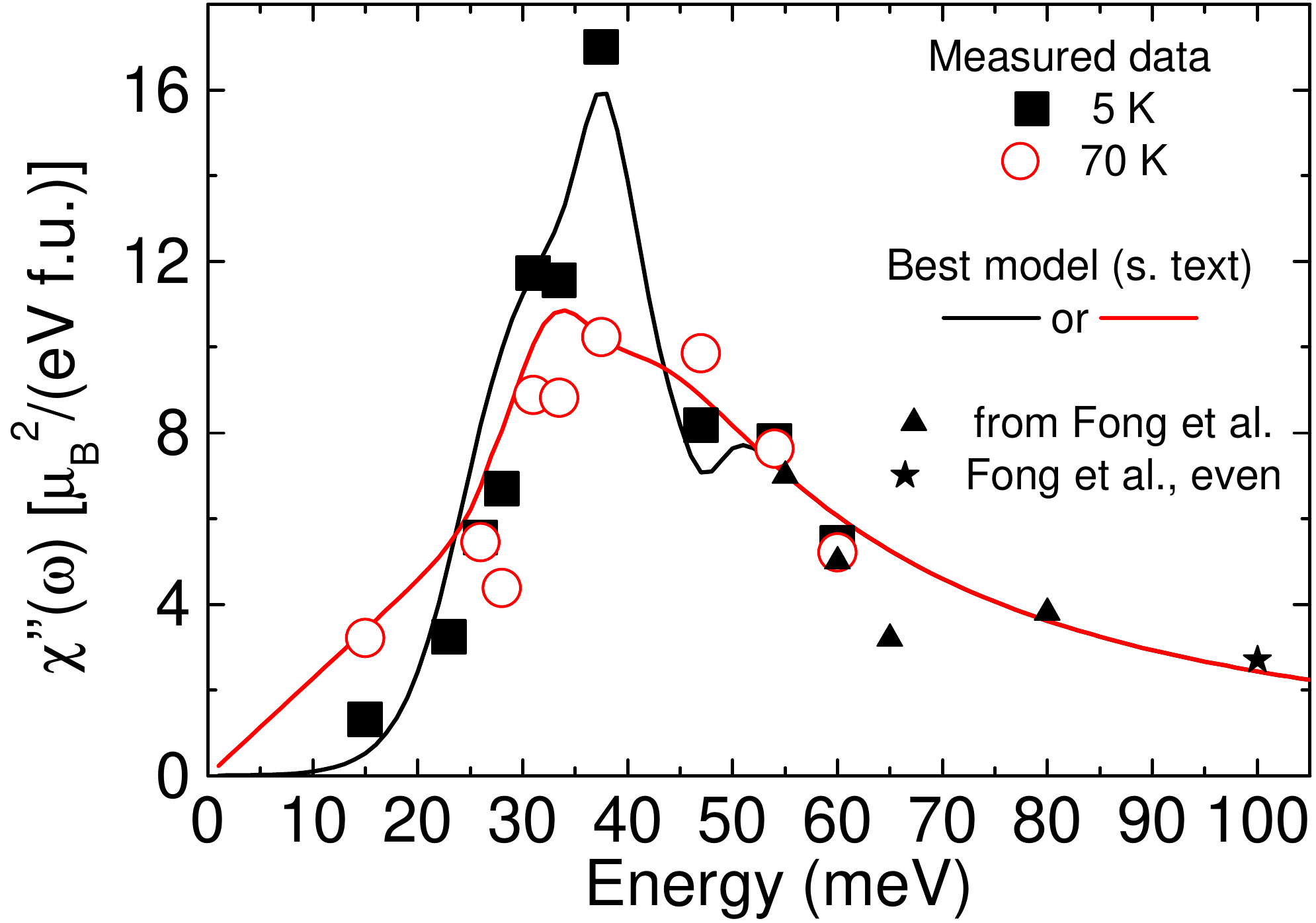}
\caption{\label{fig:qIntModel} Imaginary part of the local spin susceptibility \imChiQInt\ for the model functions described in \ref{subsubsec:scstate} (5\,K) and \ref{subsubsec:normalstate} (70\,K), compared with the experimental values from Fig.~\ref{fig:qInt}. The additional points are from Ref. \onlinecite{FongKeimer00} and were obtained from measurements on a sample with $\tc=67$\,K, similar to our $\tc=61$\,K.}
\end{figure}

Here we will first describe the fitting model for the superconducting state at 5\,K (\ref{subsubsec:scstate}) and the normal state at 70\,K (\ref{subsubsec:normalstate}) in detail. In \ref{subsubsec:modeldiscussion} we compare the models (convoluted with the resolution function) with the corresponding experimental data and discuss the limitations of our fit.

\subsubsection{Superconducting state ($T=5$\,K)}
\label{subsubsec:scstate}
The best model we found to fit the data at 5\,K with a reasonable number of parameters is based on a modified damped harmonic oscillator. We use the model on a purely empirical basis and will not try to justify it physically.

As suggested by the phenomenology of the data, we subdivide $\imChiQE$ into an upper (u) and a lower (l) branch, both described by a modified DHO function. The upper branch describes mainly the the energy range above \Eres\ and the lower branch mainly the range below \Eres. However, it should be kept in mind that the functions do not drop to zero at \Eres\ and the intensity of the two branches overlaps. It turns out that a smooth cutoff, not contained in the DHO model, is necessary to handle these ``cross contaminations'' to the respective other branch.

The total magnetic susceptibility then reads:
\begin{equation}
\label{eq:chiLowChiUp}
\imChiQE = N_0\left[\imChiQEl + N_{0u}\, \imChiQEu\right].
\end{equation}
Here $N_0$ is an overall normalization factor which is chosen such that the \QQ-integrated susceptibility \imChiQInt\ has a maximal value of $16\,\mu^2_{\scriptscriptstyle{\text B}}/\,(\text{eV\,f.u.})$ at the resonance energy (Fig.~\ref{fig:qIntModel}), which is known from previous work.\cite{Hayden04, FongKeimer00} $N_{0u}=1.35$ is a scaling factor between the upper and lower branch. We first discuss the lower branch (we remind that for the sake of a compact notation we have set $\hbar=1$ and quote $\omega$ in meV)
\begin{equation}
\label{eq:lowerSCBranch}
\imChiQEl =N_\ell(\omega) \frac{2 \omega_\text{red}\gamma_{\scriptscriptstyle{\mathbf{Q}}}}
{\left(\omega_\text{red}^2-\omega^2_{\scriptscriptstyle{\mathbf{Q}}}\right)^{2}+\left(2\omega_\text{red}\gamma_{\scriptscriptstyle{\mathbf{Q}}}\right)^2},
\end{equation}
where $\omega_\text{red}=2\omega_r-\omega$ and $\ores=38.5$\,meV is the resonance energy.\footnote{One should keep in mind that \imChiQE\ is required to be an odd function of $\omega$, which is not the case for the lower branch, eq. \eqref{eq:lowerSCBranch}. Thus, this formula is only valid for $\omega>0$, and would need to be extended correspondingly if one were interested in negative $\omega$. This is straight-forward due to the superconducting gap.} The dispersion of the mode is given by
\begin{equation}
\omega_{\scriptscriptstyle{\mathbf{Q}}} = \ores + s_a (H-H_0)^2 + s_b (K-K_0)^2,
\end{equation}
where $\qaf=(H_0, K_0)$ is the antiferromagnetic wavevector in the plane, with $s_a=280\,\text{meV\,\AA}^2$ and $s_b=455\,\text{meV\,\AA}^2$. The damping $\gamma_{\scriptscriptstyle{\mathbf{Q}}}$ has the following angular dependence:
\begin{equation}
\label{eq:gammal}
\gamma_{\scriptscriptstyle{\mathbf{Q}}}=\gamma_a+\Delta\gamma\frac{(K-K_0)^2}{|\QQ-\qaf|^2}.
\end{equation}
Here $a$ denotes the \astar-direction in the Brillouin zone and $\gamma_a=1.5$\,meV and $\Delta\gamma=\gamma_b-\gamma_a=6.5$\,meV.\footnote{In the online supplementary material of Ref. \onlinecite{DahmHinkov09}, where we first described the model, there is a typographical error in the equation corresponding to eq. \eqref{eq:gammal} here.} The prefactor $N_\ell(\omega)$ implements the superconductivity-induced spin gap below $\sim24$\,meV and the mentioned cutoff at high energies above $\sim42$\,meV:\footnote{Note that this gap function does not strictly go to 0 for $\omega\rightarrow 0$, but to a value $<10^{-3}$. This is orders of magnitude below the experimental error and good enough for all practical purposes. For a strictly analytical continuation to negative $\omega$, however, one wood need to slightly modify \eqref{eq:SCgap}. }
\begin{equation}
\label{eq:SCgap}
N_\ell(\omega)=\frac{\exp\left[ (\omega-24)/3 \right]}{1+\exp\left[ (\omega-24)/3 \right]}\; \frac{1}{1+\exp(\omega-42)}.
\end{equation}

\begin{figure}[t]
\includegraphics[width=0.3\columnwidth]{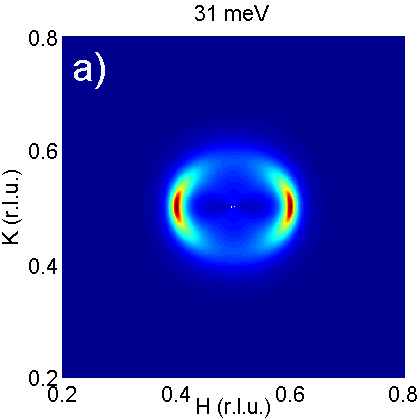}
\includegraphics[width=0.31\columnwidth]{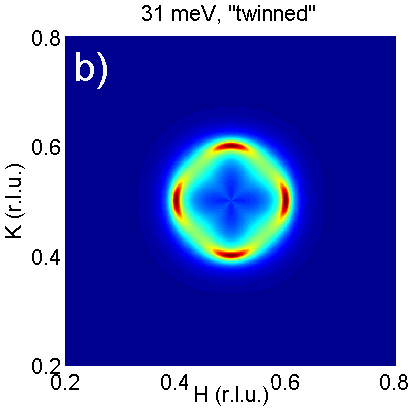}
\includegraphics[width=0.36\columnwidth]{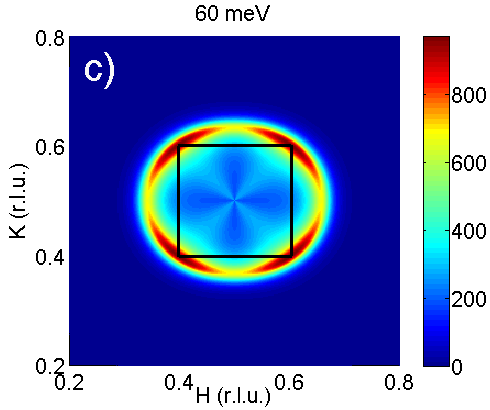}
\caption{\label{fig:modelCuts} Constant-energy maps through the model for \imChiQE\ in the superconducting state at 5\,K described in \ref{subsubsec:scstate}. a) 31\,meV, b) \imChiQE\ at 31\,meV as it would appear in a twinned sample. The map was obtained by superposing the map it a) with its transpose. c) 60\,meV. The corners of the black square define the positions of maximal intensity of the upper branch in the \{110\} directions. The shown intensity was not convoluted with the instrumental resolution function, is shown in arbitrary units and is not comparable between different panels. }
\end{figure}

The upper branch is given by
\begin{equation}
\label{eq:upperBranch}
\imChiQEu= N_u(\omega, \QQ)\, \frac{2\omega\Gamma}{\left(\omega^2-\Omega_{\scriptscriptstyle{\mathbf{Q}}}^2\right)^2+(2\omega\Gamma)^2},
\end{equation}
where in contrast to the lower branch the damping $\Gamma=11$\,meV is a constant, independent of the azimuthal angle. Also, due to the steeper dispersion at higher energies, we model $\Omega_{\scriptscriptstyle{\mathbf{Q}}}$ with a quartic power law ($p=4$):
\begin{equation}
\label{eq:upperBranchDispersion}
\Omega_{\scriptscriptstyle{\mathbf{Q}}}=\ores+\left[ S_a^{2/p}\,(H-H_0)^2 +  S_b^{2/p}\,(K-K_0)^2 \right]^{p/2}.
\end{equation}
Here $S_a=4830\,\text{meV\,\AA}^p$ and $S_a=10065\,\text{meV\,\AA}^p$. Despite the seemingly large difference between $S_a$ and $S_b$, the in-plane anisotropy is smaller than in the lower branch, and in addition there is no anisotropy in the damping or the amplitude (Fig.~\ref{fig:modelCuts}). The prefactor $N_u(\omega, \QQ)$ is somewhat more complicated than $N_\ell(\omega)$, since it accounts for three effects: the cutoff towards low energies, $N_\text{cutoff}(\omega)$, the superconductivity-induced suppression above \ores, $N_\text{dip}(\omega)$ (Fig.~\ref{fig:qIntModel}) and the azimuthal intensity distribution with peaks along the (110) directions, $N_\text{rot}(\QQ)$ (Fig.~\ref{fig:highE}),
\begin{equation}
\label{eq:upperBranchPrefactors}
N_u(\omega,\QQ)=N_\text{cutoff}(\omega) N_\text{dip}(\omega) N_\text{rot}(\QQ).
\end{equation}
The different terms are
\begin{equation}
\label{eq:upperBranchCutoff}
N_\text{cutoff}(\omega)=1-\frac{1}{1+\exp\left[ (\omega-36)/1.5 \right]},
\end{equation}
\begin{equation}
\label{eq:upperBranchDip}
N_\text{dip}(\omega) = 1-0.2\exp\left[ -(\omega-\Omega_\text{dip})^2/(2\sigma^2_\text{dip}) \right],
\end{equation}
\begin{equation}
\label{eq:upperBranchFourtyFive}
N_\text{rot}(\QQ) = 1-0.3\left( \frac{(H-H_0)^2-(K-K_0)^2}{|\QQ-\qaf|^2} \right)^2
\end{equation}
and $\Omega_\text{dip}=47$\,meV, $\sigma_\text{dip}=2$\,meV. The whole function \imChiQE\ needs to be cut at the zone center and continued periodically to account for the lattice symmetry.

\begin{figure}[b]
\begin{center}
\includegraphics[width=0.63\textwidth]{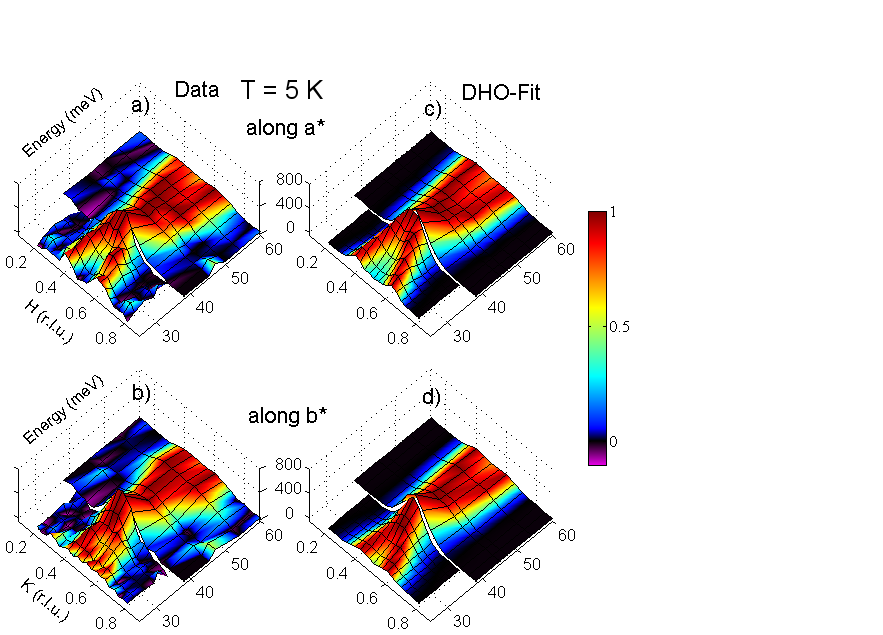}
\end{center}\vspace{-4mm}
\caption{\label{fig:SCFit} Color representation of the magnetic intensity between 26 and 60\,meV at $T=5$\,K. a), b) Measured data, identical with Fig.~\ref{fig:mapsAlt}a), b). c), d) Fit to the damped harmonic oscillator model described in eqs. \eqref{eq:chiLowChiUp}--\eqref{eq:upperBranchFourtyFive} of \ref{subsubsec:scstate}. A detailed comparison for the individual scans is shown in Fig.~\ref{fig:allFits}.}
\end{figure}

\subsubsection{Normal state at 70\,K}
\label{subsubsec:normalstate}
\begin{figure*}[]
\begin{minipage}[]{0.7\textwidth}
\includegraphics[width=0.95\textwidth]{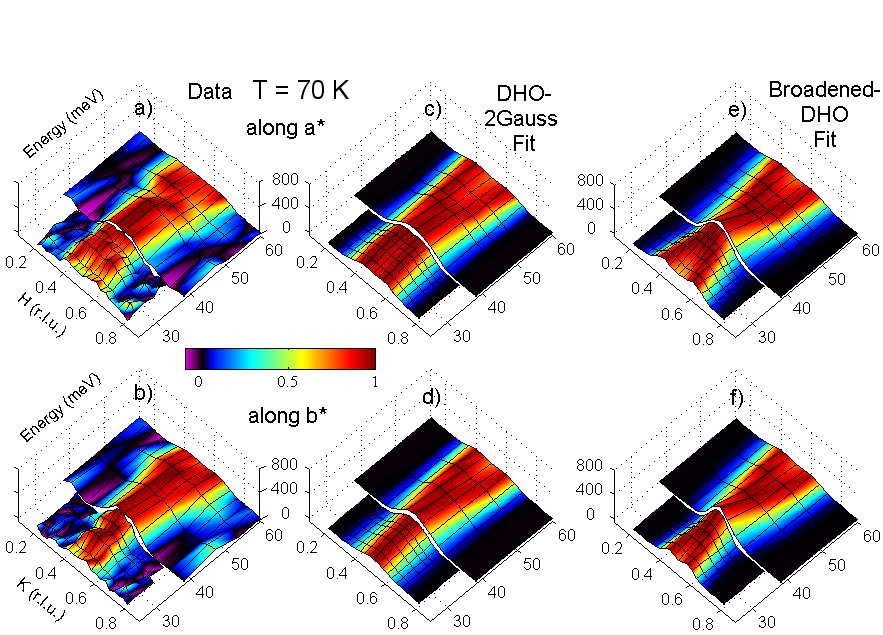}
\end{minipage}
\begin{minipage}[]{0.29\textwidth}
\caption{\label{fig:NSFit} Color representation of the magnetic intensity between 26 and 60\,meV at $T=70$\,K. a), b) Measured data, identical with Fig.~\ref{fig:mapsAlt}c), d). c), d) Fit to the model described in \ref{subsubsec:normalstate}, which best reproduces the data. A detailed comparison for the individual scans is shown in Fig.~\ref{fig:allFits}. e), f) Representation of a broadened damped harmonic oscillator model (increased damping constants) which fails to reproduce salient features of the data, like the absence of an hourglass constriction and the nearly vertical dispersion.}
\end{minipage}
\end{figure*}

\begin{figure}[b]
\begin{center}
\includegraphics[width=0.7\textwidth]{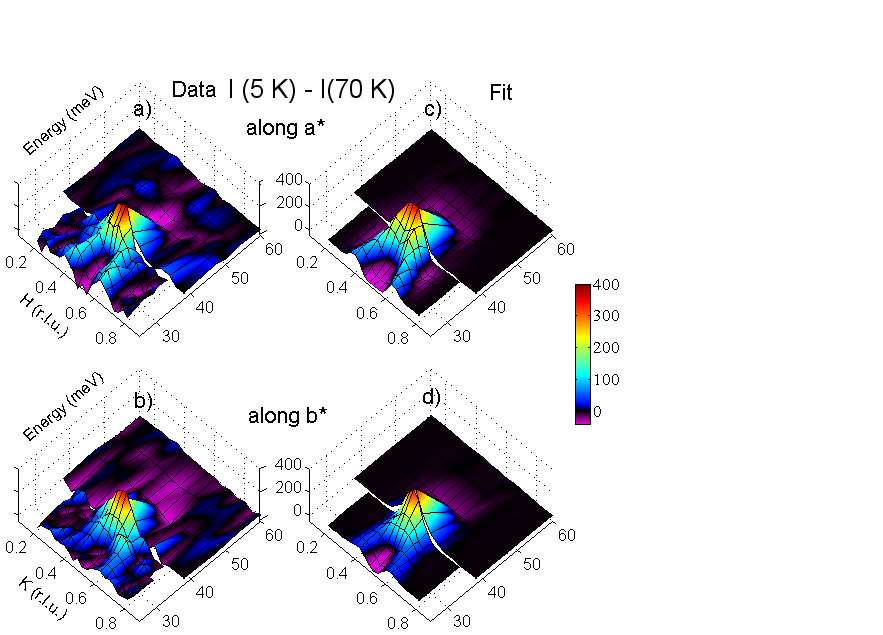}
\end{center}\vspace{-4mm}
\caption{\label{fig:SC-NSFit} Color representation of the subtracted magnetic intensity $I(5\,\text{K}) - I(70\,\text{K})$ between 26 and 60\,meV. a), b) Measured data, identical with Fig.~\ref{fig:mapsConv}e), f). c), d) Difference between the DHO model describing the data at 5\,K, \ref{subsubsec:scstate}, and the model describing the data at 70\,K, \ref{subsubsec:normalstate}. }
\end{figure}

\begin{figure*}[]
\includegraphics[width=0.9\textwidth]{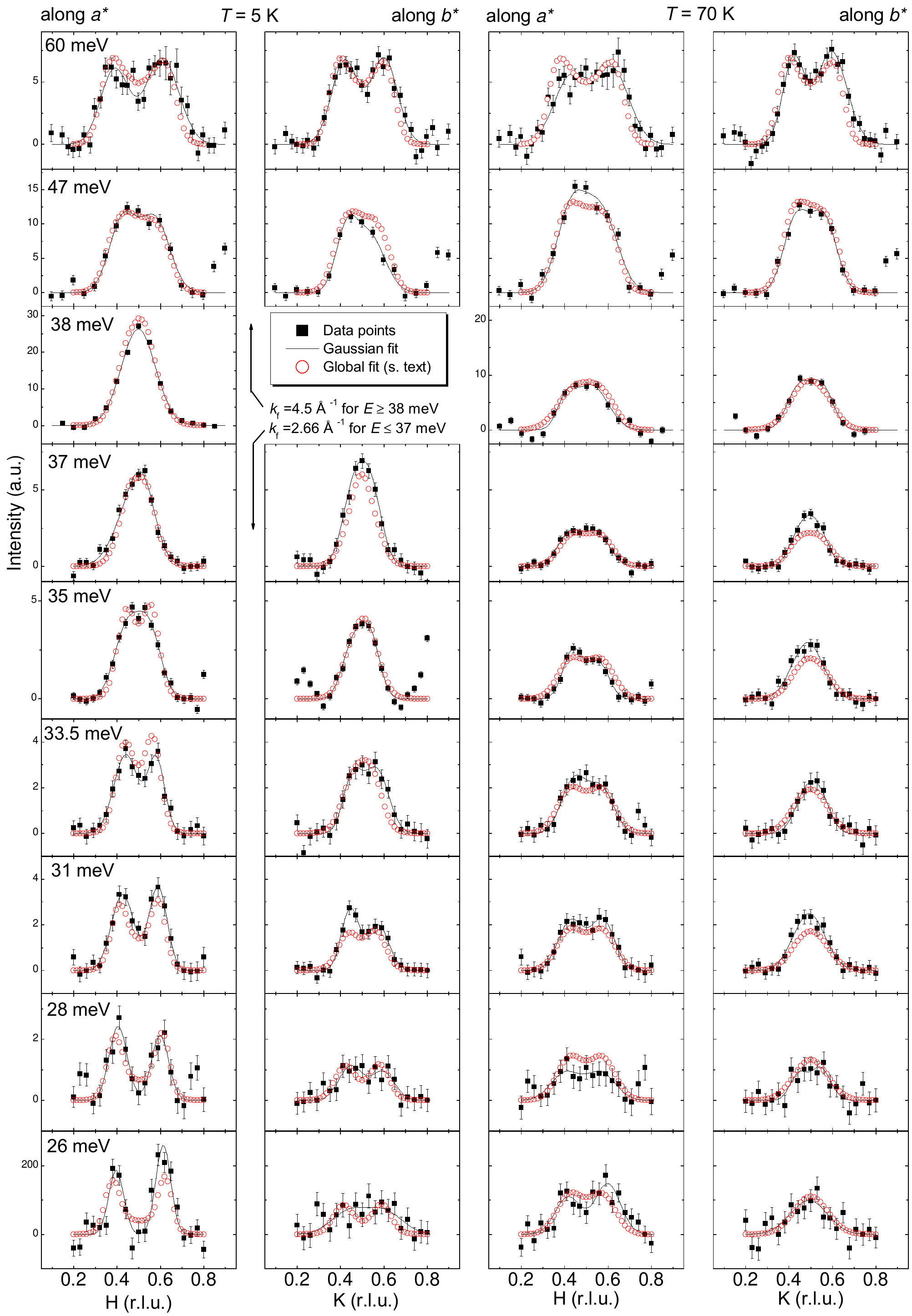}
\caption{\label{fig:allFits} Comparison between the measured data (as shown in Figs. \ref{fig:lowE}--\ref{fig:mapsConv}, full squares) and the corresponding model (open circles): The DHO model, for $T=5$\,K (\ref{subsubsec:scstate}, left two columns); and the model for $T=70$\,K (\ref{subsubsec:normalstate}, right two columns).}
\end{figure*}

At 70\,K we apply the same subdivision into a lower and upper branch like in eq. \eqref{eq:chiLowChiUp}. The prefactor $N_0$ has the same numerical value like at 5\,K. The upper branch \imChiQEu\ is described by the same model \eqref{eq:upperBranch} as at 5\,K, with the same dispersion \eqref{eq:upperBranchDispersion} and damping $\Gamma=11$\,meV. Also the prefactor $N_\text{rot}(\omega)$ is the same as in \eqref{eq:upperBranchFourtyFive}, while $N_\text{dip}$ being 1 and $N_\text{cutoff}$ being
\begin{equation}
N_\text{cutoff}(\omega)=1-\frac{1}{1+\exp\left[ (\omega-37)/2 \right]}
\end{equation}
are different from \eqref{eq:upperBranchDip} and \eqref{eq:upperBranchCutoff}, respectively.

In contrast, no satisfactory fit can be obtained for the low-energy branch at 70\,K with the DHO model used at 5\,K. The best one can do within the framework of the DHO model is to increase the damping parameters $\gamma_a$ and $\gamma_b$ significantly -- however, the hourglass character of the dispersion is not lost by this (Fig.~\ref{fig:NSFit}e, f). A much better fit is obtained if the lower branch is modeled by two  anisotropic two-dimensional Gaussian distributions, displaced from \qaf\ along the \astar-direction:
\begin{equation}
\imChiQEl= N_\ell(\omega)\sum_{i=1,2} e^{-4\ln2 \left[ \left( \frac{H-H_i}{\sigma_a}\right)^2 + \left( \frac{K-K_i}{\sigma_b}\right)^2\right]}.
\end{equation}

Within the energy range of $\sim25-38$\,meV, the experimentally observed incommensurability does not change within the error, hence we can assume $\mathbf{Q_1}=(H_1, K_1)=(0.575, 0.5)$ and $\mathbf{Q_2}=(H_2, K_2)=(0.425, 0.5)$ to be energy-independent. The Gaussian widths are $\sigma_a=0.125$ and $\sigma_b=0.17$, both in r.\,l.\,u. The normalization factor is given by
\begin{equation}
N_\ell(\omega)=N_{0\ell} N_\text{cutup} (\omega) N_\text{cutdown} (\omega) N_\text{lin} (\omega),
\end{equation}
where $N_{0\ell}=0.0181$.

\begin{equation}
N_\text{cutup} (\omega) = \frac{1}{1+\exp\left[(\omega-37)/2 \right]}
\end{equation}
provides a cutoff towards higher energies, while
\begin{equation}
N_\text{cutdown} (\omega) =\frac{0.5+\exp\left[(\omega-29.5)/2\right]}{1+\exp\left[(\omega-29.5)/2\right]}
\end{equation}
provides a partial cutoff towards lower energies. Finally,
\begin{equation}
N_\text{lin} (\omega) = \left\{ \begin{array}{cc}
1 & \text{for } \omega\geq27\,\text{meV} \\
\omega/27 & \text{for } \omega<27\,\text{meV} \\
\end{array}\right.
\end{equation}
provides a simple linear decrease below 27\,meV, consistent with previous work.\cite{StockBuyers04}

\begin{figure}[]
\includegraphics[width=0.9\columnwidth]{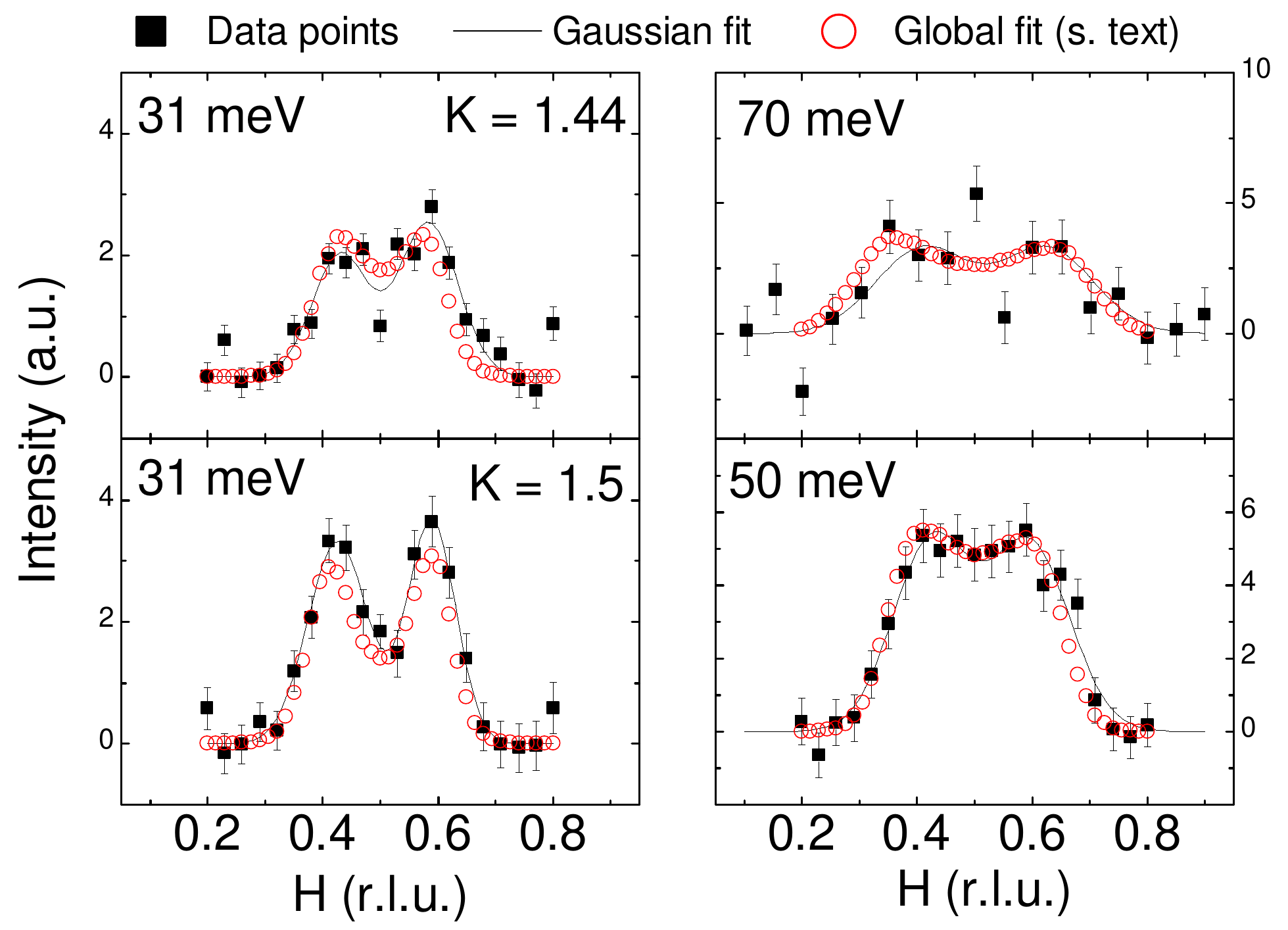}
\caption{\label{fig:allFits2} Continued from Fig.~\ref{fig:allFits}: Comparison between measured data (full squares) and the DHO model, \ref{subsubsec:scstate}, at $T=5$\,K,  (open circles).}
\end{figure}

\subsubsection{Summary of the models}
\label{subsubsec:modeldiscussion}
In Fig.~\ref{fig:qIntModel} we show the momentum-integrated spin susceptibility \imChiQE\ for both models (5\,K and 70\,K) in absolute units. The sum-rule integral $S=(\pi g^2 \mu_{\scriptscriptstyle{\text{B}}}^2)^{-1}\int_0^\infty d\omega \,d\QQ \,\imChiQE$ is identical for both models within the error bars, as required. The fact that the experimental \QQ-integrated spin susceptibility from Fig.~\ref{fig:qInt} agrees well with the model curves lends confidence to our results. The comparison with experimental values previously obtained\cite{FongKeimer00} in a sample with a  \tc\ of $67$\,K (Fig.~\ref{fig:qIntModel}) indicates that our extrapolation towards high energies is correct up to at least $100$\,meV,\footnote{At 100\,meV, only data for the even mode are reported in Ref. \onlinecite{FongKeimer00}. However, since this energy is far above the even gap of $\sim50$\,meV, even and odd values should not be substantially different.} and with a somewhat increased error tolerance probably far above these energies. This is further supported by a comparison with the evolution of the high-energy data of Hayden et al.\cite{Hayden04} The in-plane maps presented there suggest that the in-plane intensity distribution of our model, in particular the axial intensity distribution with maxima in the \{110\} directions, agrees with experiment up to the same energy. This is also apparent from a comparison with Fig.~\ref{fig:modelCuts} where we show constant-energy maps of \imChiQE\ at energies below and above \ores. When comparing it with Fig.~\ref{fig:TOF} and other measured data, it should be kept in mind that Fig.~\ref{fig:modelCuts} shows intensity that was not convoluted with the instrumental resolution function or averaged over a certain energy range: In the measured data, the intensity distribution always appears spread over a larger region. We also mention that due to the decreasing intensity and spurious contaminations of the signal it is difficult to obtain detailed data far below 20\,meV. We can clearly establish the opening of a superconducting gap and find indication for a linear decrease of the intensity. Also the strongly anisotropic incommensurability we observe at the lowest covered energies implies a natural extension towards $\EE=0$. All this is in agreement with Refs. \onlinecite{FongKeimer00} and \onlinecite{StockBuyers04} and suggests the validity of our models for energies approaching $\EE=0$. Finally, we also note that our model (Fig.~\ref{fig:modelCuts}b, artificially ``twinned'') is in remarkable agreement with the ``castle''-like structure reported by Mook \etal\ (Ref. \onlinecite{MookDai98}) for a twinned \ybcosixsix\ sample.

We will now evaluate in more detail how well our models reproduce the experiments. To do this, we compare the model functions, convoluted with the instrumental resolution function, with the experimental data using two different representations: In Figs. \ref{fig:SCFit}--\ref{fig:SC-NSFit} we use the color representation introduced in \ref{subsec:dispersion} (Figs. \ref{fig:mapsConv} and \ref{fig:mapsAlt}), which permits the assessment of the salient features like dispersion, hourglass constriction and resonance peak at a glance. Figs. \ref{fig:allFits} and \ref{fig:allFits2}, on the other hand, allow us a detailed comparison of the model with the individual constant-energy scan profiles.

Beginning in the superconducting state at 5\,K, the two representations show the excellent fit of our DHO model, both qualitatively and quantitatively. In particular, the hourglass dispersion with the typical intensity increase and \QQ-width constriction at \Eres\ is reproduced very well. The comparison in Fig.~\ref{fig:allFits} shows that this is not only true for the shape but also for the intensity. In Fig.~\ref{fig:allFits2} we compare our model with a scan profile along \astar\ which does not cross \qaf\ but $\QQ=(0.5, 1.44, 1.7)$. The good agreement confirms that our model also captures the essential features of the detailed in-plane distribution of the lower branch, as was already indicated by a qualitative comparison of Fig.~\ref{fig:TOF}a,b) and Fig.~\ref{fig:modelCuts}a).

Proceeding to 70\,K, we first note that the data cannot be reproduced by simply broadening the DHO model function used at 5\,K. This is illustrated by a comparison of the data (Fig.~\ref{fig:NSFit}a,b) with the ``best'' fitting broadened DHO model (Fig.~\ref{fig:NSFit}e,f). In particular, the DHO model still exhibits a commensurate resonance constriction, and the the lower and upper branches disperse towards the constriction in a way comparable to the situation at 5\,K. The only well-reproduced part of the spectrum is the high-energy part, which does not change significantly compared to 5\,K anyway.

A much better agreement with the data is obtained using the two-Gaussian model described in \ref{subsubsec:normalstate} for the lower branch: The steep dispersion and the absence of an hourglass constriction are reproduced very well. Also the quasi-one-dimensional character at $\EE<\Eres$ is captured. The detailed comparison with the individual scan profiles (Fig.~\ref{fig:allFits}) again confirms the good agreement. Both representations show that the signal along \bstar\ remains commensurate down to lowest covered energies.

Next, we show in Fig.~\ref{fig:SC-NSFit} that our models also reproduce the difference between the spin susceptibilities at 5 and 70\,K, where we observe just a downward dispersing mode.

\begin{figure}[t]
\includegraphics[width=0.9\columnwidth]{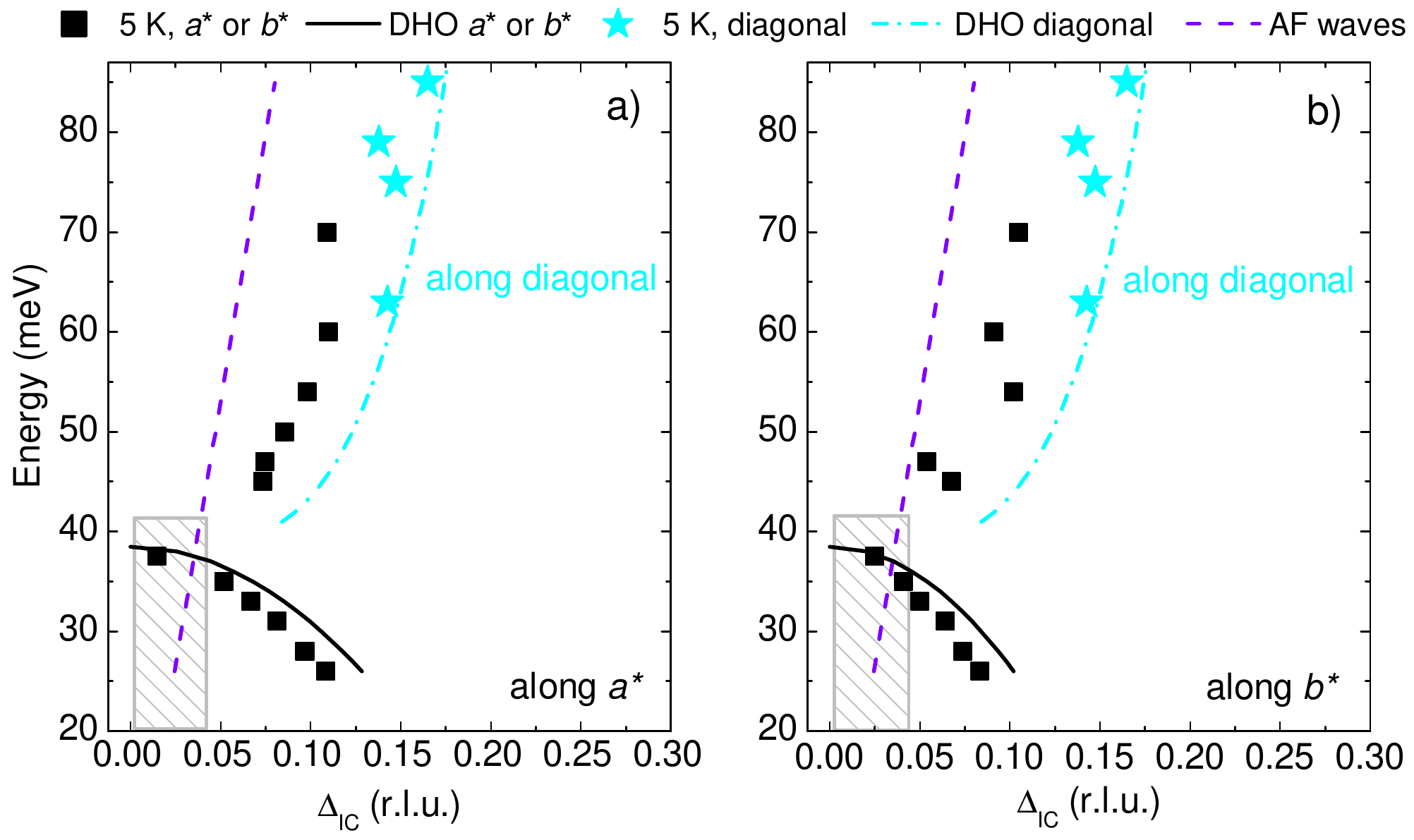}
\caption{\label{fig:dispDHO} Comparison of the dispersion relations as obtained from Gaussian fits to the data (symbols, see Fig.~\ref{fig:dispersion}) with those of the model derived in \ref{subsubsec:scstate} for $T=5$\,K (solid lines). Black symbols and lines are for the \astar-direction in a) and the \bstar-direction in b). Stars and dash-dotted lines are for the diagonal direction, where \deltaIC\ was obtained from the TOF data for the points at 63 and 75\,meV (Fig.~\ref{fig:TOF}), from the 3-axis data at 79\,meV (Fig.~\ref{fig:highE}) and from TOF measurements shown in Ref. \onlinecite{Hayden04} for 85\,meV.}
\end{figure}

Finally, we revisit a point raised in \ref{subsec:dispersion}, namely the dispersion relations of the upward- and downward-dispersing modes. In Fig.~\ref{fig:dispDHO} we compare the dispersion obtained from Gaussian fits to the scan profiles with the dispersion of the model functions. An inspection of  Fig.~\ref{fig:modelCuts} explains why the Gaussian fits always tend to yield a smaller incommensurability \deltaIC\ than the model \imChiQE: The in-plane azimuthal distribution exhibits appreciable intensity not only in the principal crystallographic directions but also between them. Due to the large resolution volume, in the course of a scan along \astar\ or \bstar\ (or along \{110\}, for that matter) intensity is still collected far beyond the maximum  of \imChiQE\ in this particular direction. This shifts the intensity maximum of the measured scan closer to \qaf, in particular at high energies where the resolution volume is especially large and the azimuthal intensity maxima are along \{110\}, affecting the scans along \astar\ and \bstar. This demonstrates that it is safer to consider the full in-plane distribution of the model \imChiQE, extracted under inclusion of the resolution function, as we did. Just showing the intensity maxima in particular directions, no matter whether extracted from simple fits to the data or from the model \imChiQE, reproduces qualitative features  but only conveys limited quantitative information.

\section{Discussion}
\label{sec:discussion}
In the previous section \ref{sec:results} we have shown that the salient features of the spin excitations in \ybcosixsix\ do not arise from resolution effects. Our analysis in \ref{subsec:model} has led to quantitative models for \imChiQE\ in both the superconducting and the normal states which capture these features, namely the continuous low-energy excitation branch at 5\,K and the abrupt topology change towards a ``Y''-shaped dispersion with no pronounced resonance feature at 70\,K; the change of the spectral distribution at low energies from weakly to strongly anisotropic upon heating above 70\,K; and the four-fold symmetry of the upper branch and its insensitivity to temperature variations across \tc.

What are the implications of these results for the scenarios invoked to explain the peculiar spin excitations and the delicate interplay between different competing phases in underdoped cuprates?

When the hourglass dispersion was reported for the \lsbco\ family,\cite{Tranquada04} the notion of bond-centered stripes in the form of weakly-coupled two-leg spin ladders was put forward as an explanation,\cite{Tranquada04} and the concomitant observation \cite{Hayden04} of a similar dispersion geometry in the superconducting state of \ybco\ seemed to establish this as a universal scenario for cuprates. This idea has triggered many subsequent theoretical papers \cite{AndersenHedegard05, YaoCarlson06, UhrigSchmidt04, VojtaUlbricht04} but it has now become clear that it is not applicable to \ybco\ without further elaboration. \cite{HinkovKeimer07} The first difficulty is the observed change from two-fold symmetry at low energies to four-fold symmetry at high energies (Figs. \ref{fig:TOF}, \ref{fig:highE} and \ref{fig:modelCuts}). In the model of two-leg spin ladders, a constant-energy cut through the upper branch should show two parallel streaks running across the ladder, since the gapped excitations disperse only along the one-dimensional ladder. Introducing weak correlations between the ladders results in two incommensurate spin-wave-like cones at low energies, displaced from \qaf\ perpendicularly to the ladder direction. The resulting spectrum is two-fold symmetric at any energy, in contrast to our data from twin-free samples, which show incommensurate peaks along both \astar\ and \bstar\ in high-energy scans through \qaf, (Fig.~\ref{fig:highE}). One might object that orthorhombicity does not provide a sufficiently strong potential to pin ladders \emph{either} along \astar\ \emph{or} along \bstar, and that domains of both orientations are present. This scenario is implicitly included in the calculations of Refs. \onlinecite{AndersenHedegard05,YaoCarlson06,UhrigSchmidt04,VojtaUlbricht04}, where the simulated data are artificially twinned. However our observation of a strong in-plane anisotropy at low energies in the normal-state data contradicts this assumption.

Different approaches have been used to address and to some degree to resolve these problems. Magnetic fluctuations of unidirectional, metallic stripes numerically computed in the framework of a Hubbard model exhibit two-fold symmetry at low energies, and high-energy branches dispersing along both orthogonal principal crystallographic directions without the necessity to invoke twinning.\cite{SeiboldLorenzana05} However, the model predicts a gap between lower and upper branches in the \astar-direction and a non-monotonic evolution in the \bstar-direction, which we do not observe experimentally. It has also been shown that the ``Y''-shaped dispersion can be understood as arising from fluctuating short-range stripe segments, and that an unequal population of the two perpendicular stripe orientations results in increasing anisotropy at low energies, in agreement with our observations. \cite{VojtaVojta06} However, this model is purely phenomenological and contains many fitting parameters that cannot be determined from first principles.

A second shortcoming of these early models of the hourglass dispersion is their failure to address the influence of superconductivity on the spin fluctuation spectrum. The neutron data we have presented here clearly document a transition from a ``Y''-shaped to an hourglass dispersion at $T\leq\tc$. We note that this change of the dispersion topology, and in particular the fact that the spectrum at 70\,K cannot be obtained from the one at 5\,K by increasing the mode damping (\ref{subsubsec:normalstate} and Fig.~\ref{fig:NSFit}), is difficult to reconcile with the notion of a collective mode above \tc,\cite{MorrPines98} which is merely concealed due to scattering by continuum excitations.

Our data rather indicate a ``resonant'' collective mode that only appears below \tc. The occurrence of a resonant mode in the $d$-wave superconducting state has been theoretically treated by many authors,\cite{DahmTewordt95a, DemlerZhang95, BulutScalapino96, MillisMonien96, AbanovChubukov99, KaoSi00, Onufrieva02, BrinckmannLee02, SegaBonca03} as reviewed in Ref. \onlinecite{Eschrig06}. In a simple Fermi-liquid-based scenario, the resonance is described as a spin exciton, \ie\ a spin-triplet bound state stabilized by electron-electron interactions below the electron-hole continuum setting in at twice the superconducting gap. The observed intensity redistribution below \tc, in which the resonant mode draws spectral weight from both lower and higher energies (Figs. \ref{fig:mapsConv}\,e),f) and \ref{fig:SC-NSFit}), as well as the formation of a downward dispersing branch with a shape roughly following the $d$-wave superconducting gap support this kind of scenario.\footnote{One might object that when comparing superconducting and normal state data, one also compares different temperatures (5\,K and 70\,K), which might be the reason for the different incommensurabilities and the topology change. Fig.~\ref{fig:tempDep}, however, shows that the superconducting signal geometry is already fully developed at 50\,K and hardly changes upon further cooling.} In the framework the weak-coupling Fermi-liquid scenarios proposed so far, \cite{ZhouLi04, EreminManske05, SchnyderManske06} hybridization of electronic states based on the CuO chains and CuO$_2$ layers in the \ybco\ crystal structure with electronic states in the CuO$_2$ layers might provide an explanation of the modest in-plane anisotropy in the spectra in the superconducting state, but appears too weak to account for the strong anisotropy of the spectra above \tc\ (Fig. \ref{fig:lowE}). We also note that the results of most weak-coupling calculations have thus far been presented in arbitrary intensity units. In view of the recently reported failure of such calculations to reproduce the absolute intensity of the observed spin excitations in an electron-doped cuprate superconductor, \cite{Kruger07} the comparisons between experimental and numerical data on an absolute intensity scale constitutes an important challenge for the future. The results presented here provide a good basis for studies designed to meet this challenge.

Recent theoretical work indicates that it may be possible to describe the spin excitation spectra above and below \tc\ in the framework of a single scenario, according to which superconductivity competes with a Pomeranchuk instability of the Fermi surface.  It is noteworthy that from a symmetry point of view, the Pomeranchuk model and the fluctuating-stripe models are equivalent: In both cases, the $C_4$ symmetry is spontaneously broken. An analysis of \imChiQE\ in the slave-boson mean-field scheme of Ref. \onlinecite{YamaseMetzner06} reproduces the transition from a ``Y''-shaped dispersion to an hourglass dispersion upon entering the SC state and (qualitatively) also the crossover from two-fold to four-fold symmetry with increasing energy. Future work will have to show if this framework can be extended to even lower doping levels, where the ground state exhibits incommensurate magnetic order and nearly gapless magnetic excitations, \cite{HinkovHaug08, HaugHinkov09} and is believed to become insulating at high magnetic fields.\cite{SunAndo04} In this regime, the localized-electron scenarios that have been discussed in the context of the spin excitations of \lsbco\ may be a more promising starting point.

For a complete understanding, some questions remain to be addressed in future investigations: First, it is interesting to determine the dispersion topology in \ybcosixfour\ and to compare it to the typical ``Y''-shaped dispersion we see here. Second, the onset temperature of the incommensurability in \ybcosixsix\ must be determined. This is particularly important in light of the fact that, based on the onset temperature of the in-plane anisotropy of the Nernst effect, $T_{\nu}$ the authors of Ref. \onlinecite{DaouTaillefer10} relate electronic liquid-crystalline order to the pseudo-gap phenomenon. Interestingly, the onset temperature of the novel magnetic order at $Q=0$, which was also related to the pseudo-gap state, is  $\sim220$\,K in our \ybcosixsix\ sample,\cite{FauqueBourges06}and thus close to $T_{\nu}$.

Finally, we expect that the full $\mathbf{Q}$- and $\omega$-resolved data set of magnetic excitations and the analytical description we have presented here will greatly facilitate quantitative comparison with spectroscopic data obtained by complementary experimental methods. Previous attempts to relate anomalous features in charge-fluctuation spectra to the spin excitations have often narrowly focused on particular regions of momentum space. For instance, because of the strong temperature dependence of the spin fluctuation spectrum near \qaf, the insensitivity of the nodal kink in ARPES spectra to changing the temperature across \tc\ had been taken as an indication that this feature originates from coupling to phonons. We have shown, however, that coupling to the weakly temperature dependent high-energy spin-excitation branch contributes substantially to the nodal kink observed in ARPES spectra.\cite{DahmHinkov09} For the same reason, the absence of renormalization effects attributable to the magnetic resonant mode in optical spectra of overdoped cuprates \cite{HwangTimusk04} does not invalidate theories based on magnetically mediated pairing mechanisms. A quantitative description of optical and STS spectra in terms of spin-fermion models requires a full energy- and momentum-integration over all spin-fluctuation mediated scattering channels. The analytical model functions presented here for \ybcosixsix\ provide a good starting point for such endeavors. As we have shown in the case of ARPES, \cite{DahmHinkov09} this approach also allows quantitative experimental tests of spin-fluctuation mediated pairing models of high-temperature superconductivity.

 \emph{Acknowledgements:} We thank  C.~T.~Lin, D.~P.~Chen, S.~Lacher, B.~Baum and M. Raichle for the sample preparation, M.~Ohl for the design of the sample holder and S. Pailh{\` e}s, B. Fauqu{\' e} and T.~G.~Perring for help during the experiment and discussions. We thank T.~Dahm, D.~S.~Inosov, W.~Metzner, O.~P.~Sushkov, J.~M.~Tranquada and H.~Yamase for fruitful discussions.

\bibliography{../../../papers/cuprates}

\begin{thebibliography}{124}%
\makeatletter
\providecommand \@ifxundefined [1]{%
 \@ifx{#1\undefined}
}%
\providecommand \@ifnum [1]{%
 \ifnum #1\expandafter \@firstoftwo
 \else \expandafter \@secondoftwo
 \fi
}%
\providecommand \@ifx [1]{%
 \ifx #1\expandafter \@firstoftwo
 \else \expandafter \@secondoftwo
 \fi
}%
\providecommand \natexlab [1]{#1}%
\providecommand \enquote  [1]{``#1''}%
\providecommand \bibnamefont  [1]{#1}%
\providecommand \bibfnamefont [1]{#1}%
\providecommand \citenamefont [1]{#1}%
\providecommand \href@noop [0]{\@secondoftwo}%
\providecommand \href [0]{\begingroup \@sanitize@url \@href}%
\providecommand \@href[1]{\@@startlink{#1}\@@href}%
\providecommand \@@href[1]{\endgroup#1\@@endlink}%
\providecommand \@sanitize@url [0]{\catcode `\\12\catcode `\$12\catcode
  `\&12\catcode `\#12\catcode `\^12\catcode `\_12\catcode `\%12\relax}%
\providecommand \@@startlink[1]{}%
\providecommand \@@endlink[0]{}%
\providecommand \url  [0]{\begingroup\@sanitize@url \@url }%
\providecommand \@url [1]{\endgroup\@href {#1}{\urlprefix }}%
\providecommand \urlprefix  [0]{URL }%
\providecommand \Eprint [0]{\href }%
\@ifxundefined \urlstyle {%
  \providecommand \doi  [0]{\begingroup \@sanitize@url \@doi}%
  \providecommand \@doi [1]{\endgroup \@@startlink {\doibase
  #1}doi:\discretionary {}{}{}#1\@@endlink }%
}{%
  \providecommand \doi  [0]{doi:\discretionary{}{}{}\begingroup
  \urlstyle{rm}\Url }%
}%
\providecommand \doibase [0]{http://dx.doi.org/}%
\providecommand \Doi [0]{\begingroup \@sanitize@url \@Doi }%
\providecommand \@Doi  [1]{\endgroup\@@startlink{\doibase#1}\@@Doi}%
\providecommand \@@Doi [1]{#1\@@endlink}%
\providecommand \selectlanguage [0]{\@gobble}%
\providecommand \bibinfo  [0]{\@secondoftwo}%
\providecommand \bibfield  [0]{\@secondoftwo}%
\providecommand \translation [1]{[#1]}%
\providecommand \BibitemOpen [0]{}%
\providecommand \bibitemStop [0]{}%
\providecommand \bibitemNoStop [0]{.\EOS\space}%
\providecommand \EOS [0]{\spacefactor3000\relax}%
\providecommand \BibitemShut  [1]{\csname bibitem#1\endcsname}%
\bibitem [{\citenamefont {Doiron-Leyraud}\ \emph {et~al.}(2007)\citenamefont
  {Doiron-Leyraud}, \citenamefont {Proust}, \citenamefont {LeBoeuf},
  \citenamefont {Levallois}, \citenamefont {Bonnemaison}, \citenamefont
  {Liang}, \citenamefont {Bonn}, \citenamefont {Hardy},\ and\ \citenamefont
  {Taillefer}}]{Doiron-LeyraudTaillefer07}%
  \BibitemOpen
  \bibfield  {author} {\bibinfo {author} {\bibfnamefont {N.}~\bibnamefont
  {Doiron-Leyraud}}, \bibinfo {author} {\bibfnamefont {C.}~\bibnamefont
  {Proust}}, \bibinfo {author} {\bibfnamefont {D.}~\bibnamefont {LeBoeuf}},
  \bibinfo {author} {\bibfnamefont {J.}~\bibnamefont {Levallois}}, \bibinfo
  {author} {\bibfnamefont {J.-B.}\ \bibnamefont {Bonnemaison}}, \bibinfo
  {author} {\bibfnamefont {R.}~\bibnamefont {Liang}}, \bibinfo {author}
  {\bibfnamefont {D.~A.}\ \bibnamefont {Bonn}}, \bibinfo {author}
  {\bibfnamefont {W.~N.}\ \bibnamefont {Hardy}}, \ and\ \bibinfo {author}
  {\bibfnamefont {L.}~\bibnamefont {Taillefer}},\ }\Doi {10.1038/nature05872}
  {\bibfield  {journal} {\bibinfo  {journal} {Nature},\ }\textbf {\bibinfo
  {volume} {447}},\ \bibinfo {pages} {565} (\bibinfo {year}
  {2007})}\BibitemShut {NoStop}%
\bibitem [{\citenamefont {Bangura}\ \emph {et~al.}(2008)\citenamefont
  {Bangura}, \citenamefont {Fletcher}, \citenamefont {Carrington},
  \citenamefont {Levallois}, \citenamefont {Nardone}, \citenamefont {Vignolle},
  \citenamefont {Heard}, \citenamefont {Doiron-Leyraud}, \citenamefont
  {LeBoeuf}, \citenamefont {Taillefer}, \citenamefont {Adachi}, \citenamefont
  {Proust},\ and\ \citenamefont {Hussey}}]{BanguraHussey08}%
  \BibitemOpen
  \bibfield  {author} {\bibinfo {author} {\bibfnamefont {A.~F.}\ \bibnamefont
  {Bangura}}, \bibinfo {author} {\bibfnamefont {J.~D.}\ \bibnamefont
  {Fletcher}}, \bibinfo {author} {\bibfnamefont {A.}~\bibnamefont
  {Carrington}}, \bibinfo {author} {\bibfnamefont {J.}~\bibnamefont
  {Levallois}}, \bibinfo {author} {\bibfnamefont {M.}~\bibnamefont {Nardone}},
  \bibinfo {author} {\bibfnamefont {B.}~\bibnamefont {Vignolle}}, \bibinfo
  {author} {\bibfnamefont {P.~J.}\ \bibnamefont {Heard}}, \bibinfo {author}
  {\bibfnamefont {N.}~\bibnamefont {Doiron-Leyraud}}, \bibinfo {author}
  {\bibfnamefont {D.}~\bibnamefont {LeBoeuf}}, \bibinfo {author} {\bibfnamefont
  {L.}~\bibnamefont {Taillefer}}, \bibinfo {author} {\bibfnamefont
  {S.}~\bibnamefont {Adachi}}, \bibinfo {author} {\bibfnamefont
  {C.}~\bibnamefont {Proust}}, \ and\ \bibinfo {author} {\bibfnamefont {N.~E.}\
  \bibnamefont {Hussey}},\ }\Doi {10.1103/PhysRevLett.100.047004} {\bibfield
  {journal} {\bibinfo  {journal} {Phys. Rev. Lett.},\ }\textbf {\bibinfo
  {volume} {100}},\ \bibinfo {pages} {047004} (\bibinfo {year}
  {2008})}\BibitemShut {NoStop}%
\bibitem [{\citenamefont {Yelland}\ \emph {et~al.}(2008)\citenamefont
  {Yelland}, \citenamefont {Singleton}, \citenamefont {Mielke}, \citenamefont
  {Harrison}, \citenamefont {Balakirev}, \citenamefont {Dabrowski},\ and\
  \citenamefont {Cooper}}]{YellandCooper08}%
  \BibitemOpen
  \bibfield  {author} {\bibinfo {author} {\bibfnamefont {E.~A.}\ \bibnamefont
  {Yelland}}, \bibinfo {author} {\bibfnamefont {J.}~\bibnamefont {Singleton}},
  \bibinfo {author} {\bibfnamefont {C.~H.}\ \bibnamefont {Mielke}}, \bibinfo
  {author} {\bibfnamefont {N.}~\bibnamefont {Harrison}}, \bibinfo {author}
  {\bibfnamefont {F.~F.}\ \bibnamefont {Balakirev}}, \bibinfo {author}
  {\bibfnamefont {B.}~\bibnamefont {Dabrowski}}, \ and\ \bibinfo {author}
  {\bibfnamefont {J.~R.}\ \bibnamefont {Cooper}},\ }\Doi
  {10.1103/PhysRevLett.100.047003} {\bibfield  {journal} {\bibinfo  {journal}
  {Phys. Rev. Lett.},\ }\textbf {\bibinfo {volume} {100}},\ \bibinfo {pages}
  {047003} (\bibinfo {year} {2008})}\BibitemShut {NoStop}%
\bibitem [{\citenamefont {Helm}\ \emph {et~al.}(2009)\citenamefont {Helm},
  \citenamefont {Kartsovnik}, \citenamefont {Bartkowiak}, \citenamefont
  {Bittner}, \citenamefont {Lambacher}, \citenamefont {Erb}, \citenamefont
  {Wosnitza},\ and\ \citenamefont {Gross}}]{HelmGross09}%
  \BibitemOpen
  \bibfield  {author} {\bibinfo {author} {\bibfnamefont {T.}~\bibnamefont
  {Helm}}, \bibinfo {author} {\bibfnamefont {M.~V.}\ \bibnamefont
  {Kartsovnik}}, \bibinfo {author} {\bibfnamefont {M.}~\bibnamefont
  {Bartkowiak}}, \bibinfo {author} {\bibfnamefont {N.}~\bibnamefont {Bittner}},
  \bibinfo {author} {\bibfnamefont {M.}~\bibnamefont {Lambacher}}, \bibinfo
  {author} {\bibfnamefont {A.}~\bibnamefont {Erb}}, \bibinfo {author}
  {\bibfnamefont {J.}~\bibnamefont {Wosnitza}}, \ and\ \bibinfo {author}
  {\bibfnamefont {R.}~\bibnamefont {Gross}},\ }\Doi
  {10.1103/PhysRevLett.103.157002} {\bibfield  {journal} {\bibinfo  {journal}
  {Phys. Rev. Lett.},\ }\textbf {\bibinfo {volume} {103}},\ \bibinfo {pages}
  {157002} (\bibinfo {year} {2009})}\BibitemShut {NoStop}%
\bibitem [{\citenamefont {Borisenko}\ \emph {et~al.}(2006)\citenamefont
  {Borisenko}, \citenamefont {Kordyuk}, \citenamefont {Zabolotnyy},
  \citenamefont {Geck}, \citenamefont {Inosov}, \citenamefont {Koitzsch},
  \citenamefont {Fink}, \citenamefont {Knupfer}, \citenamefont {B{\"u}chner},
  \citenamefont {Hinkov}, \citenamefont {Lin}, \citenamefont {Keimer},
  \citenamefont {Wolf}, \citenamefont {Chiuzbaian}, \citenamefont {Patthey},\
  and\ \citenamefont {Follath}}]{Borisenko06}%
  \BibitemOpen
  \bibfield  {author} {\bibinfo {author} {\bibfnamefont {S.~V.}\ \bibnamefont
  {Borisenko}}, \bibinfo {author} {\bibfnamefont {A.~A.}\ \bibnamefont
  {Kordyuk}}, \bibinfo {author} {\bibfnamefont {V.~B.}\ \bibnamefont
  {Zabolotnyy}}, \bibinfo {author} {\bibfnamefont {J.}~\bibnamefont {Geck}},
  \bibinfo {author} {\bibfnamefont {D.~S.}\ \bibnamefont {Inosov}}, \bibinfo
  {author} {\bibfnamefont {A.}~\bibnamefont {Koitzsch}}, \bibinfo {author}
  {\bibfnamefont {J.}~\bibnamefont {Fink}}, \bibinfo {author} {\bibfnamefont
  {M.}~\bibnamefont {Knupfer}}, \bibinfo {author} {\bibfnamefont
  {B.}~\bibnamefont {B{\"u}chner}}, \bibinfo {author} {\bibfnamefont
  {V.}~\bibnamefont {Hinkov}}, \bibinfo {author} {\bibfnamefont {C.~T.}\
  \bibnamefont {Lin}}, \bibinfo {author} {\bibfnamefont {B.}~\bibnamefont
  {Keimer}}, \bibinfo {author} {\bibfnamefont {T.}~\bibnamefont {Wolf}},
  \bibinfo {author} {\bibfnamefont {S.~G.}\ \bibnamefont {Chiuzbaian}},
  \bibinfo {author} {\bibfnamefont {L.}~\bibnamefont {Patthey}}, \ and\
  \bibinfo {author} {\bibfnamefont {R.}~\bibnamefont {Follath}},\ }\Doi
  {10.1103/PhysRevLett.96.11700} {\bibfield  {journal} {\bibinfo  {journal}
  {Phys. Rev. Lett.},\ }\textbf {\bibinfo {volume} {96}},\ \bibinfo {pages}
  {117004} (\bibinfo {year} {2006})}\BibitemShut {NoStop}%
\bibitem [{\citenamefont {Zabolotnyy}\ \emph
  {et~al.}(2007){\natexlab{a}}\citenamefont {Zabolotnyy}, \citenamefont
  {Borisenko}, \citenamefont {Kordyuk}, \citenamefont {Geck}, \citenamefont
  {Inosov}, \citenamefont {Koitzsch}, \citenamefont {Fink}, \citenamefont
  {Knupfer}, \citenamefont {B{\"u}chner}, \citenamefont {Drechsler},
  \citenamefont {Berger}, \citenamefont {Erb}, \citenamefont {Lambacher},
  \citenamefont {Patthey}, \citenamefont {Hinkov},\ and\ \citenamefont
  {Keimer}}]{ZabolotnyyBorisenko07}%
  \BibitemOpen
  \bibfield  {author} {\bibinfo {author} {\bibfnamefont {V.~B.}\ \bibnamefont
  {Zabolotnyy}}, \bibinfo {author} {\bibfnamefont {S.~V.}\ \bibnamefont
  {Borisenko}}, \bibinfo {author} {\bibfnamefont {A.~A.}\ \bibnamefont
  {Kordyuk}}, \bibinfo {author} {\bibfnamefont {J.}~\bibnamefont {Geck}},
  \bibinfo {author} {\bibfnamefont {D.~S.}\ \bibnamefont {Inosov}}, \bibinfo
  {author} {\bibfnamefont {A.}~\bibnamefont {Koitzsch}}, \bibinfo {author}
  {\bibfnamefont {J.}~\bibnamefont {Fink}}, \bibinfo {author} {\bibfnamefont
  {M.}~\bibnamefont {Knupfer}}, \bibinfo {author} {\bibfnamefont
  {B.}~\bibnamefont {B{\"u}chner}}, \bibinfo {author} {\bibfnamefont {S.~L.}\
  \bibnamefont {Drechsler}}, \bibinfo {author} {\bibfnamefont {H.}~\bibnamefont
  {Berger}}, \bibinfo {author} {\bibfnamefont {A.}~\bibnamefont {Erb}},
  \bibinfo {author} {\bibfnamefont {M.}~\bibnamefont {Lambacher}}, \bibinfo
  {author} {\bibfnamefont {L.}~\bibnamefont {Patthey}}, \bibinfo {author}
  {\bibfnamefont {V.}~\bibnamefont {Hinkov}}, \ and\ \bibinfo {author}
  {\bibfnamefont {B.}~\bibnamefont {Keimer}},\ }\Doi
  {10.1103/PhysRevB.76.064519} {\bibfield  {journal} {\bibinfo  {journal}
  {Phys. Rev. B},\ }\textbf {\bibinfo {volume} {76}},\ \bibinfo {pages}
  {064519} (\bibinfo {year} {2007}{\natexlab{a}})}\BibitemShut {NoStop}%
\bibitem [{\citenamefont {Zabolotnyy}\ \emph
  {et~al.}(2007){\natexlab{b}}\citenamefont {Zabolotnyy}, \citenamefont
  {Borisenko}, \citenamefont {Kordyuk}, \citenamefont {Inosov}, \citenamefont
  {Koitzsch}, \citenamefont {Geck}, \citenamefont {Fink}, \citenamefont
  {Knupfer}, \citenamefont {Buechner}, \citenamefont {Drechsler}, \citenamefont
  {Hinkov}, \citenamefont {Keimer},\ and\ \citenamefont
  {Patthey}}]{ZabolotnyyBorisenko07a}%
  \BibitemOpen
  \bibfield  {author} {\bibinfo {author} {\bibfnamefont {V.~B.}\ \bibnamefont
  {Zabolotnyy}}, \bibinfo {author} {\bibfnamefont {S.~V.}\ \bibnamefont
  {Borisenko}}, \bibinfo {author} {\bibfnamefont {A.~A.}\ \bibnamefont
  {Kordyuk}}, \bibinfo {author} {\bibfnamefont {D.~S.}\ \bibnamefont {Inosov}},
  \bibinfo {author} {\bibfnamefont {A.}~\bibnamefont {Koitzsch}}, \bibinfo
  {author} {\bibfnamefont {J.}~\bibnamefont {Geck}}, \bibinfo {author}
  {\bibfnamefont {J.}~\bibnamefont {Fink}}, \bibinfo {author} {\bibfnamefont
  {M.}~\bibnamefont {Knupfer}}, \bibinfo {author} {\bibfnamefont
  {B.}~\bibnamefont {Buechner}}, \bibinfo {author} {\bibfnamefont {S.-L.}\
  \bibnamefont {Drechsler}}, \bibinfo {author} {\bibfnamefont {V.}~\bibnamefont
  {Hinkov}}, \bibinfo {author} {\bibfnamefont {B.}~\bibnamefont {Keimer}}, \
  and\ \bibinfo {author} {\bibfnamefont {L.}~\bibnamefont {Patthey}},\ }\Doi
  {10.1103/PhysRevB.76.024502} {\bibfield  {journal} {\bibinfo  {journal}
  {Phys. Rev. B},\ }\textbf {\bibinfo {volume} {76}},\ \bibinfo {pages}
  {024502} (\bibinfo {year} {2007}{\natexlab{b}})}\BibitemShut {NoStop}%
\bibitem [{\citenamefont {Damascelli}\ \emph {et~al.}(2003)\citenamefont
  {Damascelli}, \citenamefont {Hussain},\ and\ \citenamefont
  {Shen}}]{DamascelliShen03}%
  \BibitemOpen
  \bibfield  {author} {\bibinfo {author} {\bibfnamefont {A.}~\bibnamefont
  {Damascelli}}, \bibinfo {author} {\bibfnamefont {Z.}~\bibnamefont {Hussain}},
  \ and\ \bibinfo {author} {\bibfnamefont {Z.-X.}\ \bibnamefont {Shen}},\
  }\href@noop {} {\bibfield  {journal} {\bibinfo  {journal} {Rev. Mod. Phys.},\
  }\textbf {\bibinfo {volume} {75}},\ \bibinfo {pages} {473} (\bibinfo {year}
  {2003})}\BibitemShut {NoStop}%
\bibitem [{\citenamefont {Lee}\ \emph {et~al.}(2006)\citenamefont {Lee},
  \citenamefont {Fujita}, \citenamefont {McElroy}, \citenamefont {Slezak},
  \citenamefont {Wang}, \citenamefont {Aiura}, \citenamefont {Bando},
  \citenamefont {Ishikado}, \citenamefont {Masui}, \citenamefont {Zhu},
  \citenamefont {Balatsky}, \citenamefont {Eisaki}, \citenamefont {Uchida},\
  and\ \citenamefont {Davis}}]{LeeDavis06}%
  \BibitemOpen
  \bibfield  {author} {\bibinfo {author} {\bibfnamefont {J.}~\bibnamefont
  {Lee}}, \bibinfo {author} {\bibfnamefont {K.}~\bibnamefont {Fujita}},
  \bibinfo {author} {\bibfnamefont {K.}~\bibnamefont {McElroy}}, \bibinfo
  {author} {\bibfnamefont {J.~A.}\ \bibnamefont {Slezak}}, \bibinfo {author}
  {\bibfnamefont {M.}~\bibnamefont {Wang}}, \bibinfo {author} {\bibfnamefont
  {Y.}~\bibnamefont {Aiura}}, \bibinfo {author} {\bibfnamefont
  {H.}~\bibnamefont {Bando}}, \bibinfo {author} {\bibfnamefont
  {M.}~\bibnamefont {Ishikado}}, \bibinfo {author} {\bibfnamefont
  {T.}~\bibnamefont {Masui}}, \bibinfo {author} {\bibfnamefont {J.~X.}\
  \bibnamefont {Zhu}}, \bibinfo {author} {\bibfnamefont {A.~V.}\ \bibnamefont
  {Balatsky}}, \bibinfo {author} {\bibfnamefont {H.}~\bibnamefont {Eisaki}},
  \bibinfo {author} {\bibfnamefont {S.}~\bibnamefont {Uchida}}, \ and\ \bibinfo
  {author} {\bibfnamefont {J.~C.}\ \bibnamefont {Davis}},\ }\Doi
  {10.1038/nature04973} {\bibfield  {journal} {\bibinfo  {journal} {Nature},\
  }\textbf {\bibinfo {volume} {442}},\ \bibinfo {pages} {546} (\bibinfo {year}
  {2006})}\BibitemShut {NoStop}%
\bibitem [{\citenamefont {Jenkins}\ \emph {et~al.}(2009)\citenamefont
  {Jenkins}, \citenamefont {Fasano}, \citenamefont {Berthod}, \citenamefont
  {Maggio-Aprile}, \citenamefont {Piriou}, \citenamefont {Giannini},
  \citenamefont {Hoogenboom}, \citenamefont {Hess}, \citenamefont {Cren},\ and\
  \citenamefont {Fischer}}]{Jenkins09}%
  \BibitemOpen
  \bibfield  {author} {\bibinfo {author} {\bibfnamefont {N.}~\bibnamefont
  {Jenkins}}, \bibinfo {author} {\bibfnamefont {Y.}~\bibnamefont {Fasano}},
  \bibinfo {author} {\bibfnamefont {C.}~\bibnamefont {Berthod}}, \bibinfo
  {author} {\bibfnamefont {I.}~\bibnamefont {Maggio-Aprile}}, \bibinfo {author}
  {\bibfnamefont {A.}~\bibnamefont {Piriou}}, \bibinfo {author} {\bibfnamefont
  {E.}~\bibnamefont {Giannini}}, \bibinfo {author} {\bibfnamefont {B.~W.}\
  \bibnamefont {Hoogenboom}}, \bibinfo {author} {\bibfnamefont
  {C.}~\bibnamefont {Hess}}, \bibinfo {author} {\bibfnamefont {T.}~\bibnamefont
  {Cren}}, \ and\ \bibinfo {author} {\bibfnamefont {O.}~\bibnamefont
  {Fischer}},\ }\Doi {10.1103/PhysRevLett.103.227001} {\bibfield  {journal}
  {\bibinfo  {journal} {Phys. Rev. Lett.},\ }\textbf {\bibinfo {volume}
  {103}},\ \bibinfo {pages} {227001} (\bibinfo {year} {2009})}\BibitemShut
  {NoStop}%
\bibitem [{\citenamefont {Hwang}\ \emph {et~al.}(2007)\citenamefont {Hwang},
  \citenamefont {Timusk}, \citenamefont {Schachinger},\ and\ \citenamefont
  {Carbotte}}]{Hwang07}%
  \BibitemOpen
  \bibfield  {author} {\bibinfo {author} {\bibfnamefont {J.}~\bibnamefont
  {Hwang}}, \bibinfo {author} {\bibfnamefont {T.}~\bibnamefont {Timusk}},
  \bibinfo {author} {\bibfnamefont {E.}~\bibnamefont {Schachinger}}, \ and\
  \bibinfo {author} {\bibfnamefont {J.~P.}\ \bibnamefont {Carbotte}},\ }\Doi
  {10.1103/PhysRevB.75.144508} {\bibfield  {journal} {\bibinfo  {journal}
  {Phys. Rev. B},\ }\textbf {\bibinfo {volume} {75}},\ \bibinfo {pages}
  {144508} (\bibinfo {year} {2007})}\BibitemShut {NoStop}%
\bibitem [{\citenamefont {van Heumen}\ \emph {et~al.}(2009)\citenamefont {van
  Heumen}, \citenamefont {Muhlethaler}, \citenamefont {Kuzmenko}, \citenamefont
  {Eisaki}, \citenamefont {Meevasana}, \citenamefont {Greven},\ and\
  \citenamefont {van~der Marel}}]{Heumen09}%
  \BibitemOpen
  \bibfield  {author} {\bibinfo {author} {\bibfnamefont {E.}~\bibnamefont {van
  Heumen}}, \bibinfo {author} {\bibfnamefont {E.}~\bibnamefont {Muhlethaler}},
  \bibinfo {author} {\bibfnamefont {A.~B.}\ \bibnamefont {Kuzmenko}}, \bibinfo
  {author} {\bibfnamefont {H.}~\bibnamefont {Eisaki}}, \bibinfo {author}
  {\bibfnamefont {W.}~\bibnamefont {Meevasana}}, \bibinfo {author}
  {\bibfnamefont {M.}~\bibnamefont {Greven}}, \ and\ \bibinfo {author}
  {\bibfnamefont {D.}~\bibnamefont {van~der Marel}},\ }\Doi
  {10.1103/PhysRevB.79.184512} {\bibfield  {journal} {\bibinfo  {journal}
  {Phys. Rev. B},\ }\textbf {\bibinfo {volume} {79}},\ \bibinfo {pages}
  {184512} (\bibinfo {year} {2009})}\BibitemShut {NoStop}%
\bibitem [{\citenamefont {Anderson}(2007)}]{Anderson07}%
  \BibitemOpen
  \bibfield  {author} {\bibinfo {author} {\bibfnamefont {P.~W.}\ \bibnamefont
  {Anderson}},\ }\Doi {10.1126/science.1140970} {\bibfield  {journal} {\bibinfo
   {journal} {Science},\ }\textbf {\bibinfo {volume} {316}},\ \bibinfo {pages}
  {1705} (\bibinfo {year} {2007})}\BibitemShut {NoStop}%
\bibitem [{\citenamefont {Khatami}\ \emph {et~al.}(2009)\citenamefont
  {Khatami}, \citenamefont {Macridin},\ and\ \citenamefont
  {Jarrell}}]{Khatami09}%
  \BibitemOpen
  \bibfield  {author} {\bibinfo {author} {\bibfnamefont {E.}~\bibnamefont
  {Khatami}}, \bibinfo {author} {\bibfnamefont {A.}~\bibnamefont {Macridin}}, \
  and\ \bibinfo {author} {\bibfnamefont {M.}~\bibnamefont {Jarrell}},\
  }\href@noop {} {\bibfield  {journal} {\bibinfo  {journal} {Phys. Rev. B},\
  }\textbf {\bibinfo {volume} {80}},\ \bibinfo {pages} {172505} (\bibinfo
  {year} {2009})}\BibitemShut {NoStop}%
\bibitem [{\citenamefont {Bobroff}\ \emph {et~al.}(2002)\citenamefont
  {Bobroff}, \citenamefont {Alloul}, \citenamefont {Ouazi}, \citenamefont
  {Mendels}, \citenamefont {Mahajan}, \citenamefont {Blanchard}, \citenamefont
  {Collin}, \citenamefont {Guillen},\ and\ \citenamefont
  {Marucco}}]{BobroffAlloul02}%
  \BibitemOpen
  \bibfield  {author} {\bibinfo {author} {\bibfnamefont {J.}~\bibnamefont
  {Bobroff}}, \bibinfo {author} {\bibfnamefont {H.}~\bibnamefont {Alloul}},
  \bibinfo {author} {\bibfnamefont {S.}~\bibnamefont {Ouazi}}, \bibinfo
  {author} {\bibfnamefont {P.}~\bibnamefont {Mendels}}, \bibinfo {author}
  {\bibfnamefont {A.}~\bibnamefont {Mahajan}}, \bibinfo {author} {\bibfnamefont
  {N.}~\bibnamefont {Blanchard}}, \bibinfo {author} {\bibfnamefont
  {G.}~\bibnamefont {Collin}}, \bibinfo {author} {\bibfnamefont
  {V.}~\bibnamefont {Guillen}}, \ and\ \bibinfo {author} {\bibfnamefont
  {J.}~\bibnamefont {Marucco}},\ }\Doi {10.1103/PhysRevLett.89.157002}
  {\bibfield  {journal} {\bibinfo  {journal} {Phys. Rev. Lett.},\ }\textbf
  {\bibinfo {volume} {89}},\ \bibinfo {pages} {157002} (\bibinfo {year}
  {2002})}\BibitemShut {NoStop}%
\bibitem [{\citenamefont {Arai}\ \emph {et~al.}(1999)\citenamefont {Arai},
  \citenamefont {Nishijima}, \citenamefont {Endoh}, \citenamefont {Egami},
  \citenamefont {Tajima}, \citenamefont {Tomimoto}, \citenamefont {Shiohara},
  \citenamefont {Takahashi}, \citenamefont {Garrett},\ and\ \citenamefont
  {Bennington}}]{Arai99}%
  \BibitemOpen
  \bibfield  {author} {\bibinfo {author} {\bibfnamefont {M.}~\bibnamefont
  {Arai}}, \bibinfo {author} {\bibfnamefont {T.}~\bibnamefont {Nishijima}},
  \bibinfo {author} {\bibfnamefont {Y.}~\bibnamefont {Endoh}}, \bibinfo
  {author} {\bibfnamefont {T.}~\bibnamefont {Egami}}, \bibinfo {author}
  {\bibfnamefont {S.}~\bibnamefont {Tajima}}, \bibinfo {author} {\bibfnamefont
  {K.}~\bibnamefont {Tomimoto}}, \bibinfo {author} {\bibfnamefont
  {Y.}~\bibnamefont {Shiohara}}, \bibinfo {author} {\bibfnamefont
  {M.}~\bibnamefont {Takahashi}}, \bibinfo {author} {\bibfnamefont
  {A.}~\bibnamefont {Garrett}}, \ and\ \bibinfo {author} {\bibfnamefont
  {S.~M.}\ \bibnamefont {Bennington}},\ }\href@noop {} {\bibfield  {journal}
  {\bibinfo  {journal} {Phys. Rev. Lett.},\ }\textbf {\bibinfo {volume} {83}},\
  \bibinfo {pages} {608} (\bibinfo {year} {1999})}\BibitemShut {NoStop}%
\bibitem [{\citenamefont {Hayden}\ \emph {et~al.}(2004)\citenamefont {Hayden},
  \citenamefont {Mook}, \citenamefont {Dai}, \citenamefont {Perring},\ and\
  \citenamefont {Do{\v g}an}}]{Hayden04}%
  \BibitemOpen
  \bibfield  {author} {\bibinfo {author} {\bibfnamefont {S.~M.}\ \bibnamefont
  {Hayden}}, \bibinfo {author} {\bibfnamefont {H.~A.}\ \bibnamefont {Mook}},
  \bibinfo {author} {\bibfnamefont {P.~C.}\ \bibnamefont {Dai}}, \bibinfo
  {author} {\bibfnamefont {T.~G.}\ \bibnamefont {Perring}}, \ and\ \bibinfo
  {author} {\bibfnamefont {F.}~\bibnamefont {Do{\v g}an}},\ }\Doi
  {10.1038/nature02576} {\bibfield  {journal} {\bibinfo  {journal} {Nature},\
  }\textbf {\bibinfo {volume} {429}},\ \bibinfo {pages} {531} (\bibinfo {year}
  {2004})}\BibitemShut {NoStop}%
\bibitem [{\citenamefont {Hinkov}\ \emph {et~al.}(2007)\citenamefont {Hinkov},
  \citenamefont {Bourges}, \citenamefont {Pailh{\` e}s}, \citenamefont {Sidis},
  \citenamefont {Ivanov}, \citenamefont {Frost}, \citenamefont {Perring},
  \citenamefont {Lin}, \citenamefont {Chen},\ and\ \citenamefont
  {Keimer}}]{HinkovKeimer07}%
  \BibitemOpen
  \bibfield  {author} {\bibinfo {author} {\bibfnamefont {V.}~\bibnamefont
  {Hinkov}}, \bibinfo {author} {\bibfnamefont {P.}~\bibnamefont {Bourges}},
  \bibinfo {author} {\bibfnamefont {S.}~\bibnamefont {Pailh{\` e}s}}, \bibinfo
  {author} {\bibfnamefont {Y.}~\bibnamefont {Sidis}}, \bibinfo {author}
  {\bibfnamefont {A.}~\bibnamefont {Ivanov}}, \bibinfo {author} {\bibfnamefont
  {C.~D.}\ \bibnamefont {Frost}}, \bibinfo {author} {\bibfnamefont {T.~G.}\
  \bibnamefont {Perring}}, \bibinfo {author} {\bibfnamefont {C.~T.}\
  \bibnamefont {Lin}}, \bibinfo {author} {\bibfnamefont {D.~P.}\ \bibnamefont
  {Chen}}, \ and\ \bibinfo {author} {\bibfnamefont {B.}~\bibnamefont
  {Keimer}},\ }\Doi {10.1038/nphys720} {\bibfield  {journal} {\bibinfo
  {journal} {Nat. Phys.},\ }\textbf {\bibinfo {volume} {3}},\ \bibinfo {pages}
  {780} (\bibinfo {year} {2007})}\BibitemShut {NoStop}%
\bibitem [{\citenamefont {Pailh{\` e}s}\ \emph {et~al.}(2003)\citenamefont
  {Pailh{\` e}s}, \citenamefont {Sidis}, \citenamefont {Bourges}, \citenamefont
  {Ulrich}, \citenamefont {Hinkov}, \citenamefont {Regnault}, \citenamefont
  {Ivanov}, \citenamefont {Liang}, \citenamefont {Lin}, \citenamefont
  {Bernhard},\ and\ \citenamefont {Keimer}}]{PailhesBourges03}%
  \BibitemOpen
  \bibfield  {author} {\bibinfo {author} {\bibfnamefont {S.}~\bibnamefont
  {Pailh{\` e}s}}, \bibinfo {author} {\bibfnamefont {Y.}~\bibnamefont {Sidis}},
  \bibinfo {author} {\bibfnamefont {P.}~\bibnamefont {Bourges}}, \bibinfo
  {author} {\bibfnamefont {C.}~\bibnamefont {Ulrich}}, \bibinfo {author}
  {\bibfnamefont {V.}~\bibnamefont {Hinkov}}, \bibinfo {author} {\bibfnamefont
  {L.-P.}\ \bibnamefont {Regnault}}, \bibinfo {author} {\bibfnamefont
  {A.}~\bibnamefont {Ivanov}}, \bibinfo {author} {\bibfnamefont
  {B.}~\bibnamefont {Liang}}, \bibinfo {author} {\bibfnamefont {C.~T.}\
  \bibnamefont {Lin}}, \bibinfo {author} {\bibfnamefont {C.}~\bibnamefont
  {Bernhard}}, \ and\ \bibinfo {author} {\bibfnamefont {B.}~\bibnamefont
  {Keimer}},\ }\Doi {10.1103/PhysRevLett.91.237002} {\bibfield  {journal}
  {\bibinfo  {journal} {Phys. Rev. Lett.},\ }\textbf {\bibinfo {volume} {91}},\
  \bibinfo {pages} {237002} (\bibinfo {year} {2003})}\BibitemShut {NoStop}%
\bibitem [{\citenamefont {Reznik}\ \emph {et~al.}(2008)\citenamefont {Reznik},
  \citenamefont {Ismer}, \citenamefont {Eremin}, \citenamefont {Pintschovius},
  \citenamefont {Wolf}, \citenamefont {Arai}, \citenamefont {Endoh},
  \citenamefont {Masui},\ and\ \citenamefont {Tajima}}]{ReznikIsmer08}%
  \BibitemOpen
  \bibfield  {author} {\bibinfo {author} {\bibfnamefont {D.}~\bibnamefont
  {Reznik}}, \bibinfo {author} {\bibfnamefont {J.~P.}\ \bibnamefont {Ismer}},
  \bibinfo {author} {\bibfnamefont {I.}~\bibnamefont {Eremin}}, \bibinfo
  {author} {\bibfnamefont {L.}~\bibnamefont {Pintschovius}}, \bibinfo {author}
  {\bibfnamefont {T.}~\bibnamefont {Wolf}}, \bibinfo {author} {\bibfnamefont
  {M.}~\bibnamefont {Arai}}, \bibinfo {author} {\bibfnamefont {Y.}~\bibnamefont
  {Endoh}}, \bibinfo {author} {\bibfnamefont {T.}~\bibnamefont {Masui}}, \ and\
  \bibinfo {author} {\bibfnamefont {S.}~\bibnamefont {Tajima}},\ }\Doi
  {10.1103/PhysRevB.78.132503} {\bibfield  {journal} {\bibinfo  {journal}
  {Phys. Rev. B},\ }\textbf {\bibinfo {volume} {78}},\ \bibinfo {pages}
  {132503} (\bibinfo {year} {2008})}\BibitemShut {NoStop}%
\bibitem [{\citenamefont {Christensen}\ \emph {et~al.}(2004)\citenamefont
  {Christensen}, \citenamefont {McMorrow}, \citenamefont {R{\o}nnow},
  \citenamefont {Lake}, \citenamefont {Hayden}, \citenamefont {Aeppli},
  \citenamefont {Perring}, \citenamefont {Mangkorntong}, \citenamefont
  {Nohara},\ and\ \citenamefont {Takagi}}]{ChristensenMcMorrow04}%
  \BibitemOpen
  \bibfield  {author} {\bibinfo {author} {\bibfnamefont {N.~B.}\ \bibnamefont
  {Christensen}}, \bibinfo {author} {\bibfnamefont {D.~F.}\ \bibnamefont
  {McMorrow}}, \bibinfo {author} {\bibfnamefont {H.~M.}\ \bibnamefont
  {R{\o}nnow}}, \bibinfo {author} {\bibfnamefont {B.}~\bibnamefont {Lake}},
  \bibinfo {author} {\bibfnamefont {S.~M.}\ \bibnamefont {Hayden}}, \bibinfo
  {author} {\bibfnamefont {G.}~\bibnamefont {Aeppli}}, \bibinfo {author}
  {\bibfnamefont {T.~G.}\ \bibnamefont {Perring}}, \bibinfo {author}
  {\bibfnamefont {M.}~\bibnamefont {Mangkorntong}}, \bibinfo {author}
  {\bibfnamefont {M.}~\bibnamefont {Nohara}}, \ and\ \bibinfo {author}
  {\bibfnamefont {H.}~\bibnamefont {Takagi}},\ }\Doi
  {10.1103/PhysRevLett.93.147002} {\bibfield  {journal} {\bibinfo  {journal}
  {Phys. Rev. Lett.},\ }\textbf {\bibinfo {volume} {93}},\ \bibinfo {pages}
  {147002} (\bibinfo {year} {2004})}\BibitemShut {NoStop}%
\bibitem [{\citenamefont {Tranquada}\ \emph {et~al.}(2004)\citenamefont
  {Tranquada}, \citenamefont {Woo}, \citenamefont {Perring}, \citenamefont
  {Goka}, \citenamefont {Gu}, \citenamefont {Xu}, \citenamefont {Fujita},\ and\
  \citenamefont {Yamada}}]{Tranquada04}%
  \BibitemOpen
  \bibfield  {author} {\bibinfo {author} {\bibfnamefont {J.~M.}\ \bibnamefont
  {Tranquada}}, \bibinfo {author} {\bibfnamefont {H.}~\bibnamefont {Woo}},
  \bibinfo {author} {\bibfnamefont {T.~G.}\ \bibnamefont {Perring}}, \bibinfo
  {author} {\bibfnamefont {H.}~\bibnamefont {Goka}}, \bibinfo {author}
  {\bibfnamefont {G.~D.}\ \bibnamefont {Gu}}, \bibinfo {author} {\bibfnamefont
  {G.}~\bibnamefont {Xu}}, \bibinfo {author} {\bibfnamefont {M.}~\bibnamefont
  {Fujita}}, \ and\ \bibinfo {author} {\bibfnamefont {K.}~\bibnamefont
  {Yamada}},\ }\Doi {10.1038/nature02574} {\bibfield  {journal} {\bibinfo
  {journal} {Nature},\ }\textbf {\bibinfo {volume} {429}},\ \bibinfo {pages}
  {534} (\bibinfo {year} {2004})}\BibitemShut {NoStop}%
\bibitem [{\citenamefont {Fauqu{\' e}}\ \emph {et~al.}(2007)\citenamefont
  {Fauqu{\' e}}, \citenamefont {Sidis}, \citenamefont {Capogna}, \citenamefont
  {Ivanov}, \citenamefont {Hradil}, \citenamefont {Ulrich}, \citenamefont
  {Rykov}, \citenamefont {Keimer},\ and\ \citenamefont
  {Bourges}}]{FauqueSidis07}%
  \BibitemOpen
  \bibfield  {author} {\bibinfo {author} {\bibfnamefont {B.}~\bibnamefont
  {Fauqu{\' e}}}, \bibinfo {author} {\bibfnamefont {Y.}~\bibnamefont {Sidis}},
  \bibinfo {author} {\bibfnamefont {L.}~\bibnamefont {Capogna}}, \bibinfo
  {author} {\bibfnamefont {A.}~\bibnamefont {Ivanov}}, \bibinfo {author}
  {\bibfnamefont {K.}~\bibnamefont {Hradil}}, \bibinfo {author} {\bibfnamefont
  {C.}~\bibnamefont {Ulrich}}, \bibinfo {author} {\bibfnamefont {A.~I.}\
  \bibnamefont {Rykov}}, \bibinfo {author} {\bibfnamefont {B.}~\bibnamefont
  {Keimer}}, \ and\ \bibinfo {author} {\bibfnamefont {P.}~\bibnamefont
  {Bourges}},\ }\Doi {10.1103/PhysRevB.76.214512} {\bibfield  {journal}
  {\bibinfo  {journal} {Phys. Rev. B},\ }\textbf {\bibinfo {volume} {76}},\
  \bibinfo {pages} {214512} (\bibinfo {year} {2007})}\BibitemShut {NoStop}%
\bibitem [{\citenamefont {Xu}\ \emph {et~al.}(2009)\citenamefont {Xu},
  \citenamefont {Gu}, \citenamefont {Huecker}, \citenamefont {Fauque},
  \citenamefont {Perring}, \citenamefont {Regnault},\ and\ \citenamefont
  {Tranquada}}]{XuGu09}%
  \BibitemOpen
  \bibfield  {author} {\bibinfo {author} {\bibfnamefont {G.}~\bibnamefont
  {Xu}}, \bibinfo {author} {\bibfnamefont {G.~D.}\ \bibnamefont {Gu}}, \bibinfo
  {author} {\bibfnamefont {M.}~\bibnamefont {Huecker}}, \bibinfo {author}
  {\bibfnamefont {B.}~\bibnamefont {Fauque}}, \bibinfo {author} {\bibfnamefont
  {T.~G.}\ \bibnamefont {Perring}}, \bibinfo {author} {\bibfnamefont {L.~P.}\
  \bibnamefont {Regnault}}, \ and\ \bibinfo {author} {\bibfnamefont {J.~M.}\
  \bibnamefont {Tranquada}},\ }\Doi {10.1038/NPHYS1360} {\bibfield  {journal}
  {\bibinfo  {journal} {Nat. Phys.},\ }\textbf {\bibinfo {volume} {5}},\
  \bibinfo {pages} {642} (\bibinfo {year} {2009})}\BibitemShut {NoStop}%
\bibitem [{\citenamefont {Inosov}\ \emph {et~al.}(2010)\citenamefont {Inosov},
  \citenamefont {Park}, \citenamefont {Bourges}, \citenamefont {Sun},
  \citenamefont {Sidis}, \citenamefont {Schneidewind}, \citenamefont {Hradil},
  \citenamefont {Haug}, \citenamefont {Lin}, \citenamefont {Keimer},\ and\
  \citenamefont {Hinkov}}]{InosovPark10}%
  \BibitemOpen
  \bibfield  {author} {\bibinfo {author} {\bibfnamefont {D.~S.}\ \bibnamefont
  {Inosov}}, \bibinfo {author} {\bibfnamefont {J.~T.}\ \bibnamefont {Park}},
  \bibinfo {author} {\bibfnamefont {P.}~\bibnamefont {Bourges}}, \bibinfo
  {author} {\bibfnamefont {D.~L.}\ \bibnamefont {Sun}}, \bibinfo {author}
  {\bibfnamefont {Y.}~\bibnamefont {Sidis}}, \bibinfo {author} {\bibfnamefont
  {A.}~\bibnamefont {Schneidewind}}, \bibinfo {author} {\bibfnamefont
  {K.}~\bibnamefont {Hradil}}, \bibinfo {author} {\bibfnamefont
  {D.}~\bibnamefont {Haug}}, \bibinfo {author} {\bibfnamefont {C.~T.}\
  \bibnamefont {Lin}}, \bibinfo {author} {\bibfnamefont {B.}~\bibnamefont
  {Keimer}}, \ and\ \bibinfo {author} {\bibfnamefont {V.}~\bibnamefont
  {Hinkov}},\ }\href@noop {} {\bibfield  {journal} {\bibinfo  {journal} {Nat.
  Phys.},\ }\textbf {\bibinfo {volume} {6}},\ \bibinfo {pages} {178} (\bibinfo
  {year} {2010})}\BibitemShut {NoStop}%
\bibitem [{\citenamefont {Hinkov}\ \emph {et~al.}(2004)\citenamefont {Hinkov},
  \citenamefont {Pailh{\` e}s}, \citenamefont {Bourges}, \citenamefont {Sidis},
  \citenamefont {Ivanov}, \citenamefont {Kulakov}, \citenamefont {Lin},
  \citenamefont {Chen}, \citenamefont {Bernhard},\ and\ \citenamefont
  {Keimer}}]{HinkovKeimer04}%
  \BibitemOpen
  \bibfield  {author} {\bibinfo {author} {\bibfnamefont {V.}~\bibnamefont
  {Hinkov}}, \bibinfo {author} {\bibfnamefont {S.}~\bibnamefont {Pailh{\`
  e}s}}, \bibinfo {author} {\bibfnamefont {P.}~\bibnamefont {Bourges}},
  \bibinfo {author} {\bibfnamefont {Y.}~\bibnamefont {Sidis}}, \bibinfo
  {author} {\bibfnamefont {A.}~\bibnamefont {Ivanov}}, \bibinfo {author}
  {\bibfnamefont {A.}~\bibnamefont {Kulakov}}, \bibinfo {author} {\bibfnamefont
  {C.~T.}\ \bibnamefont {Lin}}, \bibinfo {author} {\bibfnamefont {D.~P.}\
  \bibnamefont {Chen}}, \bibinfo {author} {\bibfnamefont {C.}~\bibnamefont
  {Bernhard}}, \ and\ \bibinfo {author} {\bibfnamefont {B.}~\bibnamefont
  {Keimer}},\ }\Doi {10.1038/nature02774} {\bibfield  {journal} {\bibinfo
  {journal} {Nature},\ }\textbf {\bibinfo {volume} {430}},\ \bibinfo {pages}
  {650} (\bibinfo {year} {2004})}\BibitemShut {NoStop}%
\bibitem [{\citenamefont {Fauqu{\' e}}\ \emph {et~al.}(2006)\citenamefont
  {Fauqu{\' e}}, \citenamefont {Sidis}, \citenamefont {Hinkov}, \citenamefont
  {Pailh{\`e}s}, \citenamefont {Lin}, \citenamefont {Chaud},\ and\
  \citenamefont {Bourges}}]{FauqueBourges06}%
  \BibitemOpen
  \bibfield  {author} {\bibinfo {author} {\bibfnamefont {B.}~\bibnamefont
  {Fauqu{\' e}}}, \bibinfo {author} {\bibfnamefont {Y.}~\bibnamefont {Sidis}},
  \bibinfo {author} {\bibfnamefont {V.}~\bibnamefont {Hinkov}}, \bibinfo
  {author} {\bibfnamefont {S.}~\bibnamefont {Pailh{\`e}s}}, \bibinfo {author}
  {\bibfnamefont {C.~T.}\ \bibnamefont {Lin}}, \bibinfo {author} {\bibfnamefont
  {X.}~\bibnamefont {Chaud}}, \ and\ \bibinfo {author} {\bibfnamefont
  {P.}~\bibnamefont {Bourges}},\ }\Doi {10.1103/PhysRevLett.96.197001}
  {\bibfield  {journal} {\bibinfo  {journal} {Phys. Rev. Lett.},\ }\textbf
  {\bibinfo {volume} {96}},\ \bibinfo {pages} {197001} (\bibinfo {year}
  {2006})}\BibitemShut {NoStop}%
\bibitem [{\citenamefont {Pailh{\` e}s}\ \emph {et~al.}(2006)\citenamefont
  {Pailh{\` e}s}, \citenamefont {Ulrich}, \citenamefont {Fauqu{\' e}},
  \citenamefont {Hinkov}, \citenamefont {Sidis}, \citenamefont {Ivanov},
  \citenamefont {Lin}, \citenamefont {Keimer},\ and\ \citenamefont
  {Bourges}}]{PailhesBourges06}%
  \BibitemOpen
  \bibfield  {author} {\bibinfo {author} {\bibfnamefont {S.}~\bibnamefont
  {Pailh{\` e}s}}, \bibinfo {author} {\bibfnamefont {C.}~\bibnamefont
  {Ulrich}}, \bibinfo {author} {\bibfnamefont {B.}~\bibnamefont {Fauqu{\' e}}},
  \bibinfo {author} {\bibfnamefont {V.}~\bibnamefont {Hinkov}}, \bibinfo
  {author} {\bibfnamefont {Y.}~\bibnamefont {Sidis}}, \bibinfo {author}
  {\bibfnamefont {A.}~\bibnamefont {Ivanov}}, \bibinfo {author} {\bibfnamefont
  {C.~T.}\ \bibnamefont {Lin}}, \bibinfo {author} {\bibfnamefont
  {B.}~\bibnamefont {Keimer}}, \ and\ \bibinfo {author} {\bibfnamefont
  {P.}~\bibnamefont {Bourges}},\ }\Doi {10.1103/PhysRevLett.96.257001}
  {\bibfield  {journal} {\bibinfo  {journal} {Phys. Rev. Lett.},\ }\textbf
  {\bibinfo {volume} {96}},\ \bibinfo {pages} {257001} (\bibinfo {year}
  {2006})}\BibitemShut {NoStop}%
\bibitem [{\citenamefont {Waske}\ \emph {et~al.}(2007)\citenamefont {Waske},
  \citenamefont {Hess}, \citenamefont {B{\"u}chner}, \citenamefont {Hinkov},\
  and\ \citenamefont {Lin}}]{WaskeHess07}%
  \BibitemOpen
  \bibfield  {author} {\bibinfo {author} {\bibfnamefont {A.}~\bibnamefont
  {Waske}}, \bibinfo {author} {\bibfnamefont {C.}~\bibnamefont {Hess}},
  \bibinfo {author} {\bibfnamefont {B.}~\bibnamefont {B{\"u}chner}}, \bibinfo
  {author} {\bibfnamefont {V.}~\bibnamefont {Hinkov}}, \ and\ \bibinfo {author}
  {\bibfnamefont {C.~T.}\ \bibnamefont {Lin}},\ }\Doi
  {10.1016/j.physc.2007.03.169} {\bibfield  {journal} {\bibinfo  {journal}
  {Physica C},\ }\textbf {\bibinfo {volume} {460}},\ \bibinfo {pages} {746}
  (\bibinfo {year} {2007})}\BibitemShut {NoStop}%
\bibitem [{\citenamefont {Inosov}\ \emph {et~al.}(2007)\citenamefont {Inosov},
  \citenamefont {Fink}, \citenamefont {Kordyuk}, \citenamefont {Borisenko},
  \citenamefont {Zabolotnyy}, \citenamefont {Schuster}, \citenamefont
  {Knupfer}, \citenamefont {B{\"u}chner}, \citenamefont {Follath},
  \citenamefont {D{\"u}rr}, \citenamefont {Eberhardt}, \citenamefont {Hinkov},
  \citenamefont {Keimer},\ and\ \citenamefont {Berger}}]{InosovBorisenko07}%
  \BibitemOpen
  \bibfield  {author} {\bibinfo {author} {\bibfnamefont {D.~S.}\ \bibnamefont
  {Inosov}}, \bibinfo {author} {\bibfnamefont {J.}~\bibnamefont {Fink}},
  \bibinfo {author} {\bibfnamefont {A.~A.}\ \bibnamefont {Kordyuk}}, \bibinfo
  {author} {\bibfnamefont {S.~V.}\ \bibnamefont {Borisenko}}, \bibinfo {author}
  {\bibfnamefont {V.~B.}\ \bibnamefont {Zabolotnyy}}, \bibinfo {author}
  {\bibfnamefont {R.}~\bibnamefont {Schuster}}, \bibinfo {author}
  {\bibfnamefont {M.}~\bibnamefont {Knupfer}}, \bibinfo {author} {\bibfnamefont
  {B.}~\bibnamefont {B{\"u}chner}}, \bibinfo {author} {\bibfnamefont
  {R.}~\bibnamefont {Follath}}, \bibinfo {author} {\bibfnamefont {H.~A.}\
  \bibnamefont {D{\"u}rr}}, \bibinfo {author} {\bibfnamefont {W.}~\bibnamefont
  {Eberhardt}}, \bibinfo {author} {\bibfnamefont {V.}~\bibnamefont {Hinkov}},
  \bibinfo {author} {\bibfnamefont {B.}~\bibnamefont {Keimer}}, \ and\ \bibinfo
  {author} {\bibfnamefont {H.}~\bibnamefont {Berger}},\ }\Doi
  {10.1103/PhysRevLett.99.237002} {\bibfield  {journal} {\bibinfo  {journal}
  {Phys. Rev. Lett.},\ }\textbf {\bibinfo {volume} {99}},\ \bibinfo {pages}
  {237002} (\bibinfo {year} {2007})}\BibitemShut {NoStop}%
\bibitem [{\citenamefont {Hinkov}\ \emph {et~al.}(2008)\citenamefont {Hinkov},
  \citenamefont {Haug}, \citenamefont {Fauqu{\' e}}, \citenamefont {Bourges},
  \citenamefont {Sidis}, \citenamefont {Ivanov}, \citenamefont {Bernhard},
  \citenamefont {Lin},\ and\ \citenamefont {Keimer}}]{HinkovHaug08}%
  \BibitemOpen
  \bibfield  {author} {\bibinfo {author} {\bibfnamefont {V.}~\bibnamefont
  {Hinkov}}, \bibinfo {author} {\bibfnamefont {D.}~\bibnamefont {Haug}},
  \bibinfo {author} {\bibfnamefont {B.}~\bibnamefont {Fauqu{\' e}}}, \bibinfo
  {author} {\bibfnamefont {P.}~\bibnamefont {Bourges}}, \bibinfo {author}
  {\bibfnamefont {Y.}~\bibnamefont {Sidis}}, \bibinfo {author} {\bibfnamefont
  {A.}~\bibnamefont {Ivanov}}, \bibinfo {author} {\bibfnamefont
  {C.}~\bibnamefont {Bernhard}}, \bibinfo {author} {\bibfnamefont {C.~T.}\
  \bibnamefont {Lin}}, \ and\ \bibinfo {author} {\bibfnamefont
  {B.}~\bibnamefont {Keimer}},\ }\Doi {10.1126/science.1152309} {\bibfield
  {journal} {\bibinfo  {journal} {Science},\ }\textbf {\bibinfo {volume}
  {319}},\ \bibinfo {pages} {597} (\bibinfo {year} {2008})}\BibitemShut
  {NoStop}%
\bibitem [{\citenamefont {White}\ \emph {et~al.}(2009)\citenamefont {White},
  \citenamefont {Hinkov}, \citenamefont {Heslop}, \citenamefont {Lycett},
  \citenamefont {Forgan}, \citenamefont {Bowell}, \citenamefont {Straessle},
  \citenamefont {Abrahamsen}, \citenamefont {Laver}, \citenamefont {Dewhurst},
  \citenamefont {Kohlbrecher}, \citenamefont {Gavilano}, \citenamefont {Mesot},
  \citenamefont {Keimer},\ and\ \citenamefont {Erb}}]{WhiteHinkov09}%
  \BibitemOpen
  \bibfield  {author} {\bibinfo {author} {\bibfnamefont {J.~S.}\ \bibnamefont
  {White}}, \bibinfo {author} {\bibfnamefont {V.}~\bibnamefont {Hinkov}},
  \bibinfo {author} {\bibfnamefont {R.~W.}\ \bibnamefont {Heslop}}, \bibinfo
  {author} {\bibfnamefont {R.~J.}\ \bibnamefont {Lycett}}, \bibinfo {author}
  {\bibfnamefont {E.~M.}\ \bibnamefont {Forgan}}, \bibinfo {author}
  {\bibfnamefont {C.}~\bibnamefont {Bowell}}, \bibinfo {author} {\bibfnamefont
  {S.}~\bibnamefont {Straessle}}, \bibinfo {author} {\bibfnamefont {A.~B.}\
  \bibnamefont {Abrahamsen}}, \bibinfo {author} {\bibfnamefont
  {M.}~\bibnamefont {Laver}}, \bibinfo {author} {\bibfnamefont {C.~D.}\
  \bibnamefont {Dewhurst}}, \bibinfo {author} {\bibfnamefont {J.}~\bibnamefont
  {Kohlbrecher}}, \bibinfo {author} {\bibfnamefont {J.~L.}\ \bibnamefont
  {Gavilano}}, \bibinfo {author} {\bibfnamefont {J.}~\bibnamefont {Mesot}},
  \bibinfo {author} {\bibfnamefont {B.}~\bibnamefont {Keimer}}, \ and\ \bibinfo
  {author} {\bibfnamefont {A.}~\bibnamefont {Erb}},\ }\Doi
  {10.1103/PhysRevLett.102.097001} {\bibfield  {journal} {\bibinfo  {journal}
  {Phys. Rev. Lett.},\ }\textbf {\bibinfo {volume} {102}},\ \bibinfo {pages}
  {097001} (\bibinfo {year} {2009})}\BibitemShut {NoStop}%
\bibitem [{\citenamefont {Dahm}\ \emph {et~al.}(2009)\citenamefont {Dahm},
  \citenamefont {Hinkov}, \citenamefont {Borisenko}, \citenamefont {Kordyuk},
  \citenamefont {Zabolotnyy}, \citenamefont {Fink}, \citenamefont
  {B{\"u}chner}, \citenamefont {Scalapino}, \citenamefont {Hanke},\ and\
  \citenamefont {Keimer}}]{DahmHinkov09}%
  \BibitemOpen
  \bibfield  {author} {\bibinfo {author} {\bibfnamefont {T.}~\bibnamefont
  {Dahm}}, \bibinfo {author} {\bibfnamefont {V.}~\bibnamefont {Hinkov}},
  \bibinfo {author} {\bibfnamefont {S.~V.}\ \bibnamefont {Borisenko}}, \bibinfo
  {author} {\bibfnamefont {A.~A.}\ \bibnamefont {Kordyuk}}, \bibinfo {author}
  {\bibfnamefont {V.~B.}\ \bibnamefont {Zabolotnyy}}, \bibinfo {author}
  {\bibfnamefont {J.}~\bibnamefont {Fink}}, \bibinfo {author} {\bibfnamefont
  {B.}~\bibnamefont {B{\"u}chner}}, \bibinfo {author} {\bibfnamefont {D.~J.}\
  \bibnamefont {Scalapino}}, \bibinfo {author} {\bibfnamefont {W.}~\bibnamefont
  {Hanke}}, \ and\ \bibinfo {author} {\bibfnamefont {B.}~\bibnamefont
  {Keimer}},\ }\Doi {10.1038/NPHYS1180} {\bibfield  {journal} {\bibinfo
  {journal} {Nat. Phys.},\ }\textbf {\bibinfo {volume} {5}},\ \bibinfo {pages}
  {217} (\bibinfo {year} {2009})}\BibitemShut {NoStop}%
\bibitem [{\citenamefont {Haug}\ \emph {et~al.}(2009)\citenamefont {Haug},
  \citenamefont {Hinkov}, \citenamefont {Suchaneck}, \citenamefont {Inosov},
  \citenamefont {Christensen}, \citenamefont {Niedermayer}, \citenamefont
  {Bourges}, \citenamefont {Sidis}, \citenamefont {Park}, \citenamefont
  {Ivanov}, \citenamefont {Lin}, \citenamefont {Mesot},\ and\ \citenamefont
  {Keimer}}]{HaugHinkov09}%
  \BibitemOpen
  \bibfield  {author} {\bibinfo {author} {\bibfnamefont {D.}~\bibnamefont
  {Haug}}, \bibinfo {author} {\bibfnamefont {V.}~\bibnamefont {Hinkov}},
  \bibinfo {author} {\bibfnamefont {A.}~\bibnamefont {Suchaneck}}, \bibinfo
  {author} {\bibfnamefont {D.~S.}\ \bibnamefont {Inosov}}, \bibinfo {author}
  {\bibfnamefont {N.~B.}\ \bibnamefont {Christensen}}, \bibinfo {author}
  {\bibfnamefont {C.}~\bibnamefont {Niedermayer}}, \bibinfo {author}
  {\bibfnamefont {P.}~\bibnamefont {Bourges}}, \bibinfo {author} {\bibfnamefont
  {Y.}~\bibnamefont {Sidis}}, \bibinfo {author} {\bibfnamefont {J.~T.}\
  \bibnamefont {Park}}, \bibinfo {author} {\bibfnamefont {A.}~\bibnamefont
  {Ivanov}}, \bibinfo {author} {\bibfnamefont {C.~T.}\ \bibnamefont {Lin}},
  \bibinfo {author} {\bibfnamefont {J.}~\bibnamefont {Mesot}}, \ and\ \bibinfo
  {author} {\bibfnamefont {B.}~\bibnamefont {Keimer}},\ }\Doi
  {10.1103/PhysRevLett.103.017001} {\bibfield  {journal} {\bibinfo  {journal}
  {Phys. Rev. Lett.},\ }\textbf {\bibinfo {volume} {103}},\ \bibinfo {pages}
  {017001} (\bibinfo {year} {2009})}\BibitemShut {NoStop}%
\bibitem [{\citenamefont {Maiti}\ \emph {et~al.}(2009)\citenamefont {Maiti},
  \citenamefont {Fink}, \citenamefont {de~Jong}, \citenamefont {Gorgoi},
  \citenamefont {Lin}, \citenamefont {Raichle}, \citenamefont {Hinkov},
  \citenamefont {Lambacher}, \citenamefont {Erb},\ and\ \citenamefont
  {Golden}}]{MaitiFink09}%
  \BibitemOpen
  \bibfield  {author} {\bibinfo {author} {\bibfnamefont {K.}~\bibnamefont
  {Maiti}}, \bibinfo {author} {\bibfnamefont {J.}~\bibnamefont {Fink}},
  \bibinfo {author} {\bibfnamefont {S.}~\bibnamefont {de~Jong}}, \bibinfo
  {author} {\bibfnamefont {M.}~\bibnamefont {Gorgoi}}, \bibinfo {author}
  {\bibfnamefont {C.~T.}\ \bibnamefont {Lin}}, \bibinfo {author} {\bibfnamefont
  {M.}~\bibnamefont {Raichle}}, \bibinfo {author} {\bibfnamefont
  {V.}~\bibnamefont {Hinkov}}, \bibinfo {author} {\bibfnamefont
  {M.}~\bibnamefont {Lambacher}}, \bibinfo {author} {\bibfnamefont
  {A.}~\bibnamefont {Erb}}, \ and\ \bibinfo {author} {\bibfnamefont {M.~S.}\
  \bibnamefont {Golden}},\ }\Doi {10.1103/PhysRevB.80.165132} {\bibfield
  {journal} {\bibinfo  {journal} {Phys. Rev. B},\ }\textbf {\bibinfo {volume}
  {80}},\ \bibinfo {pages} {165132} (\bibinfo {year} {2009})}\BibitemShut
  {NoStop}%
\bibitem [{\citenamefont {Yamase}\ and\ \citenamefont
  {Metzner}(2006)}]{YamaseMetzner06}%
  \BibitemOpen
  \bibfield  {author} {\bibinfo {author} {\bibfnamefont {H.}~\bibnamefont
  {Yamase}}\ and\ \bibinfo {author} {\bibfnamefont {W.}~\bibnamefont
  {Metzner}},\ }\Doi {10.1103/PhysRevB.73.214517} {\bibfield  {journal}
  {\bibinfo  {journal} {Phys. Rev. B},\ }\textbf {\bibinfo {volume} {73}},\
  \bibinfo {pages} {214517} (\bibinfo {year} {2006})}\BibitemShut {NoStop}%
\bibitem [{\citenamefont {Eremin}\ \emph {et~al.}(2007)\citenamefont {Eremin},
  \citenamefont {Morr}, \citenamefont {Chubukov},\ and\ \citenamefont
  {Bennemann}}]{EreminBennemann07}%
  \BibitemOpen
  \bibfield  {author} {\bibinfo {author} {\bibfnamefont {I.}~\bibnamefont
  {Eremin}}, \bibinfo {author} {\bibfnamefont {D.~K.}\ \bibnamefont {Morr}},
  \bibinfo {author} {\bibfnamefont {A.~V.}\ \bibnamefont {Chubukov}}, \ and\
  \bibinfo {author} {\bibfnamefont {K.~H.}\ \bibnamefont {Bennemann}},\ }\Doi
  {10.1103/PhysRevB.75.184534} {\bibfield  {journal} {\bibinfo  {journal}
  {Phys. Rev. B},\ }\textbf {\bibinfo {volume} {75}},\ \bibinfo {pages}
  {184534} (\bibinfo {year} {2007})}\BibitemShut {NoStop}%
\bibitem [{\citenamefont {Andersen}\ and\ \citenamefont
  {Hedeg{\aa}rd}(2005)}]{AndersenHedegard05}%
  \BibitemOpen
  \bibfield  {author} {\bibinfo {author} {\bibfnamefont {B.~M.}\ \bibnamefont
  {Andersen}}\ and\ \bibinfo {author} {\bibfnamefont {P.}~\bibnamefont
  {Hedeg{\aa}rd}},\ }\Doi {10.1103/PhysRevLett.95.037002} {\bibfield  {journal}
  {\bibinfo  {journal} {Phys. Rev. Lett.},\ }\textbf {\bibinfo {volume} {95}},\
  \bibinfo {pages} {037002} (\bibinfo {year} {2005})}\BibitemShut {NoStop}%
\bibitem [{\citenamefont {Yao}\ \emph {et~al.}(2006)\citenamefont {Yao},
  \citenamefont {Carlson},\ and\ \citenamefont {Campbell}}]{YaoCarlson06}%
  \BibitemOpen
  \bibfield  {author} {\bibinfo {author} {\bibfnamefont {D.~X.}\ \bibnamefont
  {Yao}}, \bibinfo {author} {\bibfnamefont {E.~W.}\ \bibnamefont {Carlson}}, \
  and\ \bibinfo {author} {\bibfnamefont {D.~K.}\ \bibnamefont {Campbell}},\
  }\Doi {10.1103/PhysRevLett.97.017003} {\bibfield  {journal} {\bibinfo
  {journal} {Phys. Rev. Lett.},\ }\textbf {\bibinfo {volume} {97}},\ \bibinfo
  {pages} {107006} (\bibinfo {year} {2006})}\BibitemShut {NoStop}%
\bibitem [{\citenamefont {Uhrig}\ \emph {et~al.}(2004)\citenamefont {Uhrig},
  \citenamefont {Schmidt},\ and\ \citenamefont {Gr{\"
  u}ninger}}]{UhrigSchmidt04}%
  \BibitemOpen
  \bibfield  {author} {\bibinfo {author} {\bibfnamefont {G.~S.}\ \bibnamefont
  {Uhrig}}, \bibinfo {author} {\bibfnamefont {K.~P.}\ \bibnamefont {Schmidt}},
  \ and\ \bibinfo {author} {\bibfnamefont {M.}~\bibnamefont {Gr{\" u}ninger}},\
  }\Doi {10.1103/PhysRevLett.93.267003} {\bibfield  {journal} {\bibinfo
  {journal} {Phys. Rev. Lett.},\ }\textbf {\bibinfo {volume} {93}},\ \bibinfo
  {pages} {267003} (\bibinfo {year} {2004})}\BibitemShut {NoStop}%
\bibitem [{\citenamefont {Vojta}\ and\ \citenamefont
  {Ulbricht}(2004)}]{VojtaUlbricht04}%
  \BibitemOpen
  \bibfield  {author} {\bibinfo {author} {\bibfnamefont {M.}~\bibnamefont
  {Vojta}}\ and\ \bibinfo {author} {\bibfnamefont {T.}~\bibnamefont
  {Ulbricht}},\ }\Doi {10.1103/PhysRevLett.93.127002} {\bibfield  {journal}
  {\bibinfo  {journal} {Phys. Rev. Lett.},\ }\textbf {\bibinfo {volume} {93}},\
  \bibinfo {pages} {127002} (\bibinfo {year} {2004})}\BibitemShut {NoStop}%
\bibitem [{\citenamefont {Seibold}\ and\ \citenamefont
  {Lorenzana}(2005)}]{SeiboldLorenzana05}%
  \BibitemOpen
  \bibfield  {author} {\bibinfo {author} {\bibfnamefont {G.}~\bibnamefont
  {Seibold}}\ and\ \bibinfo {author} {\bibfnamefont {J.}~\bibnamefont
  {Lorenzana}},\ }\Doi {10.1103/PhysRevLett.94.107006} {\bibfield  {journal}
  {\bibinfo  {journal} {Phys. Rev. Lett.},\ }\textbf {\bibinfo {volume} {94}},\
  \bibinfo {pages} {107006} (\bibinfo {year} {2005})}\BibitemShut {NoStop}%
\bibitem [{\citenamefont {Lindg{\aa}rd}(2005)}]{Lindgard05}%
  \BibitemOpen
  \bibfield  {author} {\bibinfo {author} {\bibfnamefont {P.~A.}\ \bibnamefont
  {Lindg{\aa}rd}},\ }\Doi {10.1103/PhysRevLett.95.217001} {\bibfield  {journal}
  {\bibinfo  {journal} {Phys. Rev. Lett.},\ }\textbf {\bibinfo {volume} {95}},\
  \bibinfo {pages} {217001} (\bibinfo {year} {2005})}\BibitemShut {NoStop}%
\bibitem [{\citenamefont {Milstein}\ and\ \citenamefont
  {Sushkov}(2008)}]{MilsteinSushkov08}%
  \BibitemOpen
  \bibfield  {author} {\bibinfo {author} {\bibfnamefont {A.~I.}\ \bibnamefont
  {Milstein}}\ and\ \bibinfo {author} {\bibfnamefont {O.~P.}\ \bibnamefont
  {Sushkov}},\ }\Doi {10.1103/PhysRevB.78.014501} {\bibfield  {journal}
  {\bibinfo  {journal} {Phys. Rev. B},\ }\textbf {\bibinfo {volume} {78}},\
  \bibinfo {pages} {014501} (\bibinfo {year} {2008})}\BibitemShut {NoStop}%
\bibitem [{\citenamefont {Pardini}\ \emph {et~al.}(2008)\citenamefont
  {Pardini}, \citenamefont {Singh}, \citenamefont {Katanin},\ and\
  \citenamefont {Sushkov}}]{PardiniSushkov08}%
  \BibitemOpen
  \bibfield  {author} {\bibinfo {author} {\bibfnamefont {T.}~\bibnamefont
  {Pardini}}, \bibinfo {author} {\bibfnamefont {R.~R.~P.}\ \bibnamefont
  {Singh}}, \bibinfo {author} {\bibfnamefont {A.}~\bibnamefont {Katanin}}, \
  and\ \bibinfo {author} {\bibfnamefont {O.~P.}\ \bibnamefont {Sushkov}},\
  }\Doi {10.1103/PhysRevB.78.024439} {\bibfield  {journal} {\bibinfo  {journal}
  {Phys. Rev. B},\ }\textbf {\bibinfo {volume} {78}},\ \bibinfo {pages}
  {024439} (\bibinfo {year} {2008})}\BibitemShut {NoStop}%
\bibitem [{\citenamefont {Sushkov}(2009)}]{Sushkov09}%
  \BibitemOpen
  \bibfield  {author} {\bibinfo {author} {\bibfnamefont {O.~P.}\ \bibnamefont
  {Sushkov}},\ }\Doi {10.1103/PhysRevB.79.174519} {\bibfield  {journal}
  {\bibinfo  {journal} {Phys. Rev. B},\ }\textbf {\bibinfo {volume} {79}},\
  \bibinfo {pages} {174519} (\bibinfo {year} {2009})}\BibitemShut {NoStop}%
\bibitem [{\citenamefont {Zeyher}(2010)}]{Zeyher10}%
  \BibitemOpen
  \bibfield  {author} {\bibinfo {author} {\bibfnamefont {R.}~\bibnamefont
  {Zeyher}},\ }\href@noop {} {\bibfield  {journal} {\bibinfo  {journal} {EPL},\
  }\textbf {\bibinfo {volume} {90}},\ \bibinfo {pages} {17006} (\bibinfo {year}
  {2010})}\BibitemShut {NoStop}%
\bibitem [{\citenamefont {Tranquada}\ \emph {et~al.}(1995)\citenamefont
  {Tranquada}, \citenamefont {Sternlieb}, \citenamefont {Axe}, \citenamefont
  {Nakamura},\ and\ \citenamefont {Uchida}}]{Tranquada95}%
  \BibitemOpen
  \bibfield  {author} {\bibinfo {author} {\bibfnamefont {J.~M.}\ \bibnamefont
  {Tranquada}}, \bibinfo {author} {\bibfnamefont {B.~J.}\ \bibnamefont
  {Sternlieb}}, \bibinfo {author} {\bibfnamefont {J.~D.}\ \bibnamefont {Axe}},
  \bibinfo {author} {\bibfnamefont {Y.}~\bibnamefont {Nakamura}}, \ and\
  \bibinfo {author} {\bibfnamefont {S.}~\bibnamefont {Uchida}},\ }\href@noop {}
  {\bibfield  {journal} {\bibinfo  {journal} {Nature},\ }\textbf {\bibinfo
  {volume} {375}},\ \bibinfo {pages} {561} (\bibinfo {year}
  {1995})}\BibitemShut {NoStop}%
\bibitem [{\citenamefont {Fujita}\ \emph {et~al.}(2002)\citenamefont {Fujita},
  \citenamefont {Goka}, \citenamefont {Yamada},\ and\ \citenamefont
  {Matsuda}}]{Fujita02}%
  \BibitemOpen
  \bibfield  {author} {\bibinfo {author} {\bibfnamefont {M.}~\bibnamefont
  {Fujita}}, \bibinfo {author} {\bibfnamefont {H.}~\bibnamefont {Goka}},
  \bibinfo {author} {\bibfnamefont {K.}~\bibnamefont {Yamada}}, \ and\ \bibinfo
  {author} {\bibfnamefont {M.}~\bibnamefont {Matsuda}},\ }\href@noop {}
  {\bibfield  {journal} {\bibinfo  {journal} {Phys. Rev. Lett.},\ }\textbf
  {\bibinfo {volume} {88}},\ \bibinfo {pages} {167008} (\bibinfo {year}
  {2002})}\BibitemShut {NoStop}%
\bibitem [{\citenamefont {Tranquada}(2007)}]{Tranquada05}%
  \BibitemOpen
  \bibfield  {author} {\bibinfo {author} {\bibfnamefont {J.~M.}\ \bibnamefont
  {Tranquada}},\ }\enquote {\bibinfo {title} {{H}andbook of high-temperature
  superconductivity},}\ \ (\bibinfo  {publisher} {Springer},\ \bibinfo {year}
  {2007})\ Chap.\ \bibinfo {chapter} {6. Neutron scattering studies of
  antiferromagnetic correlations in cuprates}, pp.\ \bibinfo {pages}
  {257--298},\ \bibinfo {note} {preprint at arXiv:cond-mat/0512115}\BibitemShut
  {NoStop}%
\bibitem [{\citenamefont {Chang}\ \emph {et~al.}(2008)\citenamefont {Chang},
  \citenamefont {Niedermayer}, \citenamefont {Gilardi}, \citenamefont
  {Christensen}, \citenamefont {R{\o}nnow}, \citenamefont {McMorrow},
  \citenamefont {Ay}, \citenamefont {Stahn}, \citenamefont {Sobolev},
  \citenamefont {Hiess}, \citenamefont {Pailhes}, \citenamefont {Baines},
  \citenamefont {Momono}, \citenamefont {Oda}, \citenamefont {Ido},\ and\
  \citenamefont {Mesot}}]{Chang08}%
  \BibitemOpen
  \bibfield  {author} {\bibinfo {author} {\bibfnamefont {J.}~\bibnamefont
  {Chang}}, \bibinfo {author} {\bibfnamefont {C.}~\bibnamefont {Niedermayer}},
  \bibinfo {author} {\bibfnamefont {R.}~\bibnamefont {Gilardi}}, \bibinfo
  {author} {\bibfnamefont {N.~B.}\ \bibnamefont {Christensen}}, \bibinfo
  {author} {\bibfnamefont {H.~M.}\ \bibnamefont {R{\o}nnow}}, \bibinfo {author}
  {\bibfnamefont {D.~F.}\ \bibnamefont {McMorrow}}, \bibinfo {author}
  {\bibfnamefont {M.}~\bibnamefont {Ay}}, \bibinfo {author} {\bibfnamefont
  {J.}~\bibnamefont {Stahn}}, \bibinfo {author} {\bibfnamefont
  {O.}~\bibnamefont {Sobolev}}, \bibinfo {author} {\bibfnamefont
  {A.}~\bibnamefont {Hiess}}, \bibinfo {author} {\bibfnamefont
  {S.}~\bibnamefont {Pailhes}}, \bibinfo {author} {\bibfnamefont
  {C.}~\bibnamefont {Baines}}, \bibinfo {author} {\bibfnamefont
  {N.}~\bibnamefont {Momono}}, \bibinfo {author} {\bibfnamefont
  {M.}~\bibnamefont {Oda}}, \bibinfo {author} {\bibfnamefont {M.}~\bibnamefont
  {Ido}}, \ and\ \bibinfo {author} {\bibfnamefont {J.}~\bibnamefont {Mesot}},\
  }\href@noop {} {\bibfield  {journal} {\bibinfo  {journal} {Phys. Rev. B},\
  }\textbf {\bibinfo {volume} {78}},\ \bibinfo {pages} {104525} (\bibinfo
  {year} {2008})}\BibitemShut {NoStop}%
\bibitem [{\citenamefont {Vojta}\ \emph {et~al.}(2006)\citenamefont {Vojta},
  \citenamefont {Vojta},\ and\ \citenamefont {Kaul}}]{VojtaVojta06}%
  \BibitemOpen
  \bibfield  {author} {\bibinfo {author} {\bibfnamefont {M.}~\bibnamefont
  {Vojta}}, \bibinfo {author} {\bibfnamefont {T.}~\bibnamefont {Vojta}}, \ and\
  \bibinfo {author} {\bibfnamefont {R.~K.}\ \bibnamefont {Kaul}},\ }\Doi
  {10.1103/PhysRevLett.97.097001} {\bibfield  {journal} {\bibinfo  {journal}
  {Phys. Rev. Lett.},\ }\textbf {\bibinfo {volume} {97}},\ \bibinfo {pages}
  {097001} (\bibinfo {year} {2006})}\BibitemShut {NoStop}%
\bibitem [{\citenamefont {Eschrig}(2006)}]{Eschrig06}%
  \BibitemOpen
  \bibfield  {author} {\bibinfo {author} {\bibfnamefont {M.}~\bibnamefont
  {Eschrig}},\ }\Doi {10.1080/00018730600645636} {\bibfield  {journal}
  {\bibinfo  {journal} {Adv. Phys.},\ }\textbf {\bibinfo {volume} {55}},\
  \bibinfo {pages} {47} (\bibinfo {year} {2006})}\BibitemShut {NoStop}%
\bibitem [{\citenamefont {Kivelson}\ \emph {et~al.}(1998)\citenamefont
  {Kivelson}, \citenamefont {Fradkin},\ and\ \citenamefont
  {Emery}}]{KivelsonFradkin98}%
  \BibitemOpen
  \bibfield  {author} {\bibinfo {author} {\bibfnamefont {S.~A.}\ \bibnamefont
  {Kivelson}}, \bibinfo {author} {\bibfnamefont {E.}~\bibnamefont {Fradkin}}, \
  and\ \bibinfo {author} {\bibfnamefont {V.~J.}\ \bibnamefont {Emery}},\
  }\href@noop {} {\bibfield  {journal} {\bibinfo  {journal} {Nature},\ }\textbf
  {\bibinfo {volume} {393}},\ \bibinfo {pages} {550} (\bibinfo {year}
  {1998})}\BibitemShut {NoStop}%
\bibitem [{\citenamefont {Kivelson}\ \emph {et~al.}(2003)\citenamefont
  {Kivelson}, \citenamefont {Bindloss}, \citenamefont {Fradkin}, \citenamefont
  {Oganesyan}, \citenamefont {Tranquada}, \citenamefont {Kapitulnik},\ and\
  \citenamefont {Howald}}]{KivelsonTranquada03}%
  \BibitemOpen
  \bibfield  {author} {\bibinfo {author} {\bibfnamefont {S.~A.}\ \bibnamefont
  {Kivelson}}, \bibinfo {author} {\bibfnamefont {I.~P.}\ \bibnamefont
  {Bindloss}}, \bibinfo {author} {\bibfnamefont {E.}~\bibnamefont {Fradkin}},
  \bibinfo {author} {\bibfnamefont {V.}~\bibnamefont {Oganesyan}}, \bibinfo
  {author} {\bibfnamefont {J.~M.}\ \bibnamefont {Tranquada}}, \bibinfo {author}
  {\bibfnamefont {A.}~\bibnamefont {Kapitulnik}}, \ and\ \bibinfo {author}
  {\bibfnamefont {C.}~\bibnamefont {Howald}},\ }\href@noop {} {\bibfield
  {journal} {\bibinfo  {journal} {Rev. Mod. Phys.},\ }\textbf {\bibinfo
  {volume} {75}},\ \bibinfo {pages} {1201} (\bibinfo {year}
  {2003})}\BibitemShut {NoStop}%
\bibitem [{\citenamefont {Yamase}\ and\ \citenamefont
  {Kohno}(2000){\natexlab{a}}}]{YamaseKohno00}%
  \BibitemOpen
  \bibfield  {author} {\bibinfo {author} {\bibfnamefont {H.}~\bibnamefont
  {Yamase}}\ and\ \bibinfo {author} {\bibfnamefont {H.}~\bibnamefont {Kohno}},\
  }\href@noop {} {\bibfield  {journal} {\bibinfo  {journal} {J. Phys. Soc.
  Jpn.},\ }\textbf {\bibinfo {volume} {69}},\ \bibinfo {pages} {332} (\bibinfo
  {year} {2000}{\natexlab{a}})}\BibitemShut {NoStop}%
\bibitem [{\citenamefont {Halboth}\ and\ \citenamefont
  {Metzner}(2000)}]{HalbothMetzner00}%
  \BibitemOpen
  \bibfield  {author} {\bibinfo {author} {\bibfnamefont {C.}~\bibnamefont
  {Halboth}}\ and\ \bibinfo {author} {\bibfnamefont {W.}~\bibnamefont
  {Metzner}},\ }\href@noop {} {\bibfield  {journal} {\bibinfo  {journal} {Phys.
  Rev. Lett.},\ }\textbf {\bibinfo {volume} {85}},\ \bibinfo {pages} {5162}
  (\bibinfo {year} {2000})}\BibitemShut {NoStop}%
\bibitem [{\citenamefont {Yamase}\ and\ \citenamefont
  {Kohno}(2000){\natexlab{b}}}]{YamaseKohno00a}%
  \BibitemOpen
  \bibfield  {author} {\bibinfo {author} {\bibfnamefont {H.}~\bibnamefont
  {Yamase}}\ and\ \bibinfo {author} {\bibfnamefont {H.}~\bibnamefont {Kohno}},\
  }\href@noop {} {\bibfield  {journal} {\bibinfo  {journal} {J. Phys. Soc.
  Jpn.},\ }\textbf {\bibinfo {volume} {69}},\ \bibinfo {pages} {2151} (\bibinfo
  {year} {2000}{\natexlab{b}})}\BibitemShut {NoStop}%
\bibitem [{\citenamefont {Oganesyan}\ \emph {et~al.}(2001)\citenamefont
  {Oganesyan}, \citenamefont {Kivelson},\ and\ \citenamefont
  {Fradkin}}]{OganesyanKivelson01}%
  \BibitemOpen
  \bibfield  {author} {\bibinfo {author} {\bibfnamefont {V.}~\bibnamefont
  {Oganesyan}}, \bibinfo {author} {\bibfnamefont {S.~A.}\ \bibnamefont
  {Kivelson}}, \ and\ \bibinfo {author} {\bibfnamefont {E.}~\bibnamefont
  {Fradkin}},\ }\href@noop {} {\bibfield  {journal} {\bibinfo  {journal} {Phys.
  Rev. B},\ }\textbf {\bibinfo {volume} {64}},\ \bibinfo {pages} {195109}
  (\bibinfo {year} {2001})}\BibitemShut {NoStop}%
\bibitem [{\citenamefont {Jakubczyk}\ \emph {et~al.}(2009)\citenamefont
  {Jakubczyk}, \citenamefont {Metzner},\ and\ \citenamefont
  {Yamase}}]{JakubczykMetzner09}%
  \BibitemOpen
  \bibfield  {author} {\bibinfo {author} {\bibfnamefont {P.}~\bibnamefont
  {Jakubczyk}}, \bibinfo {author} {\bibfnamefont {W.}~\bibnamefont {Metzner}},
  \ and\ \bibinfo {author} {\bibfnamefont {H.}~\bibnamefont {Yamase}},\ }\Doi
  {10.1103/PhysRevLett.103.220602} {\bibfield  {journal} {\bibinfo  {journal}
  {Phys. Rev. Lett.},\ }\textbf {\bibinfo {volume} {103}},\ \bibinfo {pages}
  {220602} (\bibinfo {year} {2009})}\BibitemShut {NoStop}%
\bibitem [{\citenamefont {Ando}\ \emph {et~al.}(2002)\citenamefont {Ando},
  \citenamefont {Segawa}, \citenamefont {Komiya},\ and\ \citenamefont
  {Lavrov}}]{AndoSegawa02}%
  \BibitemOpen
  \bibfield  {author} {\bibinfo {author} {\bibfnamefont {Y.}~\bibnamefont
  {Ando}}, \bibinfo {author} {\bibfnamefont {K.}~\bibnamefont {Segawa}},
  \bibinfo {author} {\bibfnamefont {S.}~\bibnamefont {Komiya}}, \ and\ \bibinfo
  {author} {\bibfnamefont {A.~N.}\ \bibnamefont {Lavrov}},\ }\Doi
  {10.1103/PhysRevLett.88.137005} {\bibfield  {journal} {\bibinfo  {journal}
  {Phys. Rev. Lett.},\ }\textbf {\bibinfo {volume} {88}},\ \bibinfo {pages}
  {137005} (\bibinfo {year} {2002})}\BibitemShut {NoStop}%
\bibitem [{\citenamefont {Lee}\ \emph {et~al.}(2004)\citenamefont {Lee},
  \citenamefont {Segawa}, \citenamefont {Ando},\ and\ \citenamefont
  {Basov}}]{LeeBasov04}%
  \BibitemOpen
  \bibfield  {author} {\bibinfo {author} {\bibfnamefont {Y.~S.}\ \bibnamefont
  {Lee}}, \bibinfo {author} {\bibfnamefont {K.}~\bibnamefont {Segawa}},
  \bibinfo {author} {\bibfnamefont {Y.}~\bibnamefont {Ando}}, \ and\ \bibinfo
  {author} {\bibfnamefont {D.~N.}\ \bibnamefont {Basov}},\ }\Doi
  {10.1103/PhysRevB.70.014518} {\bibfield  {journal} {\bibinfo  {journal}
  {Phys. Rev. B},\ }\textbf {\bibinfo {volume} {70}},\ \bibinfo {pages}
  {014518} (\bibinfo {year} {2004})}\BibitemShut {NoStop}%
\bibitem [{\citenamefont {Daou}\ \emph {et~al.}(2010)\citenamefont {Daou},
  \citenamefont {Chang}, \citenamefont {LeBoeuf}, \citenamefont
  {Cyr-Choiniere}, \citenamefont {Laliberte}, \citenamefont {Doiron-Leyraud},
  \citenamefont {Ramshaw}, \citenamefont {Liang}, \citenamefont {Bonn},
  \citenamefont {Hardy},\ and\ \citenamefont {Taillefer}}]{DaouTaillefer10}%
  \BibitemOpen
  \bibfield  {author} {\bibinfo {author} {\bibfnamefont {R.}~\bibnamefont
  {Daou}}, \bibinfo {author} {\bibfnamefont {J.}~\bibnamefont {Chang}},
  \bibinfo {author} {\bibfnamefont {D.}~\bibnamefont {LeBoeuf}}, \bibinfo
  {author} {\bibfnamefont {O.}~\bibnamefont {Cyr-Choiniere}}, \bibinfo {author}
  {\bibfnamefont {F.}~\bibnamefont {Laliberte}}, \bibinfo {author}
  {\bibfnamefont {N.}~\bibnamefont {Doiron-Leyraud}}, \bibinfo {author}
  {\bibfnamefont {B.~J.}\ \bibnamefont {Ramshaw}}, \bibinfo {author}
  {\bibfnamefont {R.}~\bibnamefont {Liang}}, \bibinfo {author} {\bibfnamefont
  {D.~A.}\ \bibnamefont {Bonn}}, \bibinfo {author} {\bibfnamefont {W.~N.}\
  \bibnamefont {Hardy}}, \ and\ \bibinfo {author} {\bibfnamefont
  {L.}~\bibnamefont {Taillefer}},\ }\Doi {10.1038/nature08716} {\bibfield
  {journal} {\bibinfo  {journal} {Nature},\ }\textbf {\bibinfo {volume}
  {463}},\ \bibinfo {pages} {519} (\bibinfo {year} {2010})}\BibitemShut
  {NoStop}%
\bibitem [{\citenamefont {Matsuda}\ \emph {et~al.}(2008)\citenamefont
  {Matsuda}, \citenamefont {Fujita}, \citenamefont {Wakimoto}, \citenamefont
  {Fernandez-Baca}, \citenamefont {Tranquada},\ and\ \citenamefont
  {Yamada}}]{Matsuda08}%
  \BibitemOpen
  \bibfield  {author} {\bibinfo {author} {\bibfnamefont {M.}~\bibnamefont
  {Matsuda}}, \bibinfo {author} {\bibfnamefont {M.}~\bibnamefont {Fujita}},
  \bibinfo {author} {\bibfnamefont {S.}~\bibnamefont {Wakimoto}}, \bibinfo
  {author} {\bibfnamefont {J.~A.}\ \bibnamefont {Fernandez-Baca}}, \bibinfo
  {author} {\bibfnamefont {J.~M.}\ \bibnamefont {Tranquada}}, \ and\ \bibinfo
  {author} {\bibfnamefont {K.}~\bibnamefont {Yamada}},\ }\Doi
  {10.1103/PhysRevLett.101.197001} {\bibfield  {journal} {\bibinfo  {journal}
  {Phys. Rev. Lett.},\ }\textbf {\bibinfo {volume} {101}},\ \bibinfo {pages}
  {197001} (\bibinfo {year} {2008})}\BibitemShut {NoStop}%
\bibitem [{\citenamefont {Kee}\ \emph {et~al.}(2003)\citenamefont {Kee},
  \citenamefont {Kim},\ and\ \citenamefont {Chung}}]{KeeKimChung03}%
  \BibitemOpen
  \bibfield  {author} {\bibinfo {author} {\bibfnamefont {H.~Y.}\ \bibnamefont
  {Kee}}, \bibinfo {author} {\bibfnamefont {E.~H.}\ \bibnamefont {Kim}}, \ and\
  \bibinfo {author} {\bibfnamefont {C.~H.}\ \bibnamefont {Chung}},\ }\Doi
  {10.1103/PhysRevB.68.245109} {\bibfield  {journal} {\bibinfo  {journal}
  {Phys. Rev. B},\ }\textbf {\bibinfo {volume} {68}},\ \bibinfo {pages}
  {245109} (\bibinfo {year} {2003})}\BibitemShut {NoStop}%
\bibitem [{\citenamefont {Kao}\ and\ \citenamefont {Kee}(2005)}]{KaoKee05}%
  \BibitemOpen
  \bibfield  {author} {\bibinfo {author} {\bibfnamefont {Y.~J.}\ \bibnamefont
  {Kao}}\ and\ \bibinfo {author} {\bibfnamefont {H.~Y.}\ \bibnamefont {Kee}},\
  }\href@noop {} {\bibfield  {journal} {\bibinfo  {journal} {Phys. Rev. B},\
  }\textbf {\bibinfo {volume} {72}} (\bibinfo {year} {2005})}\BibitemShut
  {NoStop}%
\bibitem [{\citenamefont {Kim}\ \emph {et~al.}(2008)\citenamefont {Kim},
  \citenamefont {Lawler}, \citenamefont {Oreto}, \citenamefont {Sachdev},
  \citenamefont {Fradkin},\ and\ \citenamefont {Kivelson}}]{KimKivelson08}%
  \BibitemOpen
  \bibfield  {author} {\bibinfo {author} {\bibfnamefont {E.-A.}\ \bibnamefont
  {Kim}}, \bibinfo {author} {\bibfnamefont {M.~J.}\ \bibnamefont {Lawler}},
  \bibinfo {author} {\bibfnamefont {P.}~\bibnamefont {Oreto}}, \bibinfo
  {author} {\bibfnamefont {S.}~\bibnamefont {Sachdev}}, \bibinfo {author}
  {\bibfnamefont {E.}~\bibnamefont {Fradkin}}, \ and\ \bibinfo {author}
  {\bibfnamefont {S.~A.}\ \bibnamefont {Kivelson}},\ }\Doi
  {10.1103/PhysRevB.77.184514} {\bibfield  {journal} {\bibinfo  {journal}
  {Phys. Rev. B},\ }\textbf {\bibinfo {volume} {77}},\ \bibinfo {pages}
  {184514} (\bibinfo {year} {2008})}\BibitemShut {NoStop}%
\bibitem [{\citenamefont {Yamase}(2009)}]{YamasePRB80}%
  \BibitemOpen
  \bibfield  {author} {\bibinfo {author} {\bibfnamefont {H.}~\bibnamefont
  {Yamase}},\ }\Doi {10.1103/PhysRevB.80.115102} {\bibfield  {journal}
  {\bibinfo  {journal} {Phys. Rev. B},\ }\textbf {\bibinfo {volume} {80}},\
  \bibinfo {pages} {115102} (\bibinfo {year} {2009})}\BibitemShut {NoStop}%
\bibitem [{\citenamefont {Sun}\ \emph {et~al.}(2010)\citenamefont {Sun},
  \citenamefont {Lawler},\ and\ \citenamefont {Kim}}]{SunKim10}%
  \BibitemOpen
  \bibfield  {author} {\bibinfo {author} {\bibfnamefont {K.}~\bibnamefont
  {Sun}}, \bibinfo {author} {\bibfnamefont {M.~J.}\ \bibnamefont {Lawler}}, \
  and\ \bibinfo {author} {\bibfnamefont {E.-A.}\ \bibnamefont {Kim}},\ }\Doi
  {10.1103/PhysRevLett.104.106405} {\bibfield  {journal} {\bibinfo  {journal}
  {Phys. Rev. Lett.},\ }\textbf {\bibinfo {volume} {104}},\ \bibinfo {pages}
  {106405} (\bibinfo {year} {2010})}\BibitemShut {NoStop}%
\bibitem [{\citenamefont {Vojta}(2009)}]{Vojta09}%
  \BibitemOpen
  \bibfield  {author} {\bibinfo {author} {\bibfnamefont {M.}~\bibnamefont
  {Vojta}},\ }\href@noop {} {\bibfield  {journal} {\bibinfo  {journal} {Adv.
  Phys.},\ }\textbf {\bibinfo {volume} {58}},\ \bibinfo {pages} {699} (\bibinfo
  {year} {2009})}\BibitemShut {NoStop}%
\bibitem [{\citenamefont {Dahm}\ and\ \citenamefont
  {Tewordt}(1995)}]{DahmTewordt95a}%
  \BibitemOpen
  \bibfield  {author} {\bibinfo {author} {\bibfnamefont {T.}~\bibnamefont
  {Dahm}}\ and\ \bibinfo {author} {\bibfnamefont {L.}~\bibnamefont {Tewordt}},\
  }\href@noop {} {\bibfield  {journal} {\bibinfo  {journal} {Phys. Rev.
  Lett.},\ }\textbf {\bibinfo {volume} {74}},\ \bibinfo {pages} {793} (\bibinfo
  {year} {1995})}\BibitemShut {NoStop}%
\bibitem [{\citenamefont {Demler}\ and\ \citenamefont
  {Zhang}(1995)}]{DemlerZhang95}%
  \BibitemOpen
  \bibfield  {author} {\bibinfo {author} {\bibfnamefont {E.}~\bibnamefont
  {Demler}}\ and\ \bibinfo {author} {\bibfnamefont {S.~C.}\ \bibnamefont
  {Zhang}},\ }\href@noop {} {\bibfield  {journal} {\bibinfo  {journal} {Phys.
  Rev. Lett.},\ }\textbf {\bibinfo {volume} {75}},\ \bibinfo {pages} {4126}
  (\bibinfo {year} {1995})}\BibitemShut {NoStop}%
\bibitem [{\citenamefont {Bulut}\ and\ \citenamefont
  {Scalapino}(1996)}]{BulutScalapino96}%
  \BibitemOpen
  \bibfield  {author} {\bibinfo {author} {\bibfnamefont {N.}~\bibnamefont
  {Bulut}}\ and\ \bibinfo {author} {\bibfnamefont {D.~J.}\ \bibnamefont
  {Scalapino}},\ }\href@noop {} {\bibfield  {journal} {\bibinfo  {journal}
  {Phys. Rev. B},\ }\textbf {\bibinfo {volume} {53}},\ \bibinfo {pages} {5149}
  (\bibinfo {year} {1996})}\BibitemShut {NoStop}%
\bibitem [{\citenamefont {Millis}\ and\ \citenamefont
  {Monien}(1996)}]{MillisMonien96}%
  \BibitemOpen
  \bibfield  {author} {\bibinfo {author} {\bibfnamefont {A.~J.}\ \bibnamefont
  {Millis}}\ and\ \bibinfo {author} {\bibfnamefont {H.}~\bibnamefont
  {Monien}},\ }\href@noop {} {\bibfield  {journal} {\bibinfo  {journal} {Phys.
  Rev. B},\ }\textbf {\bibinfo {volume} {54}},\ \bibinfo {pages} {16172}
  (\bibinfo {year} {1996})}\BibitemShut {NoStop}%
\bibitem [{\citenamefont {Abanov}\ and\ \citenamefont
  {Chubukov}(1999)}]{AbanovChubukov99}%
  \BibitemOpen
  \bibfield  {author} {\bibinfo {author} {\bibfnamefont {A.}~\bibnamefont
  {Abanov}}\ and\ \bibinfo {author} {\bibfnamefont {A.}~\bibnamefont
  {Chubukov}},\ }\href@noop {} {\bibfield  {journal} {\bibinfo  {journal}
  {Phys. Rev. Lett.},\ }\textbf {\bibinfo {volume} {83}},\ \bibinfo {pages}
  {1652} (\bibinfo {year} {1999})}\BibitemShut {NoStop}%
\bibitem [{\citenamefont {Kao}\ \emph {et~al.}(2000)\citenamefont {Kao},
  \citenamefont {Si},\ and\ \citenamefont {Levin}}]{KaoSi00}%
  \BibitemOpen
  \bibfield  {author} {\bibinfo {author} {\bibfnamefont {Y.}~\bibnamefont
  {Kao}}, \bibinfo {author} {\bibfnamefont {Q.}~\bibnamefont {Si}}, \ and\
  \bibinfo {author} {\bibfnamefont {K.}~\bibnamefont {Levin}},\ }\href@noop {}
  {\bibfield  {journal} {\bibinfo  {journal} {Phys. Rev. B},\ }\textbf
  {\bibinfo {volume} {61}},\ \bibinfo {pages} {11898} (\bibinfo {year}
  {2000})}\BibitemShut {NoStop}%
\bibitem [{\citenamefont {Onufrieva}\ and\ \citenamefont
  {Pfeuty}(2002)}]{Onufrieva02}%
  \BibitemOpen
  \bibfield  {author} {\bibinfo {author} {\bibfnamefont {F.}~\bibnamefont
  {Onufrieva}}\ and\ \bibinfo {author} {\bibfnamefont {P.}~\bibnamefont
  {Pfeuty}},\ }\Doi {10.1103/PhysRevB.65.054515} {\bibfield  {journal}
  {\bibinfo  {journal} {Phys. Rev. B},\ }\textbf {\bibinfo {volume} {65}},\
  \bibinfo {pages} {054515} (\bibinfo {year} {2002})}\BibitemShut {NoStop}%
\bibitem [{\citenamefont {Brinckmann}\ and\ \citenamefont
  {Lee}(2002)}]{BrinckmannLee02}%
  \BibitemOpen
  \bibfield  {author} {\bibinfo {author} {\bibfnamefont {J.}~\bibnamefont
  {Brinckmann}}\ and\ \bibinfo {author} {\bibfnamefont {P.~A.}\ \bibnamefont
  {Lee}},\ }\href@noop {} {\bibfield  {journal} {\bibinfo  {journal} {Phys.
  Rev. B},\ }\textbf {\bibinfo {volume} {65}} (\bibinfo {year}
  {2002})}\BibitemShut {NoStop}%
\bibitem [{\citenamefont {Squires}(1978)}]{Squires}%
  \BibitemOpen
  \bibfield  {author} {\bibinfo {author} {\bibfnamefont {G.~L.}\ \bibnamefont
  {Squires}},\ }\href@noop {} {\emph {\bibinfo {title} {{I}ntroduction to the
  theory of thermal neutron scattering}}}\ (\bibinfo  {publisher} {Dover
  publcations, INC},\ \bibinfo {year} {1978})\BibitemShut {NoStop}%
\bibitem [{\citenamefont {Fong}\ \emph {et~al.}(2000)\citenamefont {Fong},
  \citenamefont {Bourges}, \citenamefont {Sidis}, \citenamefont {Regnault},
  \citenamefont {Bossy}, \citenamefont {Ivanov}, \citenamefont {Milius},
  \citenamefont {Aksay},\ and\ \citenamefont {Keimer}}]{FongKeimer00}%
  \BibitemOpen
  \bibfield  {author} {\bibinfo {author} {\bibfnamefont {H.~F.}\ \bibnamefont
  {Fong}}, \bibinfo {author} {\bibfnamefont {P.}~\bibnamefont {Bourges}},
  \bibinfo {author} {\bibfnamefont {Y.}~\bibnamefont {Sidis}}, \bibinfo
  {author} {\bibfnamefont {L.~P.}\ \bibnamefont {Regnault}}, \bibinfo {author}
  {\bibfnamefont {J.}~\bibnamefont {Bossy}}, \bibinfo {author} {\bibfnamefont
  {A.}~\bibnamefont {Ivanov}}, \bibinfo {author} {\bibfnamefont {D.~L.}\
  \bibnamefont {Milius}}, \bibinfo {author} {\bibfnamefont {I.~A.}\
  \bibnamefont {Aksay}}, \ and\ \bibinfo {author} {\bibfnamefont
  {B.}~\bibnamefont {Keimer}},\ }\href@noop {} {\bibfield  {journal} {\bibinfo
  {journal} {Phys. Rev. B},\ }\textbf {\bibinfo {volume} {61}},\ \bibinfo
  {pages} {14773} (\bibinfo {year} {2000})}\BibitemShut {NoStop}%
\bibitem [{\citenamefont {Tranquada}\ \emph {et~al.}(1989)\citenamefont
  {Tranquada}, \citenamefont {Shirane}, \citenamefont {Keimer}, \citenamefont
  {Shamoto},\ and\ \citenamefont {Sato}}]{TranquadaKeimer89}%
  \BibitemOpen
  \bibfield  {author} {\bibinfo {author} {\bibfnamefont {J.~M.}\ \bibnamefont
  {Tranquada}}, \bibinfo {author} {\bibfnamefont {G.}~\bibnamefont {Shirane}},
  \bibinfo {author} {\bibfnamefont {B.}~\bibnamefont {Keimer}}, \bibinfo
  {author} {\bibfnamefont {S.}~\bibnamefont {Shamoto}}, \ and\ \bibinfo
  {author} {\bibfnamefont {M.}~\bibnamefont {Sato}},\ }\href@noop {} {\bibfield
   {journal} {\bibinfo  {journal} {Phys. Rev. B},\ }\textbf {\bibinfo {volume}
  {40}},\ \bibinfo {pages} {4503} (\bibinfo {year} {1989})}\BibitemShut
  {NoStop}%
\bibitem [{\citenamefont {Woo}\ \emph {et~al.}(2006)\citenamefont {Woo},
  \citenamefont {Dai}, \citenamefont {Hayden}, \citenamefont {Mook},
  \citenamefont {Dahm}, \citenamefont {Scalapino}, \citenamefont {Perring},\
  and\ \citenamefont {Dogan}}]{WooDai06}%
  \BibitemOpen
  \bibfield  {author} {\bibinfo {author} {\bibfnamefont {H.}~\bibnamefont
  {Woo}}, \bibinfo {author} {\bibfnamefont {P.~C.}\ \bibnamefont {Dai}},
  \bibinfo {author} {\bibfnamefont {S.~M.}\ \bibnamefont {Hayden}}, \bibinfo
  {author} {\bibfnamefont {H.~A.}\ \bibnamefont {Mook}}, \bibinfo {author}
  {\bibfnamefont {T.}~\bibnamefont {Dahm}}, \bibinfo {author} {\bibfnamefont
  {D.~J.}\ \bibnamefont {Scalapino}}, \bibinfo {author} {\bibfnamefont {T.~G.}\
  \bibnamefont {Perring}}, \ and\ \bibinfo {author} {\bibfnamefont
  {F.}~\bibnamefont {Dogan}},\ }\Doi {10.1038/nphys394} {\bibfield  {journal}
  {\bibinfo  {journal} {Nat. Phys.},\ }\textbf {\bibinfo {volume} {2}},\
  \bibinfo {pages} {600} (\bibinfo {year} {2006})}\BibitemShut {NoStop}%
\bibitem [{\citenamefont {Shirane}\ \emph {et~al.}(2002)\citenamefont
  {Shirane}, \citenamefont {Shapiro},\ and\ \citenamefont
  {Tranquada}}]{TripleAxisBook}%
  \BibitemOpen
  \bibfield  {author} {\bibinfo {author} {\bibfnamefont {G.}~\bibnamefont
  {Shirane}}, \bibinfo {author} {\bibfnamefont {S.~M.}\ \bibnamefont
  {Shapiro}}, \ and\ \bibinfo {author} {\bibfnamefont {J.~M.}\ \bibnamefont
  {Tranquada}},\ }\href@noop {} {\emph {\bibinfo {title} {{N}eutron scattering
  with a triple axis spectrometer}}}\ (\bibinfo  {publisher} {Cambridge
  University Press},\ \bibinfo {year} {2002})\BibitemShut {NoStop}%
\bibitem [{Note1()}]{Note1}%
  \BibitemOpen
  \bibinfo {note} {For instance, the highest attainable energy in the BZ around
  (0.5, 1.5, 1.7) was $\sim 60$\protect \tmspace +\thinmuskip {.1667em}meV due
  to the onset of parasitic intensity from the direct beam.}\BibitemShut
  {Stop}%
\bibitem [{\citenamefont {Shamoto}\ \emph {et~al.}(1993)\citenamefont
  {Shamoto}, \citenamefont {Sato}, \citenamefont {Tranquada}, \citenamefont
  {Sternlieb},\ and\ \citenamefont {Shirane}}]{ShamotoTranquada93}%
  \BibitemOpen
  \bibfield  {author} {\bibinfo {author} {\bibfnamefont {S.}~\bibnamefont
  {Shamoto}}, \bibinfo {author} {\bibfnamefont {M.}~\bibnamefont {Sato}},
  \bibinfo {author} {\bibfnamefont {J.~M.}\ \bibnamefont {Tranquada}}, \bibinfo
  {author} {\bibfnamefont {B.~J.}\ \bibnamefont {Sternlieb}}, \ and\ \bibinfo
  {author} {\bibfnamefont {G.}~\bibnamefont {Shirane}},\ }\href@noop {}
  {\bibfield  {journal} {\bibinfo  {journal} {Phys. Rev. B},\ }\textbf
  {\bibinfo {volume} {48}},\ \bibinfo {pages} {13817} (\bibinfo {year}
  {1993})}\BibitemShut {NoStop}%
\bibitem [{\citenamefont {Casalta}\ \emph {et~al.}(1994)\citenamefont
  {Casalta}, \citenamefont {Schleger}, \citenamefont {Brecht}, \citenamefont
  {Montfrooij}, \citenamefont {Andersen}, \citenamefont {Lebech}, \citenamefont
  {Schmahl}, \citenamefont {Fuess}, \citenamefont {Liang}, \citenamefont
  {Hardy},\ and\ \citenamefont {Wolf}}]{CasaltaWolf94}%
  \BibitemOpen
  \bibfield  {author} {\bibinfo {author} {\bibfnamefont {H.}~\bibnamefont
  {Casalta}}, \bibinfo {author} {\bibfnamefont {P.}~\bibnamefont {Schleger}},
  \bibinfo {author} {\bibfnamefont {E.}~\bibnamefont {Brecht}}, \bibinfo
  {author} {\bibfnamefont {W.}~\bibnamefont {Montfrooij}}, \bibinfo {author}
  {\bibfnamefont {N.~H.}\ \bibnamefont {Andersen}}, \bibinfo {author}
  {\bibfnamefont {B.}~\bibnamefont {Lebech}}, \bibinfo {author} {\bibfnamefont
  {W.~W.}\ \bibnamefont {Schmahl}}, \bibinfo {author} {\bibfnamefont
  {H.}~\bibnamefont {Fuess}}, \bibinfo {author} {\bibfnamefont {R.~X.}\
  \bibnamefont {Liang}}, \bibinfo {author} {\bibfnamefont {W.~N.}\ \bibnamefont
  {Hardy}}, \ and\ \bibinfo {author} {\bibfnamefont {T.}~\bibnamefont {Wolf}},\
  }\href@noop {} {\bibfield  {journal} {\bibinfo  {journal} {Phys. Rev. B},\
  }\textbf {\bibinfo {volume} {50}},\ \bibinfo {pages} {9688} (\bibinfo {year}
  {1994})}\BibitemShut {NoStop}%
\bibitem [{\citenamefont {Walters}\ \emph {et~al.}(2009)\citenamefont
  {Walters}, \citenamefont {Perring}, \citenamefont {Caux}, \citenamefont
  {Savici}, \citenamefont {Gu}, \citenamefont {Lee}, \citenamefont {Ku},\ and\
  \citenamefont {Zaliznyak}}]{Walters09}%
  \BibitemOpen
  \bibfield  {author} {\bibinfo {author} {\bibfnamefont {A.~C.}\ \bibnamefont
  {Walters}}, \bibinfo {author} {\bibfnamefont {T.~G.}\ \bibnamefont
  {Perring}}, \bibinfo {author} {\bibfnamefont {J.-S.}\ \bibnamefont {Caux}},
  \bibinfo {author} {\bibfnamefont {A.~T.}\ \bibnamefont {Savici}}, \bibinfo
  {author} {\bibfnamefont {G.~D.}\ \bibnamefont {Gu}}, \bibinfo {author}
  {\bibfnamefont {C.-C.}\ \bibnamefont {Lee}}, \bibinfo {author} {\bibfnamefont
  {W.}~\bibnamefont {Ku}}, \ and\ \bibinfo {author} {\bibfnamefont {I.~A.}\
  \bibnamefont {Zaliznyak}},\ }\Doi {10.1038/NPHYS1405} {\bibfield  {journal}
  {\bibinfo  {journal} {Nat. Phys.},\ }\textbf {\bibinfo {volume} {5}},\
  \bibinfo {pages} {867} (\bibinfo {year} {2009})},\ ISSN \bibinfo {issn}
  {1745-2473}\BibitemShut {NoStop}%
\bibitem [{Note2()}]{Note2}%
  \BibitemOpen
  \bibinfo {note} {Rescal5 is available on the web page of the ``Institut
  Laue-Langevin'', www.ill.eu.}\BibitemShut {Stop}%
\bibitem [{\citenamefont {Popovici}(1975)}]{Popovici75}%
  \BibitemOpen
  \bibfield  {author} {\bibinfo {author} {\bibfnamefont {M.}~\bibnamefont
  {Popovici}},\ }\href@noop {} {\bibfield  {journal} {\bibinfo  {journal} {Acta
  Crystallogr. Sect. A},\ }\textbf {\bibinfo {volume} {31}},\ \bibinfo {pages}
  {507} (\bibinfo {year} {1975})}\BibitemShut {NoStop}%
\bibitem [{Note3()}]{Note3}%
  \BibitemOpen
  \bibinfo {note} {A description of the program can be found on the web site
  http://neutron.ujf.cas.cz/restrax.}\BibitemShut {Stop}%
\bibitem [{\citenamefont {Lin}\ \emph {et~al.}(1992)\citenamefont {Lin},
  \citenamefont {Zhou}, \citenamefont {Liang}, \citenamefont {Sch{\" o}nherr},\
  and\ \citenamefont {Bender}}]{LinBender92}%
  \BibitemOpen
  \bibfield  {author} {\bibinfo {author} {\bibfnamefont {C.~T.}\ \bibnamefont
  {Lin}}, \bibinfo {author} {\bibfnamefont {W.}~\bibnamefont {Zhou}}, \bibinfo
  {author} {\bibfnamefont {W.~Y.}\ \bibnamefont {Liang}}, \bibinfo {author}
  {\bibfnamefont {E.}~\bibnamefont {Sch{\" o}nherr}}, \ and\ \bibinfo {author}
  {\bibfnamefont {H.}~\bibnamefont {Bender}},\ }\href@noop {} {\bibfield
  {journal} {\bibinfo  {journal} {Physica C},\ }\textbf {\bibinfo {volume}
  {195}},\ \bibinfo {pages} {291} (\bibinfo {year} {1992})}\BibitemShut
  {NoStop}%
\bibitem [{\citenamefont {Voronkova}\ and\ \citenamefont
  {Wolf}(1993)}]{Voronkova93}%
  \BibitemOpen
  \bibfield  {author} {\bibinfo {author} {\bibfnamefont {V.~I.}\ \bibnamefont
  {Voronkova}}\ and\ \bibinfo {author} {\bibfnamefont {T.}~\bibnamefont
  {Wolf}},\ }\href@noop {} {\bibfield  {journal} {\bibinfo  {journal} {Physica
  C},\ }\textbf {\bibinfo {volume} {218}},\ \bibinfo {pages} {175} (\bibinfo
  {year} {1993})}\BibitemShut {NoStop}%
\bibitem [{\citenamefont {Liang}\ \emph {et~al.}(2006)\citenamefont {Liang},
  \citenamefont {Bonn},\ and\ \citenamefont {Hardy}}]{LiangBonn06}%
  \BibitemOpen
  \bibfield  {author} {\bibinfo {author} {\bibfnamefont {R.~X.}\ \bibnamefont
  {Liang}}, \bibinfo {author} {\bibfnamefont {D.~A.}\ \bibnamefont {Bonn}}, \
  and\ \bibinfo {author} {\bibfnamefont {W.~N.}\ \bibnamefont {Hardy}},\ }\Doi
  {10.1103/PhysRevB.73.180505} {\bibfield  {journal} {\bibinfo  {journal}
  {Phys. Rev. B},\ }\textbf {\bibinfo {volume} {73}},\ \bibinfo {pages}
  {180505} (\bibinfo {year} {2006})}\BibitemShut {NoStop}%
\bibitem [{\citenamefont {Tallon}\ \emph {et~al.}(1995)\citenamefont {Tallon},
  \citenamefont {Bernhard}, \citenamefont {Shaked}, \citenamefont {Hitterman},\
  and\ \citenamefont {Jorgensen}}]{TallonBernhard95}%
  \BibitemOpen
  \bibfield  {author} {\bibinfo {author} {\bibfnamefont {J.~L.}\ \bibnamefont
  {Tallon}}, \bibinfo {author} {\bibfnamefont {C.}~\bibnamefont {Bernhard}},
  \bibinfo {author} {\bibfnamefont {H.}~\bibnamefont {Shaked}}, \bibinfo
  {author} {\bibfnamefont {R.~L.}\ \bibnamefont {Hitterman}}, \ and\ \bibinfo
  {author} {\bibfnamefont {J.~D.}\ \bibnamefont {Jorgensen}},\ }\href@noop {}
  {\bibfield  {journal} {\bibinfo  {journal} {Phys. Rev. B},\ }\textbf
  {\bibinfo {volume} {51}},\ \bibinfo {pages} {12911} (\bibinfo {year}
  {1995})}\BibitemShut {NoStop}%
\bibitem [{\citenamefont {Bourges}\ \emph {et~al.}(2000)\citenamefont
  {Bourges}, \citenamefont {Sidis}, \citenamefont {Fong}, \citenamefont
  {Regnault}, \citenamefont {Bossy}, \citenamefont {Ivanov},\ and\
  \citenamefont {Keimer}}]{BourgesFong00}%
  \BibitemOpen
  \bibfield  {author} {\bibinfo {author} {\bibfnamefont {P.}~\bibnamefont
  {Bourges}}, \bibinfo {author} {\bibfnamefont {Y.}~\bibnamefont {Sidis}},
  \bibinfo {author} {\bibfnamefont {H.~F.}\ \bibnamefont {Fong}}, \bibinfo
  {author} {\bibfnamefont {L.-P.}\ \bibnamefont {Regnault}}, \bibinfo {author}
  {\bibfnamefont {J.}~\bibnamefont {Bossy}}, \bibinfo {author} {\bibfnamefont
  {A.}~\bibnamefont {Ivanov}}, \ and\ \bibinfo {author} {\bibfnamefont
  {B.}~\bibnamefont {Keimer}},\ }\href@noop {} {\bibfield  {journal} {\bibinfo
  {journal} {Science},\ }\textbf {\bibinfo {volume} {288}},\ \bibinfo {pages}
  {1234} (\bibinfo {year} {2000})}\BibitemShut {NoStop}%
\bibitem [{\citenamefont {Dai}\ \emph {et~al.}(1998)\citenamefont {Dai},
  \citenamefont {Mook},\ and\ \citenamefont {Dogan}}]{DaiMook98}%
  \BibitemOpen
  \bibfield  {author} {\bibinfo {author} {\bibfnamefont {P.~C.}\ \bibnamefont
  {Dai}}, \bibinfo {author} {\bibfnamefont {H.~A.}\ \bibnamefont {Mook}}, \
  and\ \bibinfo {author} {\bibfnamefont {F.}~\bibnamefont {Dogan}},\
  }\href@noop {} {\bibfield  {journal} {\bibinfo  {journal} {Phys. Rev.
  Lett.},\ }\textbf {\bibinfo {volume} {80}},\ \bibinfo {pages} {1738}
  (\bibinfo {year} {1998})}\BibitemShut {NoStop}%
\bibitem [{\citenamefont {Mook}\ \emph {et~al.}(1998)\citenamefont {Mook},
  \citenamefont {Dai}, \citenamefont {Hayden}, \citenamefont {Aeppli},
  \citenamefont {Perring},\ and\ \citenamefont {Do{\v g}an}}]{MookDai98}%
  \BibitemOpen
  \bibfield  {author} {\bibinfo {author} {\bibfnamefont {H.~A.}\ \bibnamefont
  {Mook}}, \bibinfo {author} {\bibfnamefont {P.~C.}\ \bibnamefont {Dai}},
  \bibinfo {author} {\bibfnamefont {S.~M.}\ \bibnamefont {Hayden}}, \bibinfo
  {author} {\bibfnamefont {G.}~\bibnamefont {Aeppli}}, \bibinfo {author}
  {\bibfnamefont {T.~G.}\ \bibnamefont {Perring}}, \ and\ \bibinfo {author}
  {\bibfnamefont {F.}~\bibnamefont {Do{\v g}an}},\ }\href@noop {} {\bibfield
  {journal} {\bibinfo  {journal} {Nature},\ }\textbf {\bibinfo {volume}
  {395}},\ \bibinfo {pages} {580} (\bibinfo {year} {1998})}\BibitemShut
  {NoStop}%
\bibitem [{\citenamefont {Sidis}\ \emph {et~al.}(2007)\citenamefont {Sidis},
  \citenamefont {Pailh{\` e}s}, \citenamefont {Hinkov}, \citenamefont {Fauqu{\'
  e}}, \citenamefont {Ulrich}, \citenamefont {Capogna}, \citenamefont {Ivanov},
  \citenamefont {Regnault}, \citenamefont {Keimer},\ and\ \citenamefont
  {Bourges}}]{SidisBourges07}%
  \BibitemOpen
  \bibfield  {author} {\bibinfo {author} {\bibfnamefont {Y.}~\bibnamefont
  {Sidis}}, \bibinfo {author} {\bibfnamefont {S.}~\bibnamefont {Pailh{\` e}s}},
  \bibinfo {author} {\bibfnamefont {V.}~\bibnamefont {Hinkov}}, \bibinfo
  {author} {\bibfnamefont {B.}~\bibnamefont {Fauqu{\' e}}}, \bibinfo {author}
  {\bibfnamefont {C.}~\bibnamefont {Ulrich}}, \bibinfo {author} {\bibfnamefont
  {L.}~\bibnamefont {Capogna}}, \bibinfo {author} {\bibfnamefont
  {A.}~\bibnamefont {Ivanov}}, \bibinfo {author} {\bibfnamefont {L.-P.}\
  \bibnamefont {Regnault}}, \bibinfo {author} {\bibfnamefont {B.}~\bibnamefont
  {Keimer}}, \ and\ \bibinfo {author} {\bibfnamefont {P.}~\bibnamefont
  {Bourges}},\ }\Doi {10.1016/j.crhy.2007.07.003} {\bibfield  {journal}
  {\bibinfo  {journal} {C. R. Physique},\ }\textbf {\bibinfo {volume} {8}},\
  \bibinfo {pages} {745} (\bibinfo {year} {2007})}\BibitemShut {NoStop}%
\bibitem [{\citenamefont {Rossat-Mignod}\ \emph {et~al.}(1991)\citenamefont
  {Rossat-Mignod}, \citenamefont {Regnault}, \citenamefont {Vettier},
  \citenamefont {Bourges}, \citenamefont {Burlet}, \citenamefont {Bossy},
  \citenamefont {Henry},\ and\ \citenamefont
  {Lapertot}}]{Rossat-MignodBourges91}%
  \BibitemOpen
  \bibfield  {author} {\bibinfo {author} {\bibfnamefont {J.}~\bibnamefont
  {Rossat-Mignod}}, \bibinfo {author} {\bibfnamefont {L.~P.}\ \bibnamefont
  {Regnault}}, \bibinfo {author} {\bibfnamefont {C.}~\bibnamefont {Vettier}},
  \bibinfo {author} {\bibfnamefont {P.}~\bibnamefont {Bourges}}, \bibinfo
  {author} {\bibfnamefont {P.}~\bibnamefont {Burlet}}, \bibinfo {author}
  {\bibfnamefont {J.}~\bibnamefont {Bossy}}, \bibinfo {author} {\bibfnamefont
  {J.~Y.}\ \bibnamefont {Henry}}, \ and\ \bibinfo {author} {\bibfnamefont
  {G.}~\bibnamefont {Lapertot}},\ }\href@noop {} {\bibfield  {journal}
  {\bibinfo  {journal} {Physica C},\ }\textbf {\bibinfo {volume} {185}},\
  \bibinfo {pages} {86} (\bibinfo {year} {1991})}\BibitemShut {NoStop}%
\bibitem [{\citenamefont {Fong}\ \emph {et~al.}(1999)\citenamefont {Fong},
  \citenamefont {Bourges}, \citenamefont {Sidis}, \citenamefont {Regnault},
  \citenamefont {Ivanov}, \citenamefont {Gu}, \citenamefont {Koshizuka},\ and\
  \citenamefont {Keimer}}]{FongKeimer99}%
  \BibitemOpen
  \bibfield  {author} {\bibinfo {author} {\bibfnamefont {H.~F.}\ \bibnamefont
  {Fong}}, \bibinfo {author} {\bibfnamefont {P.}~\bibnamefont {Bourges}},
  \bibinfo {author} {\bibfnamefont {Y.}~\bibnamefont {Sidis}}, \bibinfo
  {author} {\bibfnamefont {L.-P.}\ \bibnamefont {Regnault}}, \bibinfo {author}
  {\bibfnamefont {A.}~\bibnamefont {Ivanov}}, \bibinfo {author} {\bibfnamefont
  {G.~D.}\ \bibnamefont {Gu}}, \bibinfo {author} {\bibfnamefont
  {N.}~\bibnamefont {Koshizuka}}, \ and\ \bibinfo {author} {\bibfnamefont
  {B.}~\bibnamefont {Keimer}},\ }\href@noop {} {\bibfield  {journal} {\bibinfo
  {journal} {Nature},\ }\textbf {\bibinfo {volume} {398}},\ \bibinfo {pages}
  {588} (\bibinfo {year} {1999})}\BibitemShut {NoStop}%
\bibitem [{\citenamefont {He}\ \emph {et~al.}(2002)\citenamefont {He},
  \citenamefont {Bourges}, \citenamefont {Sidis}, \citenamefont {Ulrich},
  \citenamefont {Regnault}, \citenamefont {Pailh{\` e}s}, \citenamefont
  {Berzigiarova}, \citenamefont {Kolesnikov},\ and\ \citenamefont
  {Keimer}}]{HeBourges02}%
  \BibitemOpen
  \bibfield  {author} {\bibinfo {author} {\bibfnamefont {H.}~\bibnamefont
  {He}}, \bibinfo {author} {\bibfnamefont {P.}~\bibnamefont {Bourges}},
  \bibinfo {author} {\bibfnamefont {Y.}~\bibnamefont {Sidis}}, \bibinfo
  {author} {\bibfnamefont {C.}~\bibnamefont {Ulrich}}, \bibinfo {author}
  {\bibfnamefont {L.-P.}\ \bibnamefont {Regnault}}, \bibinfo {author}
  {\bibfnamefont {S.}~\bibnamefont {Pailh{\` e}s}}, \bibinfo {author}
  {\bibfnamefont {N.~S.}\ \bibnamefont {Berzigiarova}}, \bibinfo {author}
  {\bibfnamefont {N.~N.}\ \bibnamefont {Kolesnikov}}, \ and\ \bibinfo {author}
  {\bibfnamefont {B.}~\bibnamefont {Keimer}},\ }\Doi {10.1126/science.1067877}
  {\bibfield  {journal} {\bibinfo  {journal} {Science},\ }\textbf {\bibinfo
  {volume} {295}},\ \bibinfo {pages} {1045} (\bibinfo {year}
  {2002})}\BibitemShut {NoStop}%
\bibitem [{\citenamefont {Yu}\ \emph {et~al.}(2010)\citenamefont {Yu},
  \citenamefont {Li}, \citenamefont {Motoyama}, \citenamefont {Zhao},
  \citenamefont {Barisic}, \citenamefont {Cho}, \citenamefont {Bourges},
  \citenamefont {Hradil}, \citenamefont {Mole},\ and\ \citenamefont
  {Greven}}]{YuGreven10}%
  \BibitemOpen
  \bibfield  {author} {\bibinfo {author} {\bibfnamefont {G.}~\bibnamefont
  {Yu}}, \bibinfo {author} {\bibfnamefont {Y.}~\bibnamefont {Li}}, \bibinfo
  {author} {\bibfnamefont {E.~M.}\ \bibnamefont {Motoyama}}, \bibinfo {author}
  {\bibfnamefont {X.}~\bibnamefont {Zhao}}, \bibinfo {author} {\bibfnamefont
  {N.}~\bibnamefont {Barisic}}, \bibinfo {author} {\bibfnamefont
  {Y.}~\bibnamefont {Cho}}, \bibinfo {author} {\bibfnamefont {P.}~\bibnamefont
  {Bourges}}, \bibinfo {author} {\bibfnamefont {K.}~\bibnamefont {Hradil}},
  \bibinfo {author} {\bibfnamefont {R.~A.}\ \bibnamefont {Mole}}, \ and\
  \bibinfo {author} {\bibfnamefont {M.}~\bibnamefont {Greven}},\ }\Doi
  {10.1103/PhysRevB.81.064518} {\bibfield  {journal} {\bibinfo  {journal}
  {PHYSICAL REVIEW B},\ }\textbf {\bibinfo {volume} {81}},\ \bibinfo {pages}
  {064518} (\bibinfo {year} {2010})}\BibitemShut {NoStop}%
\bibitem [{\citenamefont {Zhao}\ \emph {et~al.}(2007)\citenamefont {Zhao},
  \citenamefont {Dai}, \citenamefont {Li}, \citenamefont {Freeman},
  \citenamefont {Onose},\ and\ \citenamefont {Tokura}}]{ZhaoDai07}%
  \BibitemOpen
  \bibfield  {author} {\bibinfo {author} {\bibfnamefont {J.}~\bibnamefont
  {Zhao}}, \bibinfo {author} {\bibfnamefont {P.}~\bibnamefont {Dai}}, \bibinfo
  {author} {\bibfnamefont {S.}~\bibnamefont {Li}}, \bibinfo {author}
  {\bibfnamefont {P.~G.}\ \bibnamefont {Freeman}}, \bibinfo {author}
  {\bibfnamefont {Y.}~\bibnamefont {Onose}}, \ and\ \bibinfo {author}
  {\bibfnamefont {Y.}~\bibnamefont {Tokura}},\ }\Doi
  {10.1103/PhysRevLett.99.017001} {\bibfield  {journal} {\bibinfo  {journal}
  {Phys. Rev. Lett.},\ }\textbf {\bibinfo {volume} {99}},\ \bibinfo {pages}
  {017001} (\bibinfo {year} {2007})}\BibitemShut {NoStop}%
\bibitem [{\citenamefont {Yu}\ \emph {et~al.}(2008)\citenamefont {Yu},
  \citenamefont {Li}, \citenamefont {Motoyama}, \citenamefont {Hradil},
  \citenamefont {Mole},\ and\ \citenamefont {Greven}}]{YuGreven08}%
  \BibitemOpen
  \bibfield  {author} {\bibinfo {author} {\bibfnamefont {G.}~\bibnamefont
  {Yu}}, \bibinfo {author} {\bibfnamefont {Y.}~\bibnamefont {Li}}, \bibinfo
  {author} {\bibfnamefont {E.~M.}\ \bibnamefont {Motoyama}}, \bibinfo {author}
  {\bibfnamefont {K.}~\bibnamefont {Hradil}}, \bibinfo {author} {\bibfnamefont
  {R.~A.}\ \bibnamefont {Mole}}, \ and\ \bibinfo {author} {\bibfnamefont
  {M.}~\bibnamefont {Greven}},\ }\href@noop {} { (\bibinfo {year} {2008})},\
  \bibinfo {note} {unpublished; preprint at arXiv:0803.3250.}\BibitemShut
  {Stop}%
\bibitem [{\citenamefont {Stock}\ \emph {et~al.}(2005)\citenamefont {Stock},
  \citenamefont {Buyers}, \citenamefont {Cowley}, \citenamefont {Clegg},
  \citenamefont {Coldea}, \citenamefont {Frost}, \citenamefont {Liang},
  \citenamefont {Peets}, \citenamefont {Bonn}, \citenamefont {Hardy},\ and\
  \citenamefont {Birgeneau}}]{StockBuyers05}%
  \BibitemOpen
  \bibfield  {author} {\bibinfo {author} {\bibfnamefont {C.}~\bibnamefont
  {Stock}}, \bibinfo {author} {\bibfnamefont {W.~J.~L.}\ \bibnamefont
  {Buyers}}, \bibinfo {author} {\bibfnamefont {R.~A.}\ \bibnamefont {Cowley}},
  \bibinfo {author} {\bibfnamefont {P.~S.}\ \bibnamefont {Clegg}}, \bibinfo
  {author} {\bibfnamefont {R.}~\bibnamefont {Coldea}}, \bibinfo {author}
  {\bibfnamefont {C.~D.}\ \bibnamefont {Frost}}, \bibinfo {author}
  {\bibfnamefont {R.}~\bibnamefont {Liang}}, \bibinfo {author} {\bibfnamefont
  {D.}~\bibnamefont {Peets}}, \bibinfo {author} {\bibfnamefont
  {D.}~\bibnamefont {Bonn}}, \bibinfo {author} {\bibfnamefont {W.~N.}\
  \bibnamefont {Hardy}}, \ and\ \bibinfo {author} {\bibfnamefont {R.~J.}\
  \bibnamefont {Birgeneau}},\ }\Doi {10.1103/PhysRevB.71.024522} {\bibfield
  {journal} {\bibinfo  {journal} {Phys. Rev. B},\ }\textbf {\bibinfo {volume}
  {71}},\ \bibinfo {pages} {024522} (\bibinfo {year} {2005})}\BibitemShut
  {NoStop}%
\bibitem [{\citenamefont {Stock}\ \emph {et~al.}(2004)\citenamefont {Stock},
  \citenamefont {Buyers}, \citenamefont {Liang}, \citenamefont {Peets},
  \citenamefont {Tun}, \citenamefont {Bonn}, \citenamefont {Hardy},\ and\
  \citenamefont {Birgeneau}}]{StockBuyers04}%
  \BibitemOpen
  \bibfield  {author} {\bibinfo {author} {\bibfnamefont {C.}~\bibnamefont
  {Stock}}, \bibinfo {author} {\bibfnamefont {W.~J.~L.}\ \bibnamefont
  {Buyers}}, \bibinfo {author} {\bibfnamefont {R.}~\bibnamefont {Liang}},
  \bibinfo {author} {\bibfnamefont {D.}~\bibnamefont {Peets}}, \bibinfo
  {author} {\bibfnamefont {Z.}~\bibnamefont {Tun}}, \bibinfo {author}
  {\bibfnamefont {D.}~\bibnamefont {Bonn}}, \bibinfo {author} {\bibfnamefont
  {W.~N.}\ \bibnamefont {Hardy}}, \ and\ \bibinfo {author} {\bibfnamefont
  {R.~J.}\ \bibnamefont {Birgeneau}},\ }\Doi {10.1103/PhysRevB.69.014502}
  {\bibfield  {journal} {\bibinfo  {journal} {Phys. Rev. B},\ }\textbf
  {\bibinfo {volume} {69}},\ \bibinfo {pages} {014502} (\bibinfo {year}
  {2004})}\BibitemShut {NoStop}%
\bibitem [{\citenamefont {Mook}\ \emph {et~al.}(2000)\citenamefont {Mook},
  \citenamefont {Dai}, \citenamefont {Do{\v g}an},\ and\ \citenamefont
  {Hunt}}]{Mook00}%
  \BibitemOpen
  \bibfield  {author} {\bibinfo {author} {\bibfnamefont {H.~A.}\ \bibnamefont
  {Mook}}, \bibinfo {author} {\bibfnamefont {P.~C.}\ \bibnamefont {Dai}},
  \bibinfo {author} {\bibfnamefont {F.}~\bibnamefont {Do{\v g}an}}, \ and\
  \bibinfo {author} {\bibfnamefont {R.~D.}\ \bibnamefont {Hunt}},\ }\href@noop
  {} {\bibfield  {journal} {\bibinfo  {journal} {Nature},\ }\textbf {\bibinfo
  {volume} {404}},\ \bibinfo {pages} {729} (\bibinfo {year}
  {2000})}\BibitemShut {NoStop}%
\bibitem [{\citenamefont {Xu}\ \emph {et~al.}(2007)\citenamefont {Xu},
  \citenamefont {Tranquada}, \citenamefont {Perring}, \citenamefont {Gu},
  \citenamefont {Fujita},\ and\ \citenamefont {Yamada}}]{XuTranquada07}%
  \BibitemOpen
  \bibfield  {author} {\bibinfo {author} {\bibfnamefont {G.}~\bibnamefont
  {Xu}}, \bibinfo {author} {\bibfnamefont {J.~M.}\ \bibnamefont {Tranquada}},
  \bibinfo {author} {\bibfnamefont {T.~G.}\ \bibnamefont {Perring}}, \bibinfo
  {author} {\bibfnamefont {G.~D.}\ \bibnamefont {Gu}}, \bibinfo {author}
  {\bibfnamefont {M.}~\bibnamefont {Fujita}}, \ and\ \bibinfo {author}
  {\bibfnamefont {K.}~\bibnamefont {Yamada}},\ }\Doi
  {10.1103/PhysRevB.76.014508} {\bibfield  {journal} {\bibinfo  {journal}
  {Phys. Rev. B},\ }\textbf {\bibinfo {volume} {76}},\ \bibinfo {pages}
  {014508} (\bibinfo {year} {2007})}\BibitemShut {NoStop}%
\bibitem [{\citenamefont {Lipscombe}\ \emph {et~al.}(2009)\citenamefont
  {Lipscombe}, \citenamefont {Vignolle}, \citenamefont {Perring}, \citenamefont
  {Frost},\ and\ \citenamefont {Hayden}}]{Lipscombe09}%
  \BibitemOpen
  \bibfield  {author} {\bibinfo {author} {\bibfnamefont {O.~J.}\ \bibnamefont
  {Lipscombe}}, \bibinfo {author} {\bibfnamefont {B.}~\bibnamefont {Vignolle}},
  \bibinfo {author} {\bibfnamefont {T.~G.}\ \bibnamefont {Perring}}, \bibinfo
  {author} {\bibfnamefont {C.~D.}\ \bibnamefont {Frost}}, \ and\ \bibinfo
  {author} {\bibfnamefont {S.~M.}\ \bibnamefont {Hayden}},\ }\Doi
  {10.1103/PhysRevLett.102.167002} {\bibfield  {journal} {\bibinfo  {journal}
  {Phys. Rev. Lett.},\ }\textbf {\bibinfo {volume} {102}},\ \bibinfo {pages}
  {167002} (\bibinfo {year} {2009})}\BibitemShut {NoStop}%
\bibitem [{\citenamefont {Hayden}\ \emph {et~al.}(1996)\citenamefont {Hayden},
  \citenamefont {Aeppli}, \citenamefont {Perring}, \citenamefont {Mook},\ and\
  \citenamefont {Do{\v g}an}}]{HaydenAeppli96}%
  \BibitemOpen
  \bibfield  {author} {\bibinfo {author} {\bibfnamefont {S.~M.}\ \bibnamefont
  {Hayden}}, \bibinfo {author} {\bibfnamefont {G.}~\bibnamefont {Aeppli}},
  \bibinfo {author} {\bibfnamefont {T.~G.}\ \bibnamefont {Perring}}, \bibinfo
  {author} {\bibfnamefont {H.~A.}\ \bibnamefont {Mook}}, \ and\ \bibinfo
  {author} {\bibfnamefont {F.}~\bibnamefont {Do{\v g}an}},\ }\href@noop {}
  {\bibfield  {journal} {\bibinfo  {journal} {Phys. Rev. B},\ }\textbf
  {\bibinfo {volume} {54}},\ \bibinfo {pages} {R6905} (\bibinfo {year}
  {1996})}\BibitemShut {NoStop}%
\bibitem [{Note4()}]{Note4}%
  \BibitemOpen
  \bibinfo {note} {At the instrumental conditions used up to $\sim \unhbox
  \voidb@x \hbox {$\omega _{\protect \rm r}$}$, the \unhbox \voidb@x \hbox
  {$\protect \mathbf Q$}-resolution is lower than twice the hatched
  region.}\BibitemShut {Stop}%
\bibitem [{Note5()}]{Note5}%
  \BibitemOpen
  \bibinfo {note} {One should keep in mind that \unhbox \voidb@x \hbox {$\chi
  ''(\protect \mathbf Q,\omega )$}\ is required to be an odd function of
  $\omega $, which is not the case for the lower branch, eq. \protect \textup
  {\hbox {\mathsurround \z@ \protect \normalfont (\ignorespaces \ref
  {eq:lowerSCBranch}\unskip \@@italiccorr )}}. Thus, this formula is only valid
  for $\omega >0$, and would need to be extended correspondingly if one were
  interested in negative $\omega $. This is straight-forward due to the
  superconducting gap.}\BibitemShut {Stop}%
\bibitem [{Note6()}]{Note6}%
  \BibitemOpen
  \bibinfo {note} {In the online supplementary material of Ref. \protect
  \rev@citealpnum {DahmHinkov09}, where we first described the model, there is
  a typographical error in the equation corresponding to eq. \protect \textup
  {\hbox {\mathsurround \z@ \protect \normalfont (\ignorespaces \ref
  {eq:gammal}\unskip \@@italiccorr )}} here.}\BibitemShut {Stop}%
\bibitem [{Note7()}]{Note7}%
  \BibitemOpen
  \bibinfo {note} {Note that this gap function does not strictly go to 0 for
  $\omega \rightarrow 0$, but to a value $<10^{-3}$. This is orders of
  magnitude below the experimental error and good enough for all practical
  purposes. For a strictly analytical continuation to negative $\omega $,
  however, one wood need to slightly modify \protect \textup {\hbox
  {\mathsurround \z@ \protect \normalfont (\ignorespaces \ref {eq:SCgap}\unskip
  \@@italiccorr )}}.}\BibitemShut {Stop}%
\bibitem [{Note8()}]{Note8}%
  \BibitemOpen
  \bibinfo {note} {At 100\protect \tmspace +\thinmuskip {.1667em}meV, only data
  for the even mode are reported in Ref. \protect \rev@citealpnum
  {FongKeimer00}. However, since this energy is far above the even gap of $\sim
  50$\protect \tmspace +\thinmuskip {.1667em}meV, even and odd values should
  not be substantially different.}\BibitemShut {Stop}%
\bibitem [{\citenamefont {Morr}\ and\ \citenamefont
  {Pines}(1998)}]{MorrPines98}%
  \BibitemOpen
  \bibfield  {author} {\bibinfo {author} {\bibfnamefont {D.}~\bibnamefont
  {Morr}}\ and\ \bibinfo {author} {\bibfnamefont {D.}~\bibnamefont {Pines}},\
  }\href@noop {} {\bibfield  {journal} {\bibinfo  {journal} {Phys. Rev.
  Lett.},\ }\textbf {\bibinfo {volume} {81}},\ \bibinfo {pages} {1086}
  (\bibinfo {year} {1998})}\BibitemShut {NoStop}%
\bibitem [{\citenamefont {Sega}\ \emph {et~al.}(2003)\citenamefont {Sega},
  \citenamefont {Prelov{\v s}ek},\ and\ \citenamefont {Bon{\v
  c}a}}]{SegaBonca03}%
  \BibitemOpen
  \bibfield  {author} {\bibinfo {author} {\bibfnamefont {I.}~\bibnamefont
  {Sega}}, \bibinfo {author} {\bibfnamefont {P.}~\bibnamefont {Prelov{\v
  s}ek}}, \ and\ \bibinfo {author} {\bibfnamefont {J.}~\bibnamefont {Bon{\v
  c}a}},\ }\Doi {10.1103/PhysRevB.68.054524} {\bibfield  {journal} {\bibinfo
  {journal} {Phys. Rev. B},\ }\textbf {\bibinfo {volume} {68}},\ \bibinfo
  {pages} {054524} (\bibinfo {year} {2003})}\BibitemShut {NoStop}%
\bibitem [{Note9()}]{Note9}%
  \BibitemOpen
  \bibinfo {note} {One might object that when comparing superconducting and
  normal state data, one also compares different temperatures (5\protect
  \tmspace +\thinmuskip {.1667em}K and 70\protect \tmspace +\thinmuskip
  {.1667em}K), which might be the reason for the different incommensurabilities
  and the topology change. Fig.~\ref {fig:tempDep}, however, shows that the
  superconducting signal geometry is already fully developed at 50\protect
  \tmspace +\thinmuskip {.1667em}K and hardly changes upon further
  cooling.}\BibitemShut {Stop}%
\bibitem [{\citenamefont {Zhou}\ and\ \citenamefont {Li}(2004)}]{ZhouLi04}%
  \BibitemOpen
  \bibfield  {author} {\bibinfo {author} {\bibfnamefont {T.}~\bibnamefont
  {Zhou}}\ and\ \bibinfo {author} {\bibfnamefont {J.~X.}\ \bibnamefont {Li}},\
  }\Doi {10.1103/PhysRevB.69.224514} {\bibfield  {journal} {\bibinfo  {journal}
  {Phys. Rev. B},\ }\textbf {\bibinfo {volume} {69}},\ \bibinfo {pages}
  {224514} (\bibinfo {year} {2004})}\BibitemShut {NoStop}%
\bibitem [{\citenamefont {Eremin}\ and\ \citenamefont
  {Manske}(2005)}]{EreminManske05}%
  \BibitemOpen
  \bibfield  {author} {\bibinfo {author} {\bibfnamefont {I.}~\bibnamefont
  {Eremin}}\ and\ \bibinfo {author} {\bibfnamefont {D.}~\bibnamefont
  {Manske}},\ }\Doi {10.1103/PhysRevLett.94.067006} {\bibfield  {journal}
  {\bibinfo  {journal} {Phys. Rev. Lett.},\ }\textbf {\bibinfo {volume} {94}},\
  \bibinfo {pages} {067006} (\bibinfo {year} {2005})}\BibitemShut {NoStop}%
\bibitem [{\citenamefont {Schnyder}\ \emph {et~al.}(2006)\citenamefont
  {Schnyder}, \citenamefont {Manske}, \citenamefont {Mudry},\ and\
  \citenamefont {Sigrist}}]{SchnyderManske06}%
  \BibitemOpen
  \bibfield  {author} {\bibinfo {author} {\bibfnamefont {A.}~\bibnamefont
  {Schnyder}}, \bibinfo {author} {\bibfnamefont {D.}~\bibnamefont {Manske}},
  \bibinfo {author} {\bibfnamefont {C.}~\bibnamefont {Mudry}}, \ and\ \bibinfo
  {author} {\bibfnamefont {M.}~\bibnamefont {Sigrist}},\ }\Doi
  {10.1103/PhysRevB.73.224523} {\bibfield  {journal} {\bibinfo  {journal}
  {Phys. Rev. B},\ }\textbf {\bibinfo {volume} {73}},\ \bibinfo {pages}
  {224523} (\bibinfo {year} {2006})}\BibitemShut {NoStop}%
\bibitem [{\citenamefont {Kr{\"u}ger}\ \emph {et~al.}(2007)\citenamefont
  {Kr{\"u}ger}, \citenamefont {Wilson}, \citenamefont {Shan}, \citenamefont
  {Li}, \citenamefont {Huang}, \citenamefont {Wen}, \citenamefont {Zhang},
  \citenamefont {Dai},\ and\ \citenamefont {Zaanen}}]{Kruger07}%
  \BibitemOpen
  \bibfield  {author} {\bibinfo {author} {\bibfnamefont {F.}~\bibnamefont
  {Kr{\"u}ger}}, \bibinfo {author} {\bibfnamefont {S.~D.}\ \bibnamefont
  {Wilson}}, \bibinfo {author} {\bibfnamefont {L.}~\bibnamefont {Shan}},
  \bibinfo {author} {\bibfnamefont {S.}~\bibnamefont {Li}}, \bibinfo {author}
  {\bibfnamefont {Y.}~\bibnamefont {Huang}}, \bibinfo {author} {\bibfnamefont
  {H.-H.}\ \bibnamefont {Wen}}, \bibinfo {author} {\bibfnamefont {S.-C.}\
  \bibnamefont {Zhang}}, \bibinfo {author} {\bibfnamefont {P.}~\bibnamefont
  {Dai}}, \ and\ \bibinfo {author} {\bibfnamefont {J.}~\bibnamefont {Zaanen}},\
  }\Doi {10.1103/PhysRevB.76.094506} {\bibfield  {journal} {\bibinfo  {journal}
  {Phys. Rev. B},\ }\textbf {\bibinfo {volume} {76}},\ \bibinfo {pages}
  {094506} (\bibinfo {year} {2007})}\BibitemShut {NoStop}%
\bibitem [{\citenamefont {Sun}\ \emph {et~al.}(2004)\citenamefont {Sun},
  \citenamefont {Segawa},\ and\ \citenamefont {Ando}}]{SunAndo04}%
  \BibitemOpen
  \bibfield  {author} {\bibinfo {author} {\bibfnamefont {X.}~\bibnamefont
  {Sun}}, \bibinfo {author} {\bibfnamefont {K.}~\bibnamefont {Segawa}}, \ and\
  \bibinfo {author} {\bibfnamefont {Y.}~\bibnamefont {Ando}},\ }\Doi
  {10.1103/PhysRevLett.93.107001} {\bibfield  {journal} {\bibinfo  {journal}
  {Phys. Rev. Lett.},\ }\textbf {\bibinfo {volume} {93}},\ \bibinfo {pages}
  {107001} (\bibinfo {year} {2004})}\BibitemShut {NoStop}%
\bibitem [{\citenamefont {Hwang}\ \emph {et~al.}(2004)\citenamefont {Hwang},
  \citenamefont {Timusk},\ and\ \citenamefont {Gu}}]{HwangTimusk04}%
  \BibitemOpen
  \bibfield  {author} {\bibinfo {author} {\bibfnamefont {J.}~\bibnamefont
  {Hwang}}, \bibinfo {author} {\bibfnamefont {T.}~\bibnamefont {Timusk}}, \
  and\ \bibinfo {author} {\bibfnamefont {G.}~\bibnamefont {Gu}},\ }\Doi
  {10.1038/nature02347} {\bibfield  {journal} {\bibinfo  {journal} {Nature},\
  }\textbf {\bibinfo {volume} {427}},\ \bibinfo {pages} {714} (\bibinfo {year}
  {2004})}\BibitemShut {NoStop}%
\end{thebibliography}%

\end{document}